\documentclass[%
 reprint,
superscriptaddress,
 amsmath,amssymb,
 aps,
prb,
]{revtex4-2}

\usepackage{graphicx}
\usepackage{dcolumn}
\usepackage{bm}
\usepackage{mathrsfs}
\usepackage{mhchem}
\usepackage{siunitx}

\usepackage{color}

\usepackage{subfigure}
\usepackage{float}	
\usepackage{xspace}
\usepackage{url}

\begin{document}

\preprint{APS/123-QED}

\title{Formation of random singlets in the nanocrystalline quasi-one-dimensional spin-1/2 antiferromagnet \ce{Sr21Bi8Cu2(CO3)2O41}}

\author{Yanbo~Guo}
\author{Xinzhe~Hu}
\altaffiliation[Present address: ]{Institute of Physics, Chinese Academy of Sciences, Beijing 100190, China.}
\affiliation{Department of Physics, University of Florida, Gainesville, Florida 32611-8440, USA}
\author{Hasan~Siddiquee}
\altaffiliation[Present address: ]{Nokia Bell Labs, New Providence, New Jersey 07974, USA.}
\author{Kapila Kumarasinghe}
\affiliation{Department of Physics, University of Central Florida, Orlando, Florida 32816, USA}
\author{Swapnil~M.~Yadav}
\altaffiliation[Present address: ]{Walmart Global Tech, Sunnyvale, California 94086, USA.}
\affiliation{Department of Physics, University of Florida, Gainesville, Florida 32611-8440, USA}
\author{Eun Sang~Choi}
\affiliation{National High Magnetic Field Laboratory, Tallahassee, Florida 32310, USA}
\author{Yasuyuki~Nakajima}
\affiliation{Department of Physics, University of Central Florida, Orlando, Florida 32816, USA}
\author{Yasumasa~Takano}
\affiliation{Department of Physics, University of Florida, Gainesville, Florida 32611-8440, USA}
\date{\today}

\begin{abstract}
 Induced by quenched disorder, random-singlet states occur in a variety of low-dimensional spin-1/2 antiferromagnets, some of them candidates for quantum spin liquids. Here we report 
measurements of the specific heat, magnetization, and magnetic susceptibility of nanocrystalline \ce{Sr21Bi8Cu2(CO3)2O41}, a quasi-one-dimensional spin-1/2 antiferromagnet with alternating bonds. The results uncover the predominant presence of random-singlet spin pairs in this material, with a logarithmic probability distribution, \(P(J)\), of the renormalized, emergent exchange interaction, \(J\), in zero magnetic field and \(P(J)\) proportional to \(1/J\) in magnetic fields. We postulate that these unexpected \(J\) dependences, in contrast to the usual \(P(J)\propto 1/J^{\gamma}\) with 0\,\(<\)\,\(\gamma\)\,\(<\)\,1, and possibly also the dichotomy, arise from the finite size of the nanocrystals. 
Scaling functions for the specific heat and magnetization
reproduce our magnetocaloric-effect data, with no adjustable parameters. 
\end{abstract}

\maketitle

The size of a system under study plays profound roles in phase transitions and in related phenomena such as complex networks, swarming,
and jamming.
The most beautiful example of this is found in the superfluid transition of \ce{^4He} in restricted geometries, where
predictions from finite-size scaling \cite{PrivmanBook} have been tested in detail \cite{Gasparini}.

In many complex networks, such as airline networks, social networks, networks formed by linked webpages, and even the Internet itself, the node degree \(k\)---the number of direct links between a node and other nodes in the network---has a scale-free, power-law probability distribution. However, some network-growth models that reproduce such a distribution do so only before the system grows very large \cite{Krapivsky,Falkenberg}, whereas a finite-size-scaling analysis of
many naturally occurring complex networks reveals that the underlying scale-free
distribution of \(k\) is often obscured by the finite sizes of 
the systems \cite{Serafino}.

By contrast, in swarming of flying insects, an active system consisting of
self-propelled ``particles,"
the correlation of the seemingly erratic velocity fluctuations of two insects is scale-free, with the correlation length proportional to the system size, strongly suggesting that the system organizes itself so as to remain nearly critical \cite{swarm}.
In jamming of hard spheres and disks, on the other hand, as the system size increases, the probability distribution of contact forces approaches the infinite-size limit much more rapidly than that of interparticle gaps, indicating the presence of two vastly different coherence lengths \cite{Charbonneau}.

Motivated by these examples, here we present evidence for finite-size efffects on a random-singlet phase occurring in a nanocrystalline sample of the spin-1/2 alternating-bond antiferromagnet \ce{Sr21Bi8Cu2(CO3)2O41}. By means of specific-heat, magnetization, magnetic-susceptibility, and magnetocaloric-effect measurements, we show that the probability distribution \(P(J)\) of the renormalized, emergent exchange interaction \(J\) for the random singlets in this material takes on a hereto unobserved, logarithmic form in zero magnetic field and yet another unusual form in magnetic fields. 
  
Originally discovered in organic quasi-one-dimensional spin-1/2 antiferromagnets \cite{Bulaevskii,Azevedo,Duffy,Bozler,Sanny} and dilute phosphorus-doped Si \cite{Andres,Paalanen86,Paalanen88,Lakner,Hirsch}, the random-singlet (RS) state is a renormalized ground state consisting of spins paired into entangled singlets with a wide distribution of pair sizes \cite{MaRS,DasguptaRS,BhattRS,FisherRS,WesterbergRS}. Evidence has recently been mounding that RS states exist, induced by quenched disorder, also in non-dilute 
low-dimensional spin-1/2 antiferromagnets \cite{Kitagawa,KimchiPRX,Kimchi,LeeRS,Huang_noScaling,wangNMR,MurayamaRSexpt2,LiuRSexpt,MurayamaRSexpt,KenneyRS,ChoiRS,SongRSexpt,KunduRS,Nguyen,BaekRS,DoExpt,BahramiRS,Volkov,HongRS,YoonRS,KhatuaRS}, some of them candidates for quantum spin liquids. 

In an RS phase, low-temperature thermodynamic properties directly mirror \(P(J)\) \cite{KimchiPRX,Kimchi}. 
The magnetic specific heat \(C\)
is given by \(C/T\propto P(T)\) as a function of temperature \(T\) at zero magnetic field, the magnetization \(M\) by \(M/H\propto P(T)\) at low magnetic fields \(H\), and the magnetization as a function of \(H\)
by \(M(H)\propto\int_{0}^{H}P(J)dJ\) at zero temperature. The specific heat in magnetic fields obeys \(C/T\propto P(H)(T/H)^q\) \cite{Kimchi}, where the integer exponent \(q\) depends on the 
symmetry of the Dzyaloshinskii-Moriya (DM) interaction and is zero if the interaction is absent \cite{KimchiPRX}. At nonzero temperatures, the magnetization acquires an additional term, becoming \(M(H,T)\propto\int_{0}^{H}P(J)dJ[1-m_1 (T/H)^{2+q}]\), 
as dictated by a Maxwell relation \cite{Kimchi}. Here \(m_1\) is a constant.

A polycrystalline sample of \ce{Sr21Bi8Cu2(CO3)2O41} was 
synthesized by solid-state reaction of \ce{Bi2O3}, \ce{SrCO3}, \ce{SrO}, \ce{SrO2}, and \ce{CuO} in a 4:1:10:10:2 ratio \cite{Malo}. The powder X-ray diffraction pattern, shown
in the Supplemental Material \cite{suppl}, reveals that the crystallites are on average as small as \(26\)\(\pm\)1\,nm.

Specific-heat measurements were performed using a custom-built relaxation calorimeter 
at temperatures between \SI{59}{mK} and \SI{9.9}{K}, in zero magnetic field and fields up to \SI{14}{T}, 
and another custom-built relaxation calorimeter at temperatures between 7.4 and \SI{21}{K} in zero field.
Magnetocaloric-effect measurements were made using the first calorimeter at temperatures between 0.2 and \SI{1.6}{K} in magnetic fields up to \SI{4.5}{T}.  
The magnetization was measured at temperatures down to \SI{1.8}{K} and in magnetic fields up to \SI{7}{T} using a commercial SQUID magnetometer, and at two temperatures, 1.8 and \SI{4.2}{K}, in fields up to \SI{35}{T} using a vibrating-sample magnetometer (VSM). The magnetic susceptibility was obtained from the SQUID-magnetometer data taken at 0.1 and \SI{3.0}{T}. 

\begin{figure}[t]
	\centering
	\includegraphics[width=\linewidth]{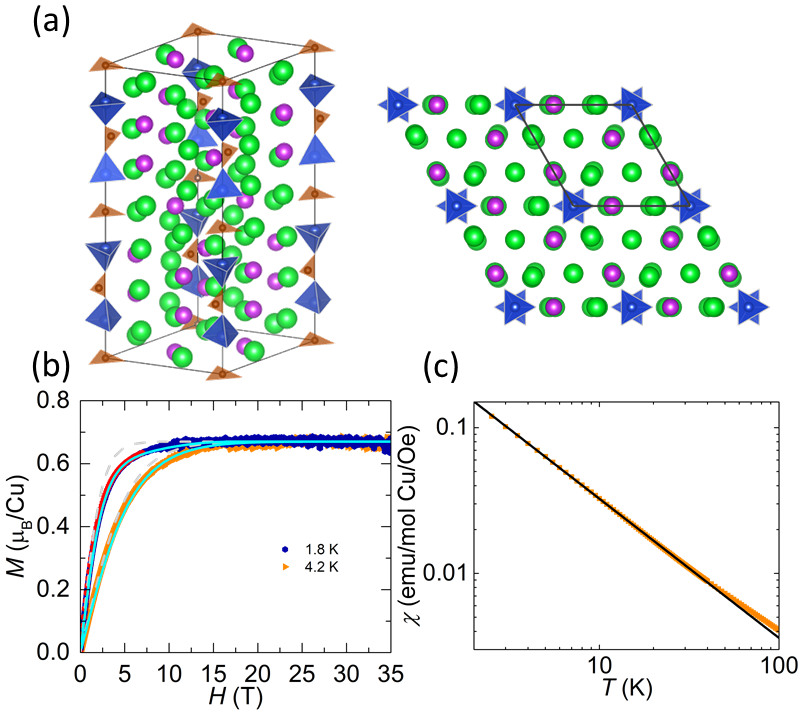}
	\caption{\label{fig:fig1}(a) Crystal structure of \ce{Sr21Bi8Cu2(CO3)2O41}, with thin lines indicating a unit cell. In the left panel the crystallographic \(c\) axis is vertical; the right panel is a view along the \(c\) axis. Blue: \ce{CuO4} tetrahedra; brown: CO\(_3^{2-}\) triangles; purple spheres: Bi; and green spheres: Sr. Oxygen atoms are not shown, but some of them are implicitly indicated as \ce{CuO4} tetrahedra and CO\(_3^{2-}\) triangles.
		The panel was produced using \textsc{vesta} \cite{VESTA}. (b) Magnetization as a function of the magnetic field 
		up to \SI{35}{T} measured using a VSM. The data have been scaled at \SI{1.8}{K} to SQUID-magnetometer data, shown in red. Dashed lines indicate the magnetization of 
		free spins. Solid lines are calculations using a spin-pair approximation with \(P(J)\propto 1/J\), 
		as described in the text. (c) Magnetic susceptibility calculated from SQUID-magnetometer data taken at \SI{0.1}{T}. Solid line is a power-law fit described in the text. }
\end{figure}

\ce{Sr21Bi8Cu2(CO3)2O41} has a hexagonal crystal structure as shown in Fig.~\ref{fig:fig1}(a), with space group \(P6_3/mcm\) (No. 193) and lattice parameters \(a=b=10.0966(3)\)\,\AA\xspace and \(c=26.3762(5)\)\,\AA\xspace \cite{Malo}. Along the crystallographic \(c\) axis, spin-1/2 Cu\(^{2+}\) ions alternate with carbonate groups, with two Cu-Cu distances: 6.44 and \SI{6.75}{\AA}. The shorter Cu-Cu distance involves a carbonate group coordinating to the two Cu\(^{2+}\) ions and two Bi\(^{5+}\) ions that form a pair of edge-shared \ce{BiO6} octahedra, whereas the longer distance is separated by a carbonate group whose oxygens do not form coordinate bonds with metal ions. The former carbonate group has three possible orientations, and the latter carbonate group has two possible orientations. The spin chains along the \(c\) axis form a triangular lattice, with the lattice constant \(a=b\),
in which the interchain exchange pathway is interrupted by a Sr\(^{2+}\) ion. These features suggest that the material is an excellent realization of an alternating-bond spin-1/2 antiferromagnet with bond strengths \(J_1\) for the shorter Cu-Cu bond and \(J_2\) for the longer Cu-Cu bond, with an extremely weak, geometrically frustrated interchain exchange. 
Due to the lack of inversion symmetry in the exchange pathways, 
DM interactions are expected to be present. Also keep in mind that the average size of the crystallites, \(26\)\(\pm\)1\,nm, in our sample implies that each spin chain may contain on average only about 40 Cu\(^{2+}\) spins. 

A Curie-Weiss fit of the magnetic susceptibility of our sample \cite{suppl} gives a Curie constant of  
0.493(7)\,emu\,K/mol\,Cu/Oe and a Curie-Weiss temperature, \(\Theta\), of \(-20.8\)\(\pm\)1.3\,K. The Curie constant yields a \(g\) factor, \(g=2.29(2)\), about 6\% smaller than 2.42 implicitly reported as an effective magnetic moment in Ref.~\cite{Malo}, which also reports \(\Theta\)\,=\,\SI{-28}{K}, about 35\% larger than our value. From our \(\Theta\) we estimate that \(J_1+J_2=83\)\(\pm\)5\,K, with a corresponding saturation magnetic field of 54\(\pm\)4\,T. 

The magnetic susceptibility of a spin-1/2 alternating-bond Heisenberg antiferromagnet should exhibit a broad maximum, indicative of an onset of short-range correlations, near the temperature \(T=(J_1+J_2)/2k_{\mathrm{B}}\), where \(k_{\mathrm{B}}\) is the Boltzmann constant 
\cite{Bulaevskii2,DuffyBarr,BonnerChi}. By contrast, the 
susceptibility of our sample only increases monotonically with decreasing temperature.
Even more remarkable are the VSM data, shown in Fig.~\ref{fig:fig1}(b). Firstly, since the ground state of a spin-1/2 alternating-bond Heisenberg antiferromagnet is dimerized, with an energy gap, 
its magnetization at zero temperature should remain zero up to a critical field, about \((J_1-J_2^2/J_1)/g\mu_{\mathrm{B}}\), where \(\mu_{\mathrm{B}}\) is the Bohr magneton \cite{Bulaevskii2,DuffyBarr,Diederix}. Instead, the measured magnetization 
increases linearly from zero with increasing magnetic field even at \SI{1.8}{K}, a temperature an order of magnitude lower than \(\Theta\), and becomes flat at about \SI{13}{T}, much lower than the expected saturation field, 54\(\pm\)4\,T. This indicates that the system is gapless and spins behave as if interactions between them are much smaller than \(k_{\mathrm{B}}\Theta\), in fact not that differently from free spins (see dashed lines in the figure\textit{}). Secondly, the value of the flat part is only \(0.672(3)\mu_{\mathrm{B}}\)/Cu, indicating that 41.3(8)\% of spins do not contribute at all to the magnetization up to \SI{35}{T}, the maximum field of our measurements.

Taken together, these features point to 58.7(8)\% of spins in our sample forming random singlets, most likely induced by quenched disorder associated with the 
nanoscale crystallite sizes. However, since the ground state of an alternating-bond antiferromagnet is dimerized, 
with an energy gap, it should be robust against quenched disorder. The disorder must exceed a critical value, 
which is larger than required to close the gap, in order for the system to enter an RS phase \cite{Hyman}. Our sample evidently satisfies this condition. Incidentally, on assumption that those spins that do not form RSs behave as a non-renormalized alternating-bond antiferromagnet, the complete absence of their contribution to magnetization up to \SI{35}{T} indicates that \(J_1\) is at least \(65\)\(\pm\)4\,K and \(J_2\) at most \(19\)\(\pm\)1\,K \cite{Bulaevskii2,DuffyBarr,Diederix}. It is suggestive that this upper estimate of \(J_2\) is comparable to the cutoff energies \(J_0\) and \(J_H\), which is extracted later.

\begin{figure}[t]
	\centering
	\includegraphics[width=\linewidth]{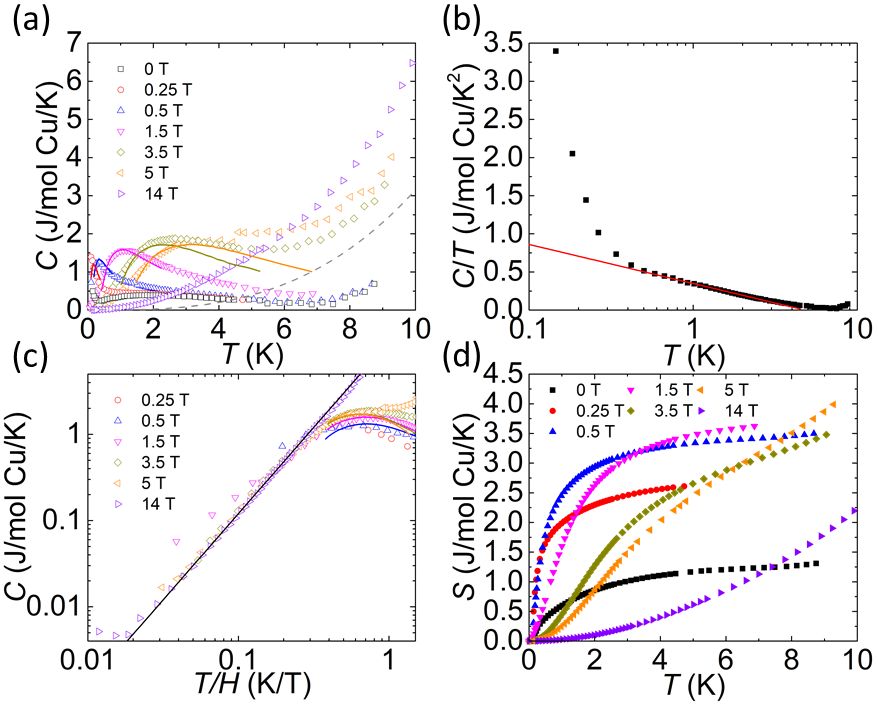}
	\caption{\label{fig:fig2}
		(a) Magnetic specific heat as a function of temperature. 
		Solid lines are calculated from the scaled magnetization data presented in Fig.~\ref{fig:fig3}(c). Phonon contribution, measured by using a separate calorimeter and indicated by a dashed line, has been subtracted. (b) Zero-field magnetic specific heat divided by temperature, with line showing a fit to \(C/T\propto \ln(T_0/T)\), where 
	\(T_0=\mathrm{\SI{4.52(9)}{K}}\). (c) Magnetic specific heat in magnetic fields as a function of \(T/H\), demonstrating scaling. Curved lines are calculated from the scaled magnetization data presented in Fig.~\ref{fig:fig3}(c). Calculated line is not shown for \SI{0.25}{T}, since it deviates from the data as can be seen in panel (a). (d) Magnetic entropy obtained from the specific heat.} 
\end{figure}

The magnetic specific heat of our sample is shown in Fig.\,\ref{fig:fig2}. At zero field, the magnetic specific heat is proportional to \(T\ln(T_0/T)\) at temperatures between 
0.5 and \SI{1.9}{K}, as shown in panel~(b), with \(T_0=\mathrm{\SI{4.52(9)}{K}}\). Fitting the logarithmic part of the specific heat to a spin-pair approximation \cite{Andres,SinghAppendix,KimchiPRX}, which treats an RS state as a product state comprising independent spin pairs that experience only intra-pair exchanges, we find that 21.0(3)\% of spins form RS pairs with \(P(J)=\ln(J_0/J)\), with the cutoff energy \(J_0=\mathrm{\SI{14.6(3)}{K}}\) \cite{suppl}. The small excess at temperatures above about \SI{2}{K} can be ascribed to a Schottky anomaly due to 1.5(2)\% of spins forming singlet pairs with \(J=\mathrm{\SI{14.5(5)}{K}}\), which within the combined uncertainty is the same as \(J_0\). The larger excess at temperatures below about \SI{0.5}{K} is proportional to \(1/T^2\), suggesting that it is not a precursor to magnetic ordering, but far too large to be a high-temperature tail of a nuclear-quadrupolar Schottky specific heat \cite{suppl}. Assuming that \(58.7(8)-21.0(3)-1.5(2)=36.2(9)\)\% of spins form singlet pairs responsible for this excess, the best fit is obtained for a \(\delta\)-function distribution of those singlet pairs with \(J=\mathrm{\SI{0.150(2)}{K}}\) \cite{suppl}.
This abundance of singlet pairs with small \(J\) is very likely due to the finite lengths of the spin chains.

 We propose that the unusual logarithmic \(P(J)\) 
 is also a finite-size effect. Alternatively, or additionally, the logarithmic specific heat may indicate that the system is close to the quantum critical point between the RS phase and a Griffiths phase---the intermediate gapless phase between the RS phase and the dimerized phase \cite{Hyman}. 
 Logarithmically singular specific heat has been observed in a variety of strongly correlated systems, including non-Fermi-liquid \(f\)-electron compounds \cite{Gegenwart, Lohneysen, Fritsch, NatPhysZhao, YangCL, Tripathi, NatureShen, Singh, Freeman, Bernal, Weber, Bauer, Huy, Matsumoto}, ferromagnetic 3\(d\)-electron metals \cite{Nicklas, YangJ, Huang, Brando, Moroni, Waki, Brando2, Wu, Schoop, Jia}, cuprate superconductors 
\cite{Michon, Girod}, \({\mathrm{Na}}_{x}{\mathrm{CoO}}_{2}\) \cite{Bruhwiler, Balicas, Okamoto}, and \ce{Sr3Ru2O7} \cite{PNASQCrit11, PRBQCrit18}, all near a quantum critical point. 

Magnetic entropy, \(S\), is obtained from the magnetic specific heat, \(C\), by integrating \(C/T\), and shown in Fig.~\ref{fig:fig2} (d). At and above \SI{0.5}{T}, where \(C\) can be extrapolated accurately 
to zero temperature, \(C/T\) is integrated from zero temperature, whereas at zero field and \SI{0.25}{T} the integration is from the lowest temperature, resulting in an underestimate of the entropy. Since \(58.7(8)\)\% of spins form random singlets according to the VSM magnetization data, we expect a high-temperature limiting entropy of \(S\)\,=\,\(0.587(8)R\ln 2\)\,=\,3.38(5)\,J/mol\,Cu/K for those spins, where \(R\) is the gas constant. The data at 0.5 and \SI{1.5}{T} tend to become flat roughly at this value, providing further evidence for the presence of an RS state in our sample. Such a trend is not observed at 3.5 and \SI{5}{T}, suggesting that dividing spins into two groups---random singlets and others---may become less valid at relatively high temperatures as the field increases.  

\begin{figure*}[t]
	\centering
	\includegraphics[width=\textwidth]{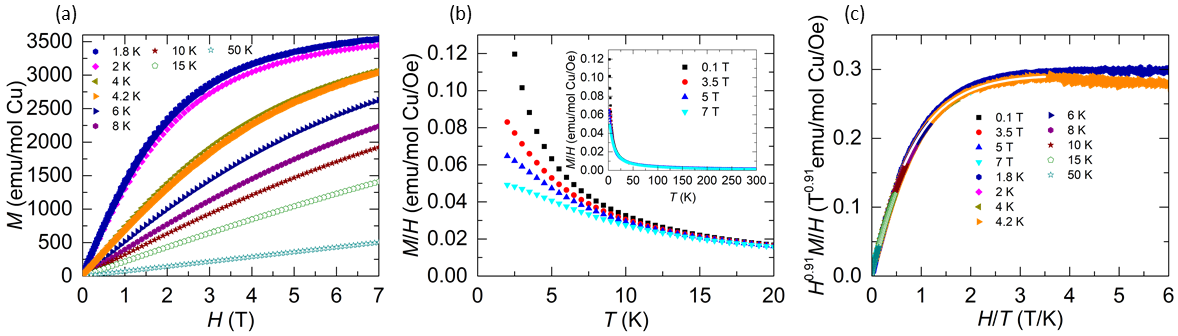}
	\caption{\label{fig:fig3}
		(a) Magnetization as a function of the magnetic field. 
		(b) Magnetization as a function of temperature. 
		(c) Scaling collapse of the magnetization data shown in panels (a) and (b), and in Fig.~\ref{fig:fig1}(b). White line is a power-polynomial fit, as described in the text.
	}
\end{figure*}

We now turn to the magnetic susceptibility, \(M/H\), measured at \SI{0.1}{T} and shown in Fig.~\ref{fig:fig1}(c). The best fit to the power law \(M/H \propto 1/T^{\gamma}\), shown as a solid line, at temperatures below \SI{30}{K} yields the exponent \(\gamma=0.955(1)\). This may appear to contradict the logarithmic \(P(J)\) extracted from the zero-field specific-heat data, since \(M/H\) of random singlets should be proportional to \(P(T)\) near zero field as described earlier, but the discrepancy is most likely due to the relatively high temperatures of the susceptibility data. The susceptibility was measured at temperatures above \SI{2}{K}, whereas the logarithmic specific heat is found below \SI{2}{K}. Indeed, a calculation using the spin-pair approximation suggests that the power-law susceptibility 
is consistent with the probability distribution of \(J\) obtained from the zero-field specific-heat data \cite{suppl}.

As shown in Fig.~\ref{fig:fig2}(c), the magnetic specific heat at 3.5, 5, and \SI{14}{T} scale as \(C\propto (T/H)^2\) for \(T/H\) less than about \SI{0.3}{K/T}, implying that \(q=1\) and \(P(J)\propto 1/J\), neither the unusual \(P(J)=\ln (J_0/J)\) found in zero field nor the usual fractional power law \(P(J)\propto 1/J^{\gamma}\) with \(0 < \gamma < 1\). 
The nonzero value of the integer exponent \(q\) indicates the presence of DM interactions, as expected, and the specific value, \(q\)\,=1, is consistent with our sample being a polycrystal \cite{Kimchi}. 
We postulate that the emergence of the yet another unusual \(P(J)\), proportional to \(1/J\), is also a finite-size effect. 

 The probability distribution \(P(J)\propto 1/J\) must have a low-energy cutoff \(\lambda\), in addition to a high-energy cutoff \(J_H\), since the integral \(\int 1/JdJ\) diverges, logarithmically. We find the cutoffs by fitting the VSM data, shown in Fig.~\ref{fig:fig1}(b), to a calculation using the spin-pair approximation, although the validity of the approximation is questionable at high fields where spins are highly polarized. 
 The best fit, shown as solid lines in Fig.~\ref{fig:fig1}(b), yields \(J_H=22.0(5)\)\,K and \(\lambda=10^{-6}J_H\), which is practically zero \cite{suppl}. The cutoff \(J_H\) differs somewhat from \(J_0\), found from the specific heat at zero field, but the difference may be non-essential, given the limitation of the spin-pair approximation.

In Fig.~\ref{fig:fig2}(c), the 0.5 T data point deviates from the \((T/H)^2\) fit at the lowest temperature, which corresponds to \(T/H \sim 0.2\,\mathrm{K}/\mathrm{T}\). Moreover, the \SI{1.5}{T} data do so for \(T/H\) below about \SI{0.13}{K/T}, where they are roughly proportional to \(T^{1.38(4)}\). This non-integer exponent cannot be explained in terms of random singlets, suggesting that fields up to \SI{1.5}{T} may be in a crossover regime where \(P(J)\) changes from the logarithmic \(J\) dependence at zero field to the \(1/J\) dependence at and above \SI{3.5}{T}.

The figure also shows that at all fields, except possibly at \SI{14}{T}, scaling breaks down when \(T/H > 0.3\)\,K/T, which in the dimensionless unit is \(T/H > 0.2\). One might expect that the data should also 
scale in this region, since \(C/T\propto T^{-\gamma}[1+c_0(H/T)^2]\) when \(T/H\gg 1\) and if \(P(J)\propto 1/J^{\gamma}\) 
\cite{Kimchi}. Here \(c_0\) is a constant. This relation, in conjunction with \(P(J)\propto 1/J\), would lead to \(C\propto 1+c_0(H/T)^2\)---scaling of \(C\) with \(T/H\). The reasoning behind the expectation is that since \(C\) scales with \(T/H\) when \(T/H\ll 1\) as well as \(T/H\gg 1\), it is natural that it will over the entire region of \(T/H\) \cite{Kimchi}. However, the first relation, which holds only when \(P(J)\) obeys a power law, consists of \emph{zero-field} low-temperature specific heat plus 
a term dictated by the Maxwell relation \((\partial M/\partial T)_H=(\partial S/\partial H)_T\). Since in the present case \(P(J)\) at \SI{3.5}{T} and above differs from \(P(J)\) at zero field, the first relation is invalid. It is therefore not 
surprising that the scaling breaks down already in the intermediate region of \(T/H\).

We now show that surprisingly, the breakdown of the scaling of the specific heat for \(T/H > 0.3\)\,K/T is accompanied by, and consistent with, a scaling of magnetization. Magnetization data over the entire temperature range of our measurements are shown in Figs.~\ref{fig:fig3}(a) and (b) as a function of field and temperature, 
along with a scaling plot in panel (c) which includes the VSM data shown in Fig.~\ref{fig:fig1}(b). The choice of \(H^{\gamma}M/H\) for the vertical axis of the scaling plot 
is motivated by that this quantity scales 
with \(T/H\) when \(P(J)\propto 1/J^{\gamma}\) \cite{Kimchi}. 
The best scaling is obtained with \(\gamma\)\,=\,0.91(2), but the scaling of course becomes poor for \(H/T\)\,\(>\)\,\SI{3.5}{T/K}, where \(M\) starts to approach a constant [see Fig.~\ref{fig:fig1}(b)]. 
The closeness of this \(\gamma\) value to 
0.955(1) found from the power-law temperature dependence of \(M/H\) at \SI{0.1}{T} may seem to imply that \(P(J)\propto 1/J^{\gamma}\) holds for our sample. 
However, as discussed earlier, this interpretation is incorrect.

We fit the scaled plot to the fifth-order power polynomial \(\sum^5_{n=0} a_n(H/T)^n\) for \(H/T\) between 0.66 and \SI{3.5}{T/K}, excluding the VSM data, which are noisier than the SQUID-magnetometer data. From this fit, the Maxwell relation leads to a formula for magnetic specific heat in magnetic fields: \(C(H,T)=C(0,T)+\sum_{n=1}^5 a_n \frac{n(n+1)}{n+2-\gamma} \frac{H^{n+2-\gamma}}{T^{n+1}}\), where we take the zero-field magnetic specific heat \(C(0,T)\) from Fig.~\ref{fig:fig2}(b). Calculated specific heat using this expression is compared with the actual data in Figs.~\ref{fig:fig2}(a) and (c) for fields up to \SI{5}{T}, excluding \SI{14}{T} for which we do not have magnetization data other than the VSM data at 1.8 and \SI{4.2}{K}. Overall, the agreement between the calculation and data is satisfactory, given that much of the calculation covers temperatures below \SI{1.8}{K}, the region in which there is no magnetization data. The agreement shows that the breakdown of scaling of the specific heat for \(T/H\)\,\(>\)\,0.3\,K/T is thermodynamically consistent with the scaling of the magnetization. We emphasize, however, that the scaling of the magnetization does not at all imply that \(P(J)\propto 1/J^{\gamma}\) with \(\gamma\)\,=\,0.91(2).

The Maxwell relation also dictates that the scaling of the specific heat at and above \SI{3.5}{T} found in the low-temperature region, \(T/H\)\,\(<\)\,0.3\,K/T, must have a counterpart in the magnetization obeying the relation \(M(H,T)=M(H,0)[1-m_1(T/H)^{2+q}]\) with \(q\)\,=\,1, as described earlier slightly implicitly. 
However, this prediction cannot be tested against the present magnetization data set, which does not extend below \(T/H=0.4\)\,K/T even at \SI{5}{T} and is scarce at \SI{14}{T}.

Finally, the magnetocaloric effect, measured at temperatures between 0.2 and \SI{1.6}{K}, exhibit excellent consistency with the magnetization and specific-heat data \cite{suppl}. 
When \(H/T\) is less than \SI{3.5}{T/K}, the magnetocaloric-effect data are in agreement, with no adjustable parameters, with the power-polynomial fit to the scaled magnetization data in the corresponding region of \(H/T\), 
shown in Fig.~\ref{fig:fig3}; for larger \(H/T\), they are in agreement with the scaled specific heat proportional to \((T/H)^2\), shown in Fig.~\ref{fig:fig2}(c), again with no adjustable parameters. These agreements provide additional evidence that only the spins that form RSs contribute to thermodynamic properties in the temperature and field regions of our experiment, except temperatures above about \SI{100}{K} where all spins contribute to the magnetic susceptibility.

	In summary, we have presented evidence that a random-singlet phase forms in nanocrystalline \ce{Sr21Bi8Cu2(CO3)2O41}, a spin-1/2 alternating-bond antiferromagnet. The probability distribution of the emergent exchange interaction, \(J\), for singlet spin pairs is logarithmic at zero field, and is replaced 
	by a \(1/J\) dependence 
	in magnetic fields. We postulate that these unusual forms of the probability distribution, and possibly the dichotomy, are finite-size effects.
	 
\begin{acknowledgments}
	We thank M.J. Beazley, D.R. Talham, and A. Trucco for valuable discussions, and A. Bangura and G.E. Jones for assistance. A portion of this work was performed at the National High Magnetic Field Laboratory (NHMFL), which is supported by the NSF Cooperative Agreement No. DMR-1644779 and the State of Florida. Y.G., X.H., and Y.T. were supported by the NHMFL UCGP Program, and H.S. and Y.N. by the NSF CAREER DMR-1944975 and a start-up fund from the University of Central Florida. 
\end{acknowledgments}

\bibliography{Ref}

\begin{thebibliography}{83}%
\makeatletter
\providecommand \@ifxundefined [1]{%
 \@ifx{#1\undefined}
}%
\providecommand \@ifnum [1]{%
 \ifnum #1\expandafter \@firstoftwo
 \else \expandafter \@secondoftwo
 \fi
}%
\providecommand \@ifx [1]{%
 \ifx #1\expandafter \@firstoftwo
 \else \expandafter \@secondoftwo
 \fi
}%
\providecommand \natexlab [1]{#1}%
\providecommand \enquote  [1]{``#1''}%
\providecommand \bibnamefont  [1]{#1}%
\providecommand \bibfnamefont [1]{#1}%
\providecommand \citenamefont [1]{#1}%
\providecommand \href@noop [0]{\@secondoftwo}%
\providecommand \href [0]{\begingroup \@sanitize@url \@href}%
\providecommand \@href[1]{\@@startlink{#1}\@@href}%
\providecommand \@@href[1]{\endgroup#1\@@endlink}%
\providecommand \@sanitize@url [0]{\catcode `\\12\catcode `\$12\catcode
  `\&12\catcode `\#12\catcode `\^12\catcode `\_12\catcode `\%12\relax}%
\providecommand \@@startlink[1]{}%
\providecommand \@@endlink[0]{}%
\providecommand \url  [0]{\begingroup\@sanitize@url \@url }%
\providecommand \@url [1]{\endgroup\@href {#1}{\urlprefix }}%
\providecommand \urlprefix  [0]{URL }%
\providecommand \Eprint [0]{\href }%
\providecommand \doibase [0]{https://doi.org/}%
\providecommand \selectlanguage [0]{\@gobble}%
\providecommand \bibinfo  [0]{\@secondoftwo}%
\providecommand \bibfield  [0]{\@secondoftwo}%
\providecommand \translation [1]{[#1]}%
\providecommand \BibitemOpen [0]{}%
\providecommand \bibitemStop [0]{}%
\providecommand \bibitemNoStop [0]{.\EOS\space}%
\providecommand \EOS [0]{\spacefactor3000\relax}%
\providecommand \BibitemShut  [1]{\csname bibitem#1\endcsname}%
\let\auto@bib@innerbib\@empty
\bibitem [{\citenamefont {Privman}(1990)}]{PrivmanBook}%
  \BibitemOpen
  \bibfield  {author} {\bibinfo {author} {\bibfnamefont {V.}~\bibnamefont
  {Privman}},\ }\href {https://doi.org/10.1142/1011} {\emph {\bibinfo {title}
  {Finite Size Scaling and Numerical Simulation of Statistical Systems}}}\
  (\bibinfo  {publisher} {World Scientific},\ \bibinfo {address} {Singapore},\
  \bibinfo {year} {1990})\BibitemShut {NoStop}%
\bibitem [{\citenamefont {Gasparini}\ \emph {et~al.}(2008)\citenamefont
  {Gasparini}, \citenamefont {Kimball}, \citenamefont {Mooney},\ and\
  \citenamefont {Diaz-{A}vila}}]{Gasparini}%
  \BibitemOpen
  \bibfield  {author} {\bibinfo {author} {\bibfnamefont {F.~M.}\ \bibnamefont
  {Gasparini}}, \bibinfo {author} {\bibfnamefont {M.~O.}\ \bibnamefont
  {Kimball}}, \bibinfo {author} {\bibfnamefont {K.~P.}\ \bibnamefont
  {Mooney}},\ and\ \bibinfo {author} {\bibfnamefont {M.}~\bibnamefont
  {Diaz-{A}vila}},\ }\bibfield  {title} {\bibinfo {title} {Finite-size scaling
  of $^{4}\mathrm{He}$ at the superfluid transition},\ }\href
  {https://doi.org/10.1103/RevModPhys.80.1009} {\bibfield  {journal} {\bibinfo
  {journal} {Rev. Mod. Phys.}\ }\textbf {\bibinfo {volume} {80}},\ \bibinfo
  {pages} {1009} (\bibinfo {year} {2008})}\BibitemShut {NoStop}%
\bibitem [{\citenamefont {Krapivsky}\ and\ \citenamefont
  {Krioukov}(2008)}]{Krapivsky}%
  \BibitemOpen
  \bibfield  {author} {\bibinfo {author} {\bibfnamefont {P.}~\bibnamefont
  {Krapivsky}}\ and\ \bibinfo {author} {\bibfnamefont {D.}~\bibnamefont
  {Krioukov}},\ }\bibfield  {title} {\bibinfo {title} {Scale-free networks as
  preasymptotic regimes of superlinear preferential attachment},\ }\href
  {https://doi.org/10.1103/PhysRevE.78.026114} {\bibfield  {journal} {\bibinfo
  {journal} {Phys. Rev. E}\ }\textbf {\bibinfo {volume} {78}},\ \bibinfo
  {pages} {026114} (\bibinfo {year} {2008})}\BibitemShut {NoStop}%
\bibitem [{\citenamefont {Falkenberg}\ \emph {et~al.}(2020)\citenamefont
  {Falkenberg}, \citenamefont {Lee}, \citenamefont {Amano}, \citenamefont
  {Ogawa}, \citenamefont {Yano}, \citenamefont {Miyake}, \citenamefont
  {Evans},\ and\ \citenamefont {Christensen}}]{Falkenberg}%
  \BibitemOpen
  \bibfield  {author} {\bibinfo {author} {\bibfnamefont {M.}~\bibnamefont
  {Falkenberg}}, \bibinfo {author} {\bibfnamefont {J.-H.}\ \bibnamefont {Lee}},
  \bibinfo {author} {\bibfnamefont {S.}~\bibnamefont {Amano}}, \bibinfo
  {author} {\bibfnamefont {K.}~\bibnamefont {Ogawa}}, \bibinfo {author}
  {\bibfnamefont {K.}~\bibnamefont {Yano}}, \bibinfo {author} {\bibfnamefont
  {Y.}~\bibnamefont {Miyake}}, \bibinfo {author} {\bibfnamefont {T.~S.}\
  \bibnamefont {Evans}},\ and\ \bibinfo {author} {\bibfnamefont
  {K.}~\bibnamefont {Christensen}},\ }\bibfield  {title} {\bibinfo {title}
  {Identifying time dependence in network growth},\ }\href
  {https://doi.org/10.1103/PhysRevResearch.2.023352} {\bibfield  {journal}
  {\bibinfo  {journal} {Phys. Rev. Res.}\ }\textbf {\bibinfo {volume} {2}},\
  \bibinfo {pages} {023352} (\bibinfo {year} {2020})}\BibitemShut {NoStop}%
\bibitem [{\citenamefont {Serafino}\ \emph {et~al.}(2021)\citenamefont
  {Serafino}, \citenamefont {Cimini}, \citenamefont {Maritan}, \citenamefont
  {Rinaldo}, \citenamefont {Suweis}, \citenamefont {Banavar},\ and\
  \citenamefont {Caldarelli}}]{Serafino}%
  \BibitemOpen
  \bibfield  {author} {\bibinfo {author} {\bibfnamefont {M.}~\bibnamefont
  {Serafino}}, \bibinfo {author} {\bibfnamefont {G.}~\bibnamefont {Cimini}},
  \bibinfo {author} {\bibfnamefont {A.}~\bibnamefont {Maritan}}, \bibinfo
  {author} {\bibfnamefont {A.}~\bibnamefont {Rinaldo}}, \bibinfo {author}
  {\bibfnamefont {S.}~\bibnamefont {Suweis}}, \bibinfo {author} {\bibfnamefont
  {J.~R.}\ \bibnamefont {Banavar}},\ and\ \bibinfo {author} {\bibfnamefont
  {G.}~\bibnamefont {Caldarelli}},\ }\bibfield  {title} {\bibinfo {title} {True
  scale-free networks hidden by finite size effects},\ }\href
  {https://doi.org/10.1073/pnas.2013825118} {\bibfield  {journal} {\bibinfo
  {journal} {Proc. Natl. Acad. Sci. USA}\ }\textbf {\bibinfo {volume} {118}},\
  \bibinfo {pages} {e2013825118} (\bibinfo {year} {2021})}\BibitemShut
  {NoStop}%
\bibitem [{\citenamefont {Attanasi}\ \emph {et~al.}(2014)\citenamefont
  {Attanasi}, \citenamefont {Cavagna}, \citenamefont {Del~Castello},
  \citenamefont {Giardina}, \citenamefont {Melillo}, \citenamefont {Parisi},
  \citenamefont {Pohl}, \citenamefont {Rossaro}, \citenamefont {Shen},
  \citenamefont {Silvestri},\ and\ \citenamefont {Viale}}]{swarm}%
  \BibitemOpen
  \bibfield  {author} {\bibinfo {author} {\bibfnamefont {A.}~\bibnamefont
  {Attanasi}}, \bibinfo {author} {\bibfnamefont {A.}~\bibnamefont {Cavagna}},
  \bibinfo {author} {\bibfnamefont {L.}~\bibnamefont {Del~Castello}}, \bibinfo
  {author} {\bibfnamefont {I.}~\bibnamefont {Giardina}}, \bibinfo {author}
  {\bibfnamefont {S.}~\bibnamefont {Melillo}}, \bibinfo {author} {\bibfnamefont
  {L.}~\bibnamefont {Parisi}}, \bibinfo {author} {\bibfnamefont
  {O.}~\bibnamefont {Pohl}}, \bibinfo {author} {\bibfnamefont {B.}~\bibnamefont
  {Rossaro}}, \bibinfo {author} {\bibfnamefont {E.}~\bibnamefont {Shen}},
  \bibinfo {author} {\bibfnamefont {E.}~\bibnamefont {Silvestri}},\ and\
  \bibinfo {author} {\bibfnamefont {M.}~\bibnamefont {Viale}},\ }\bibfield
  {title} {\bibinfo {title} {Finite-size scaling as a way to probe
  near-criticality in natural swarms},\ }\href
  {https://doi.org/10.1103/PhysRevLett.113.238102} {\bibfield  {journal}
  {\bibinfo  {journal} {Phys. Rev. Lett.}\ }\textbf {\bibinfo {volume} {113}},\
  \bibinfo {pages} {238102} (\bibinfo {year} {2014})}\BibitemShut {NoStop}%
\bibitem [{\citenamefont {Charbonneau}\ \emph {et~al.}(2021)\citenamefont
  {Charbonneau}, \citenamefont {Corwin}, \citenamefont {Dennis}, \citenamefont
  {D\'{\i}az Hern\'andez~Rojas}, \citenamefont {Ikeda}, \citenamefont
  {Parisi},\ and\ \citenamefont {Ricci-Tersenghi}}]{Charbonneau}%
  \BibitemOpen
  \bibfield  {author} {\bibinfo {author} {\bibfnamefont {P.}~\bibnamefont
  {Charbonneau}}, \bibinfo {author} {\bibfnamefont {E.~I.}\ \bibnamefont
  {Corwin}}, \bibinfo {author} {\bibfnamefont {R.~C.}\ \bibnamefont {Dennis}},
  \bibinfo {author} {\bibfnamefont {R.}~\bibnamefont {D\'{\i}az
  Hern\'andez~Rojas}}, \bibinfo {author} {\bibfnamefont {H.}~\bibnamefont
  {Ikeda}}, \bibinfo {author} {\bibfnamefont {G.}~\bibnamefont {Parisi}},\ and\
  \bibinfo {author} {\bibfnamefont {F.}~\bibnamefont {Ricci-Tersenghi}},\
  }\bibfield  {title} {\bibinfo {title} {Finite-size effects in the microscopic
  critical properties of jammed configurations: A comprehensive study of the
  effects of different types of disorder},\ }\href
  {https://doi.org/10.1103/PhysRevE.104.014102} {\bibfield  {journal} {\bibinfo
   {journal} {Phys. Rev. E}\ }\textbf {\bibinfo {volume} {104}},\ \bibinfo
  {pages} {014102} (\bibinfo {year} {2021})}\BibitemShut {NoStop}%
\bibitem [{\citenamefont {Bulaevskii}\ \emph {et~al.}(1972)\citenamefont
  {Bulaevskii}, \citenamefont {Zvarykina}, \citenamefont {Karimov},
  \citenamefont {Lyubovskii},\ and\ \citenamefont {Shchegolev}}]{Bulaevskii}%
  \BibitemOpen
  \bibfield  {author} {\bibinfo {author} {\bibfnamefont {L.}~\bibnamefont
  {Bulaevskii}}, \bibinfo {author} {\bibfnamefont {A.}~\bibnamefont
  {Zvarykina}}, \bibinfo {author} {\bibfnamefont {Y.~S.}\ \bibnamefont
  {Karimov}}, \bibinfo {author} {\bibfnamefont {R.}~\bibnamefont
  {Lyubovskii}},\ and\ \bibinfo {author} {\bibfnamefont {I.}~\bibnamefont
  {Shchegolev}},\ }\bibfield  {title} {\bibinfo {title} {Magnetic properties of
  linear conducting chains},\ }\href@noop {} {\bibfield  {journal} {\bibinfo
  {journal} {Sov. Phys. JETP}\ }\textbf {\bibinfo {volume} {35}},\ \bibinfo
  {pages} {384} (\bibinfo {year} {1972})}\BibitemShut {NoStop}%
\bibitem [{\citenamefont {Azevedo}\ and\ \citenamefont
  {Clark}(1977)}]{Azevedo}%
  \BibitemOpen
  \bibfield  {author} {\bibinfo {author} {\bibfnamefont {L.~J.}\ \bibnamefont
  {Azevedo}}\ and\ \bibinfo {author} {\bibfnamefont {W.~G.}\ \bibnamefont
  {Clark}},\ }\bibfield  {title} {\bibinfo {title} {Very-low-temperature
  specific heat of quinolinium ${(\mathrm{TCNQ})}_{2}$, a random-exchange
  {H}eisenberg antiferromagnetic chain},\ }\href
  {https://doi.org/10.1103/PhysRevB.16.3252} {\bibfield  {journal} {\bibinfo
  {journal} {Phys. Rev. B}\ }\textbf {\bibinfo {volume} {16}},\ \bibinfo
  {pages} {3252} (\bibinfo {year} {1977})}\BibitemShut {NoStop}%
\bibitem [{\citenamefont {Duffy}\ \emph {et~al.}(1979)\citenamefont {Duffy},
  \citenamefont {Weinhaus}, \citenamefont {Strandburg},\ and\ \citenamefont
  {Deck}}]{Duffy}%
  \BibitemOpen
  \bibfield  {author} {\bibinfo {author} {\bibfnamefont {W.}~\bibnamefont
  {Duffy}}, \bibinfo {author} {\bibfnamefont {F.~M.}\ \bibnamefont {Weinhaus}},
  \bibinfo {author} {\bibfnamefont {D.~L.}\ \bibnamefont {Strandburg}},\ and\
  \bibinfo {author} {\bibfnamefont {J.~F.}\ \bibnamefont {Deck}},\ }\bibfield
  {title} {\bibinfo {title} {Low-temperature heat capacity of acridinium
  ditetracyanoquinodimethanide [acridinium ${(\mathrm{TCNQ})}_{2}$]},\ }\href
  {https://doi.org/10.1103/PhysRevB.20.1164} {\bibfield  {journal} {\bibinfo
  {journal} {Phys. Rev. B}\ }\textbf {\bibinfo {volume} {20}},\ \bibinfo
  {pages} {1164} (\bibinfo {year} {1979})}\BibitemShut {NoStop}%
\bibitem [{\citenamefont {Bozler}\ \emph {et~al.}(1980)\citenamefont {Bozler},
  \citenamefont {Gould},\ and\ \citenamefont {Clark}}]{Bozler}%
  \BibitemOpen
  \bibfield  {author} {\bibinfo {author} {\bibfnamefont {H.~M.}\ \bibnamefont
  {Bozler}}, \bibinfo {author} {\bibfnamefont {C.~M.}\ \bibnamefont {Gould}},\
  and\ \bibinfo {author} {\bibfnamefont {W.~G.}\ \bibnamefont {Clark}},\
  }\bibfield  {title} {\bibinfo {title} {Crossover behavior of a
  random-exchange {H}eisenberg antiferromagnetic chain at ultralow
  temperatures},\ }\href {https://doi.org/10.1103/PhysRevLett.45.1303}
  {\bibfield  {journal} {\bibinfo  {journal} {Phys. Rev. Lett.}\ }\textbf
  {\bibinfo {volume} {45}},\ \bibinfo {pages} {1303} (\bibinfo {year}
  {1980})}\BibitemShut {NoStop}%
\bibitem [{\citenamefont {Sanny}\ \emph {et~al.}(1980)\citenamefont {Sanny},
  \citenamefont {Gr{\"u}ner},\ and\ \citenamefont {Clark}}]{Sanny}%
  \BibitemOpen
  \bibfield  {author} {\bibinfo {author} {\bibfnamefont {J.}~\bibnamefont
  {Sanny}}, \bibinfo {author} {\bibfnamefont {G.}~\bibnamefont {Gr{\"u}ner}},\
  and\ \bibinfo {author} {\bibfnamefont {W.~G.}\ \bibnamefont {Clark}},\
  }\bibfield  {title} {\bibinfo {title} {Observation of quasi-universal
  magnetic behavior in a random exchange {H}eisenberg antiferromagnetic chain:
  Neutron irradiated quinolinium ({TCNQ})$_2$},\ }\href
  {https://doi.org/10.1016/0038-1098(80)90868-6} {\bibfield  {journal}
  {\bibinfo  {journal} {Solid State Commun.}\ }\textbf {\bibinfo {volume}
  {35}},\ \bibinfo {pages} {657} (\bibinfo {year} {1980})}\BibitemShut
  {NoStop}%
\bibitem [{\citenamefont {Andres}\ \emph {et~al.}(1981)\citenamefont {Andres},
  \citenamefont {Bhatt}, \citenamefont {Goalwin}, \citenamefont {Rice},\ and\
  \citenamefont {Walstedt}}]{Andres}%
  \BibitemOpen
  \bibfield  {author} {\bibinfo {author} {\bibfnamefont {K.}~\bibnamefont
  {Andres}}, \bibinfo {author} {\bibfnamefont {R.~N.}\ \bibnamefont {Bhatt}},
  \bibinfo {author} {\bibfnamefont {P.}~\bibnamefont {Goalwin}}, \bibinfo
  {author} {\bibfnamefont {T.~M.}\ \bibnamefont {Rice}},\ and\ \bibinfo
  {author} {\bibfnamefont {R.~E.}\ \bibnamefont {Walstedt}},\ }\bibfield
  {title} {\bibinfo {title} {Low-temperature magnetic susceptibility of {S}i:
  {P} in the nonmetallic region},\ }\href
  {https://doi.org/10.1103/PhysRevB.24.244} {\bibfield  {journal} {\bibinfo
  {journal} {Phys. Rev. B}\ }\textbf {\bibinfo {volume} {24}},\ \bibinfo
  {pages} {244} (\bibinfo {year} {1981})}\BibitemShut {NoStop}%
\bibitem [{\citenamefont {Paalanen}\ \emph {et~al.}(1986)\citenamefont
  {Paalanen}, \citenamefont {Sachdev}, \citenamefont {Bhatt},\ and\
  \citenamefont {Ruckenstein}}]{Paalanen86}%
  \BibitemOpen
  \bibfield  {author} {\bibinfo {author} {\bibfnamefont {M.~A.}\ \bibnamefont
  {Paalanen}}, \bibinfo {author} {\bibfnamefont {S.}~\bibnamefont {Sachdev}},
  \bibinfo {author} {\bibfnamefont {R.~N.}\ \bibnamefont {Bhatt}},\ and\
  \bibinfo {author} {\bibfnamefont {A.~E.}\ \bibnamefont {Ruckenstein}},\
  }\bibfield  {title} {\bibinfo {title} {Spin dynamics of nearly localized
  electrons},\ }\href {https://doi.org/10.1103/PhysRevLett.57.2061} {\bibfield
  {journal} {\bibinfo  {journal} {Phys. Rev. Lett.}\ }\textbf {\bibinfo
  {volume} {57}},\ \bibinfo {pages} {2061} (\bibinfo {year}
  {1986})}\BibitemShut {NoStop}%
\bibitem [{\citenamefont {Paalanen}\ \emph {et~al.}(1988)\citenamefont
  {Paalanen}, \citenamefont {Graebner}, \citenamefont {Bhatt},\ and\
  \citenamefont {Sachdev}}]{Paalanen88}%
  \BibitemOpen
  \bibfield  {author} {\bibinfo {author} {\bibfnamefont {M.~A.}\ \bibnamefont
  {Paalanen}}, \bibinfo {author} {\bibfnamefont {J.~E.}\ \bibnamefont
  {Graebner}}, \bibinfo {author} {\bibfnamefont {R.~N.}\ \bibnamefont
  {Bhatt}},\ and\ \bibinfo {author} {\bibfnamefont {S.}~\bibnamefont
  {Sachdev}},\ }\bibfield  {title} {\bibinfo {title} {Thermodynamic behavior
  near a metal-insulator transition},\ }\href
  {https://doi.org/10.1103/PhysRevLett.61.597} {\bibfield  {journal} {\bibinfo
  {journal} {Phys. Rev. Lett.}\ }\textbf {\bibinfo {volume} {61}},\ \bibinfo
  {pages} {597} (\bibinfo {year} {1988})}\BibitemShut {NoStop}%
\bibitem [{\citenamefont {Lakner}\ and\ \citenamefont
  {L\"ohneysen}(1989)}]{Lakner}%
  \BibitemOpen
  \bibfield  {author} {\bibinfo {author} {\bibfnamefont {M.}~\bibnamefont
  {Lakner}}\ and\ \bibinfo {author} {\bibfnamefont {H.~v.}\ \bibnamefont
  {L\"ohneysen}},\ }\bibfield  {title} {\bibinfo {title} {Localized versus
  itinerant electrons at the metal-insulator transition in {S}i:{P}},\ }\href
  {https://doi.org/10.1103/PhysRevLett.63.648} {\bibfield  {journal} {\bibinfo
  {journal} {Phys. Rev. Lett.}\ }\textbf {\bibinfo {volume} {63}},\ \bibinfo
  {pages} {648} (\bibinfo {year} {1989})}\BibitemShut {NoStop}%
\bibitem [{\citenamefont {Hirsch}\ \emph {et~al.}(1992)\citenamefont {Hirsch},
  \citenamefont {Holcomb}, \citenamefont {Bhatt},\ and\ \citenamefont
  {Paalanen}}]{Hirsch}%
  \BibitemOpen
  \bibfield  {author} {\bibinfo {author} {\bibfnamefont {M.~J.}\ \bibnamefont
  {Hirsch}}, \bibinfo {author} {\bibfnamefont {D.~F.}\ \bibnamefont {Holcomb}},
  \bibinfo {author} {\bibfnamefont {R.~N.}\ \bibnamefont {Bhatt}},\ and\
  \bibinfo {author} {\bibfnamefont {M.~A.}\ \bibnamefont {Paalanen}},\
  }\bibfield  {title} {\bibinfo {title} {E{S}{R} studies of compensated
  {S}i:{P},{B} near the metal-insulator transition},\ }\href
  {https://doi.org/10.1103/PhysRevLett.68.1418} {\bibfield  {journal} {\bibinfo
   {journal} {Phys. Rev. Lett.}\ }\textbf {\bibinfo {volume} {68}},\ \bibinfo
  {pages} {1418} (\bibinfo {year} {1992})}\BibitemShut {NoStop}%
\bibitem [{\citenamefont {Ma}\ \emph {et~al.}(1979)\citenamefont {Ma},
  \citenamefont {Dasgupta},\ and\ \citenamefont {Hu}}]{MaRS}%
  \BibitemOpen
  \bibfield  {author} {\bibinfo {author} {\bibfnamefont {S.-k.}\ \bibnamefont
  {Ma}}, \bibinfo {author} {\bibfnamefont {C.}~\bibnamefont {Dasgupta}},\ and\
  \bibinfo {author} {\bibfnamefont {C.-k.}\ \bibnamefont {Hu}},\ }\bibfield
  {title} {\bibinfo {title} {Random antiferromagnetic chain},\ }\href
  {https://doi.org/10.1103/PhysRevLett.43.1434} {\bibfield  {journal} {\bibinfo
   {journal} {Phys. Rev. Lett.}\ }\textbf {\bibinfo {volume} {43}},\ \bibinfo
  {pages} {1434} (\bibinfo {year} {1979})}\BibitemShut {NoStop}%
\bibitem [{\citenamefont {Dasgupta}\ and\ \citenamefont
  {Ma}(1980)}]{DasguptaRS}%
  \BibitemOpen
  \bibfield  {author} {\bibinfo {author} {\bibfnamefont {C.}~\bibnamefont
  {Dasgupta}}\ and\ \bibinfo {author} {\bibfnamefont {S.-k.}\ \bibnamefont
  {Ma}},\ }\bibfield  {title} {\bibinfo {title} {Low-temperature properties of
  the random {H}eisenberg antiferromagnetic chain},\ }\href
  {https://doi.org/10.1103/PhysRevB.22.1305} {\bibfield  {journal} {\bibinfo
  {journal} {Phys. Rev. B}\ }\textbf {\bibinfo {volume} {22}},\ \bibinfo
  {pages} {1305} (\bibinfo {year} {1980})}\BibitemShut {NoStop}%
\bibitem [{\citenamefont {Bhatt}\ and\ \citenamefont {Lee}(1982)}]{BhattRS}%
  \BibitemOpen
  \bibfield  {author} {\bibinfo {author} {\bibfnamefont {R.~N.}\ \bibnamefont
  {Bhatt}}\ and\ \bibinfo {author} {\bibfnamefont {P.~A.}\ \bibnamefont
  {Lee}},\ }\bibfield  {title} {\bibinfo {title} {Scaling studies of highly
  disordered spin-1/2 antiferromagnetic systems},\ }\href
  {https://doi.org/10.1103/PhysRevLett.48.344} {\bibfield  {journal} {\bibinfo
  {journal} {Phys. Rev. Lett.}\ }\textbf {\bibinfo {volume} {48}},\ \bibinfo
  {pages} {344} (\bibinfo {year} {1982})}\BibitemShut {NoStop}%
\bibitem [{\citenamefont {Fisher}(1994)}]{FisherRS}%
  \BibitemOpen
  \bibfield  {author} {\bibinfo {author} {\bibfnamefont {D.~S.}\ \bibnamefont
  {Fisher}},\ }\bibfield  {title} {\bibinfo {title} {Random antiferromagnetic
  quantum spin chains},\ }\href {https://doi.org/10.1103/PhysRevB.50.3799}
  {\bibfield  {journal} {\bibinfo  {journal} {Phys. Rev. B}\ }\textbf {\bibinfo
  {volume} {50}},\ \bibinfo {pages} {3799} (\bibinfo {year}
  {1994})}\BibitemShut {NoStop}%
\bibitem [{\citenamefont {Westerberg}\ \emph {et~al.}(1997)\citenamefont
  {Westerberg}, \citenamefont {Furusaki}, \citenamefont {Sigrist},\ and\
  \citenamefont {Lee}}]{WesterbergRS}%
  \BibitemOpen
  \bibfield  {author} {\bibinfo {author} {\bibfnamefont {E.}~\bibnamefont
  {Westerberg}}, \bibinfo {author} {\bibfnamefont {A.}~\bibnamefont
  {Furusaki}}, \bibinfo {author} {\bibfnamefont {M.}~\bibnamefont {Sigrist}},\
  and\ \bibinfo {author} {\bibfnamefont {P.~A.}\ \bibnamefont {Lee}},\
  }\bibfield  {title} {\bibinfo {title} {Low-energy fixed points of random
  quantum spin chains},\ }\href {https://doi.org/10.1103/PhysRevB.55.12578}
  {\bibfield  {journal} {\bibinfo  {journal} {Phys. Rev. B}\ }\textbf {\bibinfo
  {volume} {55}},\ \bibinfo {pages} {12578} (\bibinfo {year}
  {1997})}\BibitemShut {NoStop}%
\bibitem [{\citenamefont {Kitagawa}\ \emph {et~al.}(2018)\citenamefont
  {Kitagawa}, \citenamefont {Takayama}, \citenamefont {Matsumoto},
  \citenamefont {Kato}, \citenamefont {Takano}, \citenamefont {Kishimoto},
  \citenamefont {Bette}, \citenamefont {Dinnebier}, \citenamefont {Jackeli},\
  and\ \citenamefont {Takagi}}]{Kitagawa}%
  \BibitemOpen
  \bibfield  {author} {\bibinfo {author} {\bibfnamefont {K.}~\bibnamefont
  {Kitagawa}}, \bibinfo {author} {\bibfnamefont {T.}~\bibnamefont {Takayama}},
  \bibinfo {author} {\bibfnamefont {Y.}~\bibnamefont {Matsumoto}}, \bibinfo
  {author} {\bibfnamefont {A.}~\bibnamefont {Kato}}, \bibinfo {author}
  {\bibfnamefont {R.}~\bibnamefont {Takano}}, \bibinfo {author} {\bibfnamefont
  {Y.}~\bibnamefont {Kishimoto}}, \bibinfo {author} {\bibfnamefont
  {S.}~\bibnamefont {Bette}}, \bibinfo {author} {\bibfnamefont
  {R.}~\bibnamefont {Dinnebier}}, \bibinfo {author} {\bibfnamefont
  {G.}~\bibnamefont {Jackeli}},\ and\ \bibinfo {author} {\bibfnamefont
  {H.}~\bibnamefont {Takagi}},\ }\bibfield  {title} {\bibinfo {title} {A
  spin–orbital-entangled quantum liquid on a honeycomb lattice},\ }\href
  {https://doi.org/https://doi.org/10.1038/nature25482} {\bibfield  {journal}
  {\bibinfo  {journal} {Nature (London)}\ }\textbf {\bibinfo {volume} {554}},\
  \bibinfo {pages} {341} (\bibinfo {year} {2018})}\BibitemShut {NoStop}%
\bibitem [{\citenamefont {Kimchi}\ \emph
  {et~al.}(2018{\natexlab{a}})\citenamefont {Kimchi}, \citenamefont {Nahum},\
  and\ \citenamefont {Senthil}}]{KimchiPRX}%
  \BibitemOpen
  \bibfield  {author} {\bibinfo {author} {\bibfnamefont {I.}~\bibnamefont
  {Kimchi}}, \bibinfo {author} {\bibfnamefont {A.}~\bibnamefont {Nahum}},\ and\
  \bibinfo {author} {\bibfnamefont {T.}~\bibnamefont {Senthil}},\ }\bibfield
  {title} {\bibinfo {title} {Valence bonds in random quantum magnets: Theory
  and application to \ce{YbMgGaO4}},\ }\href
  {https://doi.org/10.1103/PhysRevX.8.031028} {\bibfield  {journal} {\bibinfo
  {journal} {Phys. Rev. X}\ }\textbf {\bibinfo {volume} {8}},\ \bibinfo {pages}
  {031028} (\bibinfo {year} {2018}{\natexlab{a}})}\BibitemShut {NoStop}%
\bibitem [{\citenamefont {Kimchi}\ \emph
  {et~al.}(2018{\natexlab{b}})\citenamefont {Kimchi}, \citenamefont
  {Sheckelton}, \citenamefont {McQueen},\ and\ \citenamefont {Lee}}]{Kimchi}%
  \BibitemOpen
  \bibfield  {author} {\bibinfo {author} {\bibfnamefont {I.}~\bibnamefont
  {Kimchi}}, \bibinfo {author} {\bibfnamefont {J.~P.}\ \bibnamefont
  {Sheckelton}}, \bibinfo {author} {\bibfnamefont {T.~M.}\ \bibnamefont
  {McQueen}},\ and\ \bibinfo {author} {\bibfnamefont {P.~A.}\ \bibnamefont
  {Lee}},\ }\bibfield  {title} {\bibinfo {title} {Scaling and data collapse
  from local moments in frustrated disordered quantum spin systems},\ }\href
  {https://doi.org/https://doi.org/10.1038/s41467-018-06800-2} {\bibfield
  {journal} {\bibinfo  {journal} {Nat. Commun.}\ }\textbf {\bibinfo {volume}
  {9}},\ \bibinfo {pages} {4367} (\bibinfo {year}
  {2018}{\natexlab{b}})}\BibitemShut {NoStop}%
\bibitem [{\citenamefont {Lee}\ \emph {et~al.}(2023)\citenamefont {Lee},
  \citenamefont {Lee}, \citenamefont {Choi}, \citenamefont {Wang},
  \citenamefont {Luetkens}, \citenamefont {Shiroka}, \citenamefont {Jang},
  \citenamefont {Yoon},\ and\ \citenamefont {Choi}}]{LeeRS}%
  \BibitemOpen
  \bibfield  {author} {\bibinfo {author} {\bibfnamefont {C.}~\bibnamefont
  {Lee}}, \bibinfo {author} {\bibfnamefont {S.}~\bibnamefont {Lee}}, \bibinfo
  {author} {\bibfnamefont {Y.}~\bibnamefont {Choi}}, \bibinfo {author}
  {\bibfnamefont {C.}~\bibnamefont {Wang}}, \bibinfo {author} {\bibfnamefont
  {H.}~\bibnamefont {Luetkens}}, \bibinfo {author} {\bibfnamefont
  {T.}~\bibnamefont {Shiroka}}, \bibinfo {author} {\bibfnamefont
  {Z.}~\bibnamefont {Jang}}, \bibinfo {author} {\bibfnamefont {Y.-G.}\
  \bibnamefont {Yoon}},\ and\ \bibinfo {author} {\bibfnamefont {K.-Y.}\
  \bibnamefont {Choi}},\ }\bibfield  {title} {\bibinfo {title} {Coexistence of
  random singlets and disordered {K}itaev spin liquid in \ce{H3LiIr2O6}},\
  }\href {https://doi.org/10.1103/PhysRevB.107.014424} {\bibfield  {journal}
  {\bibinfo  {journal} {Phys. Rev. B}\ }\textbf {\bibinfo {volume} {107}},\
  \bibinfo {pages} {014424} (\bibinfo {year} {2023})}\BibitemShut {NoStop}%
\bibitem [{\citenamefont {Huang}\ \emph {et~al.}(2021)\citenamefont {Huang},
  \citenamefont {Xu}, \citenamefont {Wang}, \citenamefont {Zhao}, \citenamefont
  {Tu}, \citenamefont {Ni}, \citenamefont {Wang}, \citenamefont {Pan},
  \citenamefont {Fu}, \citenamefont {Hao}, \citenamefont {Liu}, \citenamefont
  {Mei},\ and\ \citenamefont {Li}}]{Huang_noScaling}%
  \BibitemOpen
  \bibfield  {author} {\bibinfo {author} {\bibfnamefont {Y.~Y.}\ \bibnamefont
  {Huang}}, \bibinfo {author} {\bibfnamefont {Y.}~\bibnamefont {Xu}}, \bibinfo
  {author} {\bibfnamefont {L.}~\bibnamefont {Wang}}, \bibinfo {author}
  {\bibfnamefont {C.~C.}\ \bibnamefont {Zhao}}, \bibinfo {author}
  {\bibfnamefont {C.~P.}\ \bibnamefont {Tu}}, \bibinfo {author} {\bibfnamefont
  {J.~M.}\ \bibnamefont {Ni}}, \bibinfo {author} {\bibfnamefont {L.~S.}\
  \bibnamefont {Wang}}, \bibinfo {author} {\bibfnamefont {B.~L.}\ \bibnamefont
  {Pan}}, \bibinfo {author} {\bibfnamefont {Y.}~\bibnamefont {Fu}}, \bibinfo
  {author} {\bibfnamefont {Z.}~\bibnamefont {Hao}}, \bibinfo {author}
  {\bibfnamefont {C.}~\bibnamefont {Liu}}, \bibinfo {author} {\bibfnamefont
  {J.-W.}\ \bibnamefont {Mei}},\ and\ \bibinfo {author} {\bibfnamefont {S.~Y.}\
  \bibnamefont {Li}},\ }\bibfield  {title} {\bibinfo {title} {Heat transport in
  herbertsmithite: Can a quantum spin liquid survive disorder?},\ }\href
  {https://doi.org/10.1103/PhysRevLett.127.267202} {\bibfield  {journal}
  {\bibinfo  {journal} {Phys. Rev. Lett.}\ }\textbf {\bibinfo {volume} {127}},\
  \bibinfo {pages} {267202} (\bibinfo {year} {2021})}\BibitemShut {NoStop}%
\bibitem [{\citenamefont {Wang}\ \emph {et~al.}(2021)\citenamefont {Wang},
  \citenamefont {Yuan}, \citenamefont {Singer}, \citenamefont {Smaha},
  \citenamefont {He}, \citenamefont {Wen}, \citenamefont {Lee},\ and\
  \citenamefont {Imai}}]{wangNMR}%
  \BibitemOpen
  \bibfield  {author} {\bibinfo {author} {\bibfnamefont {J.}~\bibnamefont
  {Wang}}, \bibinfo {author} {\bibfnamefont {W.}~\bibnamefont {Yuan}}, \bibinfo
  {author} {\bibfnamefont {P.~M.}\ \bibnamefont {Singer}}, \bibinfo {author}
  {\bibfnamefont {R.~W.}\ \bibnamefont {Smaha}}, \bibinfo {author}
  {\bibfnamefont {W.}~\bibnamefont {He}}, \bibinfo {author} {\bibfnamefont
  {J.}~\bibnamefont {Wen}}, \bibinfo {author} {\bibfnamefont {Y.~S.}\
  \bibnamefont {Lee}},\ and\ \bibinfo {author} {\bibfnamefont {T.}~\bibnamefont
  {Imai}},\ }\bibfield  {title} {\bibinfo {title} {Emergence of spin singlets
  with inhomogeneous gaps in the kagome lattice {H}eisenberg antiferromagnets
  {Z}n-barlowite and herbertsmithite},\ }\href
  {https://doi.org/10.1038/s41567-021-01310-3} {\bibfield  {journal} {\bibinfo
  {journal} {Nat. Phys.}\ }\textbf {\bibinfo {volume} {17}},\ \bibinfo {pages}
  {1109} (\bibinfo {year} {2021})}\BibitemShut {NoStop}%
\bibitem [{\citenamefont {Murayama}\ \emph {et~al.}(2022)\citenamefont
  {Murayama}, \citenamefont {Tominaga}, \citenamefont {Asaba}, \citenamefont
  {de~Oliveira~Silva}, \citenamefont {Sato}, \citenamefont {Suzuki},
  \citenamefont {Ukai}, \citenamefont {Suetsugu}, \citenamefont {Kasahara},
  \citenamefont {Okuma}, \citenamefont {Kimchi},\ and\ \citenamefont
  {Matsuda}}]{MurayamaRSexpt2}%
  \BibitemOpen
  \bibfield  {author} {\bibinfo {author} {\bibfnamefont {H.}~\bibnamefont
  {Murayama}}, \bibinfo {author} {\bibfnamefont {T.}~\bibnamefont {Tominaga}},
  \bibinfo {author} {\bibfnamefont {T.}~\bibnamefont {Asaba}}, \bibinfo
  {author} {\bibfnamefont {A.}~\bibnamefont {de~Oliveira~Silva}}, \bibinfo
  {author} {\bibfnamefont {Y.}~\bibnamefont {Sato}}, \bibinfo {author}
  {\bibfnamefont {H.}~\bibnamefont {Suzuki}}, \bibinfo {author} {\bibfnamefont
  {Y.}~\bibnamefont {Ukai}}, \bibinfo {author} {\bibfnamefont {S.}~\bibnamefont
  {Suetsugu}}, \bibinfo {author} {\bibfnamefont {Y.}~\bibnamefont {Kasahara}},
  \bibinfo {author} {\bibfnamefont {R.}~\bibnamefont {Okuma}}, \bibinfo
  {author} {\bibfnamefont {I.}~\bibnamefont {Kimchi}},\ and\ \bibinfo {author}
  {\bibfnamefont {Y.}~\bibnamefont {Matsuda}},\ }\bibfield  {title} {\bibinfo
  {title} {Universal scaling of specific heat in the $s=\frac{1}{2}$ quantum
  kagome antiferromagnet herbertsmithite},\ }\href
  {https://doi.org/10.1103/PhysRevB.106.174406} {\bibfield  {journal} {\bibinfo
   {journal} {Phys. Rev. B}\ }\textbf {\bibinfo {volume} {106}},\ \bibinfo
  {pages} {174406} (\bibinfo {year} {2022})}\BibitemShut {NoStop}%
\bibitem [{\citenamefont {Liu}\ \emph {et~al.}(2022)\citenamefont {Liu},
  \citenamefont {Yuan}, \citenamefont {Li}, \citenamefont {Li}, \citenamefont
  {Zhao}, \citenamefont {Liao},\ and\ \citenamefont {Li}}]{LiuRSexpt}%
  \BibitemOpen
  \bibfield  {author} {\bibinfo {author} {\bibfnamefont {J.}~\bibnamefont
  {Liu}}, \bibinfo {author} {\bibfnamefont {L.}~\bibnamefont {Yuan}}, \bibinfo
  {author} {\bibfnamefont {X.}~\bibnamefont {Li}}, \bibinfo {author}
  {\bibfnamefont {B.}~\bibnamefont {Li}}, \bibinfo {author} {\bibfnamefont
  {K.}~\bibnamefont {Zhao}}, \bibinfo {author} {\bibfnamefont {H.}~\bibnamefont
  {Liao}},\ and\ \bibinfo {author} {\bibfnamefont {Y.}~\bibnamefont {Li}},\
  }\bibfield  {title} {\bibinfo {title} {Gapless spin liquid behavior in a
  kagome {H}eisenberg antiferromagnet with randomly distributed hexagons of
  alternate bonds},\ }\href {https://doi.org/10.1103/PhysRevB.105.024418}
  {\bibfield  {journal} {\bibinfo  {journal} {Phys. Rev. B}\ }\textbf {\bibinfo
  {volume} {105}},\ \bibinfo {pages} {024418} (\bibinfo {year}
  {2022})}\BibitemShut {NoStop}%
\bibitem [{\citenamefont {Murayama}\ \emph {et~al.}(2020)\citenamefont
  {Murayama}, \citenamefont {Sato}, \citenamefont {Taniguchi}, \citenamefont
  {Kurihara}, \citenamefont {Xing}, \citenamefont {Huang}, \citenamefont
  {Kasahara}, \citenamefont {Kasahara}, \citenamefont {Kimchi}, \citenamefont
  {Yoshida}, \citenamefont {Iwasa}, \citenamefont {Mizukami}, \citenamefont
  {Shibauchi}, \citenamefont {Konczykowski},\ and\ \citenamefont
  {Matsuda}}]{MurayamaRSexpt}%
  \BibitemOpen
  \bibfield  {author} {\bibinfo {author} {\bibfnamefont {H.}~\bibnamefont
  {Murayama}}, \bibinfo {author} {\bibfnamefont {Y.}~\bibnamefont {Sato}},
  \bibinfo {author} {\bibfnamefont {T.}~\bibnamefont {Taniguchi}}, \bibinfo
  {author} {\bibfnamefont {R.}~\bibnamefont {Kurihara}}, \bibinfo {author}
  {\bibfnamefont {X.~Z.}\ \bibnamefont {Xing}}, \bibinfo {author}
  {\bibfnamefont {W.}~\bibnamefont {Huang}}, \bibinfo {author} {\bibfnamefont
  {S.}~\bibnamefont {Kasahara}}, \bibinfo {author} {\bibfnamefont
  {Y.}~\bibnamefont {Kasahara}}, \bibinfo {author} {\bibfnamefont
  {I.}~\bibnamefont {Kimchi}}, \bibinfo {author} {\bibfnamefont
  {M.}~\bibnamefont {Yoshida}}, \bibinfo {author} {\bibfnamefont
  {Y.}~\bibnamefont {Iwasa}}, \bibinfo {author} {\bibfnamefont
  {Y.}~\bibnamefont {Mizukami}}, \bibinfo {author} {\bibfnamefont
  {T.}~\bibnamefont {Shibauchi}}, \bibinfo {author} {\bibfnamefont
  {M.}~\bibnamefont {Konczykowski}},\ and\ \bibinfo {author} {\bibfnamefont
  {Y.}~\bibnamefont {Matsuda}},\ }\bibfield  {title} {\bibinfo {title} {Effect
  of quenched disorder on the quantum spin liquid state of the
  triangular-lattice antiferromagnet 1${T}$--{T}a{S}$_2$},\ }\href
  {https://doi.org/10.1103/PhysRevResearch.2.013099} {\bibfield  {journal}
  {\bibinfo  {journal} {Phys. Rev. Res.}\ }\textbf {\bibinfo {volume} {2}},\
  \bibinfo {pages} {013099} (\bibinfo {year} {2020})}\BibitemShut {NoStop}%
\bibitem [{\citenamefont {Kenney}\ \emph {et~al.}(2019)\citenamefont {Kenney},
  \citenamefont {Segre}, \citenamefont {Lafargue-Dit-Hauret}, \citenamefont
  {Lebedev}, \citenamefont {Abramchuk}, \citenamefont {Berlie}, \citenamefont
  {Cottrell}, \citenamefont {Simutis}, \citenamefont {Bahrami}, \citenamefont
  {Mordvinova}, \citenamefont {Fabbris}, \citenamefont {McChesney},
  \citenamefont {Haskel}, \citenamefont {Rocquefelte}, \citenamefont {Graf},\
  and\ \citenamefont {Tafti}}]{KenneyRS}%
  \BibitemOpen
  \bibfield  {author} {\bibinfo {author} {\bibfnamefont {E.~M.}\ \bibnamefont
  {Kenney}}, \bibinfo {author} {\bibfnamefont {C.~U.}\ \bibnamefont {Segre}},
  \bibinfo {author} {\bibfnamefont {W.}~\bibnamefont {Lafargue-Dit-Hauret}},
  \bibinfo {author} {\bibfnamefont {O.~I.}\ \bibnamefont {Lebedev}}, \bibinfo
  {author} {\bibfnamefont {M.}~\bibnamefont {Abramchuk}}, \bibinfo {author}
  {\bibfnamefont {A.}~\bibnamefont {Berlie}}, \bibinfo {author} {\bibfnamefont
  {S.~P.}\ \bibnamefont {Cottrell}}, \bibinfo {author} {\bibfnamefont
  {G.}~\bibnamefont {Simutis}}, \bibinfo {author} {\bibfnamefont
  {F.}~\bibnamefont {Bahrami}}, \bibinfo {author} {\bibfnamefont {N.~E.}\
  \bibnamefont {Mordvinova}}, \bibinfo {author} {\bibfnamefont
  {G.}~\bibnamefont {Fabbris}}, \bibinfo {author} {\bibfnamefont {J.~L.}\
  \bibnamefont {McChesney}}, \bibinfo {author} {\bibfnamefont {D.}~\bibnamefont
  {Haskel}}, \bibinfo {author} {\bibfnamefont {X.}~\bibnamefont {Rocquefelte}},
  \bibinfo {author} {\bibfnamefont {M.~J.}\ \bibnamefont {Graf}},\ and\
  \bibinfo {author} {\bibfnamefont {F.}~\bibnamefont {Tafti}},\ }\bibfield
  {title} {\bibinfo {title} {Coexistence of static and dynamic magnetism in the
  {K}itaev spin liquid material \ce{Cu2IrO3}},\ }\href
  {https://doi.org/10.1103/PhysRevB.100.094418} {\bibfield  {journal} {\bibinfo
   {journal} {Phys. Rev. B}\ }\textbf {\bibinfo {volume} {100}},\ \bibinfo
  {pages} {094418} (\bibinfo {year} {2019})}\BibitemShut {NoStop}%
\bibitem [{\citenamefont {Choi}\ \emph {et~al.}(2019)\citenamefont {Choi},
  \citenamefont {Lee}, \citenamefont {Lee}, \citenamefont {Yoon}, \citenamefont
  {Lee}, \citenamefont {Park}, \citenamefont {Ali}, \citenamefont {Singh},
  \citenamefont {Orain}, \citenamefont {Kim}, \citenamefont {Rhyee},
  \citenamefont {Chen}, \citenamefont {Chou},\ and\ \citenamefont
  {Choi}}]{ChoiRS}%
  \BibitemOpen
  \bibfield  {author} {\bibinfo {author} {\bibfnamefont {Y.~S.}\ \bibnamefont
  {Choi}}, \bibinfo {author} {\bibfnamefont {C.~H.}\ \bibnamefont {Lee}},
  \bibinfo {author} {\bibfnamefont {S.}~\bibnamefont {Lee}}, \bibinfo {author}
  {\bibfnamefont {S.}~\bibnamefont {Yoon}}, \bibinfo {author} {\bibfnamefont
  {W.-J.}\ \bibnamefont {Lee}}, \bibinfo {author} {\bibfnamefont
  {J.}~\bibnamefont {Park}}, \bibinfo {author} {\bibfnamefont {A.}~\bibnamefont
  {Ali}}, \bibinfo {author} {\bibfnamefont {Y.}~\bibnamefont {Singh}}, \bibinfo
  {author} {\bibfnamefont {J.-C.}\ \bibnamefont {Orain}}, \bibinfo {author}
  {\bibfnamefont {G.}~\bibnamefont {Kim}}, \bibinfo {author} {\bibfnamefont
  {J.-S.}\ \bibnamefont {Rhyee}}, \bibinfo {author} {\bibfnamefont {W.-T.}\
  \bibnamefont {Chen}}, \bibinfo {author} {\bibfnamefont {F.}~\bibnamefont
  {Chou}},\ and\ \bibinfo {author} {\bibfnamefont {K.-Y.}\ \bibnamefont
  {Choi}},\ }\bibfield  {title} {\bibinfo {title} {Exotic low-energy
  excitations emergent in the random {K}itaev magnet \ce{Cu2IrO3}},\ }\href
  {https://doi.org/10.1103/PhysRevLett.122.167202} {\bibfield  {journal}
  {\bibinfo  {journal} {Phys. Rev. Lett.}\ }\textbf {\bibinfo {volume} {122}},\
  \bibinfo {pages} {167202} (\bibinfo {year} {2019})}\BibitemShut {NoStop}%
\bibitem [{\citenamefont {Song}\ \emph {et~al.}(2021)\citenamefont {Song},
  \citenamefont {Zhu}, \citenamefont {Yang}, \citenamefont {Wei}, \citenamefont
  {Zhang}, \citenamefont {Yang}, \citenamefont {Sheng}, \citenamefont {Qi},
  \citenamefont {Ni}, \citenamefont {Li}, \citenamefont {Li}, \citenamefont
  {Cao}, \citenamefont {Meng}, \citenamefont {Li}, \citenamefont {Shi},\ and\
  \citenamefont {Li}}]{SongRSexpt}%
  \BibitemOpen
  \bibfield  {author} {\bibinfo {author} {\bibfnamefont {P.}~\bibnamefont
  {Song}}, \bibinfo {author} {\bibfnamefont {K.}~\bibnamefont {Zhu}}, \bibinfo
  {author} {\bibfnamefont {F.}~\bibnamefont {Yang}}, \bibinfo {author}
  {\bibfnamefont {Y.}~\bibnamefont {Wei}}, \bibinfo {author} {\bibfnamefont
  {L.}~\bibnamefont {Zhang}}, \bibinfo {author} {\bibfnamefont
  {H.}~\bibnamefont {Yang}}, \bibinfo {author} {\bibfnamefont {X.-L.}\
  \bibnamefont {Sheng}}, \bibinfo {author} {\bibfnamefont {Y.}~\bibnamefont
  {Qi}}, \bibinfo {author} {\bibfnamefont {J.}~\bibnamefont {Ni}}, \bibinfo
  {author} {\bibfnamefont {S.}~\bibnamefont {Li}}, \bibinfo {author}
  {\bibfnamefont {Y.}~\bibnamefont {Li}}, \bibinfo {author} {\bibfnamefont
  {G.}~\bibnamefont {Cao}}, \bibinfo {author} {\bibfnamefont {Z.~Y.}\
  \bibnamefont {Meng}}, \bibinfo {author} {\bibfnamefont {W.}~\bibnamefont
  {Li}}, \bibinfo {author} {\bibfnamefont {Y.}~\bibnamefont {Shi}},\ and\
  \bibinfo {author} {\bibfnamefont {S.}~\bibnamefont {Li}},\ }\bibfield
  {title} {\bibinfo {title} {Evidence for the random singlet phase in the
  honeycomb iridate \ce{SrIr2O6}},\ }\href
  {https://doi.org/10.1103/PhysRevB.103.L241114} {\bibfield  {journal}
  {\bibinfo  {journal} {Phys. Rev. B}\ }\textbf {\bibinfo {volume} {103}},\
  \bibinfo {pages} {L241114} (\bibinfo {year} {2021})}\BibitemShut {NoStop}%
\bibitem [{\citenamefont {Kundu}\ \emph {et~al.}(2020)\citenamefont {Kundu},
  \citenamefont {Hossain}, \citenamefont {S.}, \citenamefont {Das},
  \citenamefont {Baenitz}, \citenamefont {Baker}, \citenamefont {Orain},
  \citenamefont {Joshi}, \citenamefont {Mathieu}, \citenamefont {Mahadevan},
  \citenamefont {Pujari}, \citenamefont {Bhattacharjee}, \citenamefont
  {Mahajan},\ and\ \citenamefont {Sarma}}]{KunduRS}%
  \BibitemOpen
  \bibfield  {author} {\bibinfo {author} {\bibfnamefont {S.}~\bibnamefont
  {Kundu}}, \bibinfo {author} {\bibfnamefont {A.}~\bibnamefont {Hossain}},
  \bibinfo {author} {\bibfnamefont {P.~K.}\ \bibnamefont {S.}}, \bibinfo
  {author} {\bibfnamefont {R.}~\bibnamefont {Das}}, \bibinfo {author}
  {\bibfnamefont {M.}~\bibnamefont {Baenitz}}, \bibinfo {author} {\bibfnamefont
  {P.~J.}\ \bibnamefont {Baker}}, \bibinfo {author} {\bibfnamefont {J.-C.}\
  \bibnamefont {Orain}}, \bibinfo {author} {\bibfnamefont {D.~C.}\ \bibnamefont
  {Joshi}}, \bibinfo {author} {\bibfnamefont {R.}~\bibnamefont {Mathieu}},
  \bibinfo {author} {\bibfnamefont {P.}~\bibnamefont {Mahadevan}}, \bibinfo
  {author} {\bibfnamefont {S.}~\bibnamefont {Pujari}}, \bibinfo {author}
  {\bibfnamefont {S.}~\bibnamefont {Bhattacharjee}}, \bibinfo {author}
  {\bibfnamefont {A.~V.}\ \bibnamefont {Mahajan}},\ and\ \bibinfo {author}
  {\bibfnamefont {D.~D.}\ \bibnamefont {Sarma}},\ }\bibfield  {title} {\bibinfo
  {title} {Signatures of a spin-$\frac{1}{2}$ cooperative paramagnet in the
  diluted triangular lattice of \ce{Y2CuTiO6}},\ }\href
  {https://doi.org/10.1103/PhysRevLett.125.117206} {\bibfield  {journal}
  {\bibinfo  {journal} {Phys. Rev. Lett.}\ }\textbf {\bibinfo {volume} {125}},\
  \bibinfo {pages} {117206} (\bibinfo {year} {2020})}\BibitemShut {NoStop}%
\bibitem [{\citenamefont {Nguyen}\ \emph {et~al.}(2021)\citenamefont {Nguyen},
  \citenamefont {Straus}, \citenamefont {Zhang},\ and\ \citenamefont
  {Cava}}]{Nguyen}%
  \BibitemOpen
  \bibfield  {author} {\bibinfo {author} {\bibfnamefont {L.~T.}\ \bibnamefont
  {Nguyen}}, \bibinfo {author} {\bibfnamefont {D.~B.}\ \bibnamefont {Straus}},
  \bibinfo {author} {\bibfnamefont {Q.}~\bibnamefont {Zhang}},\ and\ \bibinfo
  {author} {\bibfnamefont {R.~J.}\ \bibnamefont {Cava}},\ }\bibfield  {title}
  {\bibinfo {title} {Widely spaced planes of magnetic dimers in the
  \ce{Ba6Y2Rh2Ti2O_{17-\delta}} hexagonal perovskite},\ }\href
  {https://doi.org/10.1103/PhysRevMaterials.5.034419} {\bibfield  {journal}
  {\bibinfo  {journal} {Phys. Rev. Mater.}\ }\textbf {\bibinfo {volume} {5}},\
  \bibinfo {pages} {034419} (\bibinfo {year} {2021})}\BibitemShut {NoStop}%
\bibitem [{\citenamefont {Baek}\ \emph {et~al.}(2020)\citenamefont {Baek},
  \citenamefont {Yeo}, \citenamefont {Do}, \citenamefont {Choi}, \citenamefont
  {Janssen}, \citenamefont {Vojta},\ and\ \citenamefont {B\"uchner}}]{BaekRS}%
  \BibitemOpen
  \bibfield  {author} {\bibinfo {author} {\bibfnamefont {S.-H.}\ \bibnamefont
  {Baek}}, \bibinfo {author} {\bibfnamefont {H.~W.}\ \bibnamefont {Yeo}},
  \bibinfo {author} {\bibfnamefont {S.-H.}\ \bibnamefont {Do}}, \bibinfo
  {author} {\bibfnamefont {K.-Y.}\ \bibnamefont {Choi}}, \bibinfo {author}
  {\bibfnamefont {L.}~\bibnamefont {Janssen}}, \bibinfo {author} {\bibfnamefont
  {M.}~\bibnamefont {Vojta}},\ and\ \bibinfo {author} {\bibfnamefont
  {B.}~\bibnamefont {B\"uchner}},\ }\bibfield  {title} {\bibinfo {title}
  {Observation of a random singlet state in a diluted {K}itaev honeycomb
  material},\ }\href {https://doi.org/10.1103/PhysRevB.102.094407} {\bibfield
  {journal} {\bibinfo  {journal} {Phys. Rev. B}\ }\textbf {\bibinfo {volume}
  {102}},\ \bibinfo {pages} {094407} (\bibinfo {year} {2020})}\BibitemShut
  {NoStop}%
\bibitem [{\citenamefont {Do}\ \emph {et~al.}(2020)\citenamefont {Do},
  \citenamefont {Lee}, \citenamefont {Kihara}, \citenamefont {Choi},
  \citenamefont {Yoon}, \citenamefont {Kim}, \citenamefont {Cheong},
  \citenamefont {Chen}, \citenamefont {Chou}, \citenamefont {Nojiri},\ and\
  \citenamefont {Choi}}]{DoExpt}%
  \BibitemOpen
  \bibfield  {author} {\bibinfo {author} {\bibfnamefont {S.-H.}\ \bibnamefont
  {Do}}, \bibinfo {author} {\bibfnamefont {C.~H.}\ \bibnamefont {Lee}},
  \bibinfo {author} {\bibfnamefont {T.}~\bibnamefont {Kihara}}, \bibinfo
  {author} {\bibfnamefont {Y.~S.}\ \bibnamefont {Choi}}, \bibinfo {author}
  {\bibfnamefont {S.}~\bibnamefont {Yoon}}, \bibinfo {author} {\bibfnamefont
  {K.}~\bibnamefont {Kim}}, \bibinfo {author} {\bibfnamefont {H.}~\bibnamefont
  {Cheong}}, \bibinfo {author} {\bibfnamefont {W.-T.}\ \bibnamefont {Chen}},
  \bibinfo {author} {\bibfnamefont {F.}~\bibnamefont {Chou}}, \bibinfo {author}
  {\bibfnamefont {H.}~\bibnamefont {Nojiri}},\ and\ \bibinfo {author}
  {\bibfnamefont {K.-Y.}\ \bibnamefont {Choi}},\ }\bibfield  {title} {\bibinfo
  {title} {Randomly hopping {M}ajorana fermions in the diluted {K}itaev system
  $\ensuremath{\alpha}$-{R}u$_{0.8}${I}r$_{0.2}$\ce{Cl3}},\ }\href
  {https://doi.org/10.1103/PhysRevLett.124.047204} {\bibfield  {journal}
  {\bibinfo  {journal} {Phys. Rev. Lett.}\ }\textbf {\bibinfo {volume} {124}},\
  \bibinfo {pages} {047204} (\bibinfo {year} {2020})}\BibitemShut {NoStop}%
\bibitem [{\citenamefont {Bahrami}\ \emph {et~al.}(2019)\citenamefont
  {Bahrami}, \citenamefont {Lafargue-Dit-Hauret}, \citenamefont {Lebedev},
  \citenamefont {Movshovich}, \citenamefont {Yang}, \citenamefont {Broido},
  \citenamefont {Rocquefelte},\ and\ \citenamefont {Tafti}}]{BahramiRS}%
  \BibitemOpen
  \bibfield  {author} {\bibinfo {author} {\bibfnamefont {F.}~\bibnamefont
  {Bahrami}}, \bibinfo {author} {\bibfnamefont {W.}~\bibnamefont
  {Lafargue-Dit-Hauret}}, \bibinfo {author} {\bibfnamefont {O.~I.}\
  \bibnamefont {Lebedev}}, \bibinfo {author} {\bibfnamefont {R.}~\bibnamefont
  {Movshovich}}, \bibinfo {author} {\bibfnamefont {H.-Y.}\ \bibnamefont
  {Yang}}, \bibinfo {author} {\bibfnamefont {D.}~\bibnamefont {Broido}},
  \bibinfo {author} {\bibfnamefont {X.}~\bibnamefont {Rocquefelte}},\ and\
  \bibinfo {author} {\bibfnamefont {F.}~\bibnamefont {Tafti}},\ }\bibfield
  {title} {\bibinfo {title} {Thermodynamic evidence of proximity to a {K}itaev
  spin liquid in \ce{Ag3LiIr2O6}},\ }\href
  {https://doi.org/10.1103/PhysRevLett.123.237203} {\bibfield  {journal}
  {\bibinfo  {journal} {Phys. Rev. Lett.}\ }\textbf {\bibinfo {volume} {123}},\
  \bibinfo {pages} {237203} (\bibinfo {year} {2019})}\BibitemShut {NoStop}%
\bibitem [{\citenamefont {Volkov}\ \emph {et~al.}(2020)\citenamefont {Volkov},
  \citenamefont {Won}, \citenamefont {Gorbunov}, \citenamefont {Kim},
  \citenamefont {Ye}, \citenamefont {Kim}, \citenamefont {Pixley},
  \citenamefont {Cheong},\ and\ \citenamefont {Blumberg}}]{Volkov}%
  \BibitemOpen
  \bibfield  {author} {\bibinfo {author} {\bibfnamefont {P.~A.}\ \bibnamefont
  {Volkov}}, \bibinfo {author} {\bibfnamefont {C.-J.}\ \bibnamefont {Won}},
  \bibinfo {author} {\bibfnamefont {D.~I.}\ \bibnamefont {Gorbunov}}, \bibinfo
  {author} {\bibfnamefont {J.}~\bibnamefont {Kim}}, \bibinfo {author}
  {\bibfnamefont {M.}~\bibnamefont {Ye}}, \bibinfo {author} {\bibfnamefont
  {H.-S.}\ \bibnamefont {Kim}}, \bibinfo {author} {\bibfnamefont {J.~H.}\
  \bibnamefont {Pixley}}, \bibinfo {author} {\bibfnamefont {S.-W.}\
  \bibnamefont {Cheong}},\ and\ \bibinfo {author} {\bibfnamefont
  {G.}~\bibnamefont {Blumberg}},\ }\bibfield  {title} {\bibinfo {title} {Random
  singlet state in \ce{Ba5CuIr3O12} single crystals},\ }\href
  {https://doi.org/10.1103/PhysRevB.101.020406} {\bibfield  {journal} {\bibinfo
   {journal} {Phys. Rev. B}\ }\textbf {\bibinfo {volume} {101}},\ \bibinfo
  {pages} {020406} (\bibinfo {year} {2020})}\BibitemShut {NoStop}%
\bibitem [{\citenamefont {Hong}\ \emph {et~al.}(2021)\citenamefont {Hong},
  \citenamefont {Liu}, \citenamefont {Liu}, \citenamefont {Ma}, \citenamefont
  {Koda}, \citenamefont {Li}, \citenamefont {Song}, \citenamefont {Yang},
  \citenamefont {Yang}, \citenamefont {Cheng}, \citenamefont {Zhang},
  \citenamefont {Bao}, \citenamefont {Ma}, \citenamefont {Chen}, \citenamefont
  {Sun}, \citenamefont {Guo}, \citenamefont {Luo}, \citenamefont {Sandvik},\
  and\ \citenamefont {Li}}]{HongRS}%
  \BibitemOpen
  \bibfield  {author} {\bibinfo {author} {\bibfnamefont {W.}~\bibnamefont
  {Hong}}, \bibinfo {author} {\bibfnamefont {L.}~\bibnamefont {Liu}}, \bibinfo
  {author} {\bibfnamefont {C.}~\bibnamefont {Liu}}, \bibinfo {author}
  {\bibfnamefont {X.}~\bibnamefont {Ma}}, \bibinfo {author} {\bibfnamefont
  {A.}~\bibnamefont {Koda}}, \bibinfo {author} {\bibfnamefont {X.}~\bibnamefont
  {Li}}, \bibinfo {author} {\bibfnamefont {J.}~\bibnamefont {Song}}, \bibinfo
  {author} {\bibfnamefont {W.}~\bibnamefont {Yang}}, \bibinfo {author}
  {\bibfnamefont {J.}~\bibnamefont {Yang}}, \bibinfo {author} {\bibfnamefont
  {P.}~\bibnamefont {Cheng}}, \bibinfo {author} {\bibfnamefont
  {H.}~\bibnamefont {Zhang}}, \bibinfo {author} {\bibfnamefont
  {W.}~\bibnamefont {Bao}}, \bibinfo {author} {\bibfnamefont {X.}~\bibnamefont
  {Ma}}, \bibinfo {author} {\bibfnamefont {D.}~\bibnamefont {Chen}}, \bibinfo
  {author} {\bibfnamefont {K.}~\bibnamefont {Sun}}, \bibinfo {author}
  {\bibfnamefont {W.}~\bibnamefont {Guo}}, \bibinfo {author} {\bibfnamefont
  {H.}~\bibnamefont {Luo}}, \bibinfo {author} {\bibfnamefont {A.~W.}\
  \bibnamefont {Sandvik}},\ and\ \bibinfo {author} {\bibfnamefont
  {S.}~\bibnamefont {Li}},\ }\bibfield  {title} {\bibinfo {title} {Extreme
  suppression of antiferromagnetic order and critical scaling in a
  two-dimensional random quantum magnet},\ }\href
  {https://doi.org/10.1103/PhysRevLett.126.037201} {\bibfield  {journal}
  {\bibinfo  {journal} {Phys. Rev. Lett.}\ }\textbf {\bibinfo {volume} {126}},\
  \bibinfo {pages} {037201} (\bibinfo {year} {2021})}\BibitemShut {NoStop}%
\bibitem [{\citenamefont {Yoon}\ \emph {et~al.}(2021)\citenamefont {Yoon},
  \citenamefont {Lee}, \citenamefont {Lee}, \citenamefont {Park}, \citenamefont
  {Lee}, \citenamefont {Choi}, \citenamefont {Do}, \citenamefont {Choi},
  \citenamefont {Chen}, \citenamefont {Chou}, \citenamefont {Gorbunov},
  \citenamefont {Oshima}, \citenamefont {Ali}, \citenamefont {Singh},
  \citenamefont {Berlie}, \citenamefont {Watanabe},\ and\ \citenamefont
  {Choi}}]{YoonRS}%
  \BibitemOpen
  \bibfield  {author} {\bibinfo {author} {\bibfnamefont {S.}~\bibnamefont
  {Yoon}}, \bibinfo {author} {\bibfnamefont {W.}~\bibnamefont {Lee}}, \bibinfo
  {author} {\bibfnamefont {S.}~\bibnamefont {Lee}}, \bibinfo {author}
  {\bibfnamefont {J.}~\bibnamefont {Park}}, \bibinfo {author} {\bibfnamefont
  {C.~H.}\ \bibnamefont {Lee}}, \bibinfo {author} {\bibfnamefont {Y.~S.}\
  \bibnamefont {Choi}}, \bibinfo {author} {\bibfnamefont {S.-H.}\ \bibnamefont
  {Do}}, \bibinfo {author} {\bibfnamefont {W.-J.}\ \bibnamefont {Choi}},
  \bibinfo {author} {\bibfnamefont {W.-T.}\ \bibnamefont {Chen}}, \bibinfo
  {author} {\bibfnamefont {F.}~\bibnamefont {Chou}}, \bibinfo {author}
  {\bibfnamefont {D.~I.}\ \bibnamefont {Gorbunov}}, \bibinfo {author}
  {\bibfnamefont {Y.}~\bibnamefont {Oshima}}, \bibinfo {author} {\bibfnamefont
  {A.}~\bibnamefont {Ali}}, \bibinfo {author} {\bibfnamefont {Y.}~\bibnamefont
  {Singh}}, \bibinfo {author} {\bibfnamefont {A.}~\bibnamefont {Berlie}},
  \bibinfo {author} {\bibfnamefont {I.}~\bibnamefont {Watanabe}},\ and\
  \bibinfo {author} {\bibfnamefont {K.-Y.}\ \bibnamefont {Choi}},\ }\bibfield
  {title} {\bibinfo {title} {Quantum disordered state in the
  ${J}_{1}\text{\ensuremath{-}}{J}_{2}$ square-lattice antiferromagnet
  \ce{Sr2Cu}({Te}$_{0.95}${W}$_{0.05}$)\ce{O6}},\ }\href
  {https://doi.org/10.1103/PhysRevMaterials.5.014411} {\bibfield  {journal}
  {\bibinfo  {journal} {Phys. Rev. Mater.}\ }\textbf {\bibinfo {volume} {5}},\
  \bibinfo {pages} {014411} (\bibinfo {year} {2021})}\BibitemShut {NoStop}%
\bibitem [{\citenamefont {Khatua}\ \emph {et~al.}(2022)\citenamefont {Khatua},
  \citenamefont {Gomil\ifmmode~\check{s}\else \v{s}\fi{}ek}, \citenamefont
  {Orain}, \citenamefont {Strydom}, \citenamefont {Jagli\ifmmode \check{s}\else
  \v{s}\fi{}i\ifmmode~\acute{c}\else \'{c}\fi{}}, \citenamefont {Colin},
  \citenamefont {Petit}, \citenamefont {Ozarowski}, \citenamefont
  {Mangin-Thro}, \citenamefont {Sethupathi}, \citenamefont {Rao}, \citenamefont
  {Zorko},\ and\ \citenamefont {Khuntia}}]{KhatuaRS}%
  \BibitemOpen
  \bibfield  {author} {\bibinfo {author} {\bibfnamefont {J.}~\bibnamefont
  {Khatua}}, \bibinfo {author} {\bibfnamefont {M.}~\bibnamefont
  {Gomil\ifmmode~\check{s}\else \v{s}\fi{}ek}}, \bibinfo {author}
  {\bibfnamefont {J.~C.}\ \bibnamefont {Orain}}, \bibinfo {author}
  {\bibfnamefont {A.~M.}\ \bibnamefont {Strydom}}, \bibinfo {author}
  {\bibfnamefont {Z.}~\bibnamefont {Jagli\ifmmode \check{s}\else
  \v{s}\fi{}i\ifmmode~\acute{c}\else \'{c}\fi{}}}, \bibinfo {author}
  {\bibfnamefont {C.~V.}\ \bibnamefont {Colin}}, \bibinfo {author}
  {\bibfnamefont {S.}~\bibnamefont {Petit}}, \bibinfo {author} {\bibfnamefont
  {A.}~\bibnamefont {Ozarowski}}, \bibinfo {author} {\bibfnamefont
  {L.}~\bibnamefont {Mangin-Thro}}, \bibinfo {author} {\bibfnamefont
  {K.}~\bibnamefont {Sethupathi}}, \bibinfo {author} {\bibfnamefont {M.~S.~R.}\
  \bibnamefont {Rao}}, \bibinfo {author} {\bibfnamefont {A.}~\bibnamefont
  {Zorko}},\ and\ \bibinfo {author} {\bibfnamefont {P.}~\bibnamefont
  {Khuntia}},\ }\bibfield  {title} {\bibinfo {title} {Signature of a
  randomness-driven spin-liquid state in a frustrated magnet},\ }\href
  {https://doi.org/10.1038/s42005-022-00879-2} {\bibfield  {journal} {\bibinfo
  {journal} {Commun. Phys.}\ }\textbf {\bibinfo {volume} {5}},\ \bibinfo
  {pages} {99} (\bibinfo {year} {2022})}\BibitemShut {NoStop}%
\bibitem [{\citenamefont {Malo}\ \emph {et~al.}(2014)\citenamefont {Malo},
  \citenamefont {Abakumov}, \citenamefont {Daturi}, \citenamefont {Pelloquin},
  \citenamefont {Van~Tendeloo}, \citenamefont {Guesdon},\ and\ \citenamefont
  {Hervieu}}]{Malo}%
  \BibitemOpen
  \bibfield  {author} {\bibinfo {author} {\bibfnamefont {S.}~\bibnamefont
  {Malo}}, \bibinfo {author} {\bibfnamefont {A.~M.}\ \bibnamefont {Abakumov}},
  \bibinfo {author} {\bibfnamefont {M.}~\bibnamefont {Daturi}}, \bibinfo
  {author} {\bibfnamefont {D.}~\bibnamefont {Pelloquin}}, \bibinfo {author}
  {\bibfnamefont {G.}~\bibnamefont {Van~Tendeloo}}, \bibinfo {author}
  {\bibfnamefont {A.}~\bibnamefont {Guesdon}},\ and\ \bibinfo {author}
  {\bibfnamefont {M.}~\bibnamefont {Hervieu}},\ }\bibfield  {title} {\bibinfo
  {title} {\ce{Sr21Bi8Cu2(CO3)2O41}, a \(\mathrm{Bi}^{5+}\) oxycarbonate with
  an original 10{L} structure},\ }\href {https://doi.org/10.1021/ic501322w}
  {\bibfield  {journal} {\bibinfo  {journal} {Inorg. Chem.}\ }\textbf {\bibinfo
  {volume} {53}},\ \bibinfo {pages} {10266} (\bibinfo {year}
  {2014})}\BibitemShut {NoStop}%
\bibitem [{sup()}]{suppl}%
  \BibitemOpen
  \href@noop {} {}\bibinfo {note} {See {S}upplemental {M}aterial for additional
  information and supporting evidence.}\BibitemShut {Stop}%
\bibitem [{\citenamefont {Momma}\ and\ \citenamefont {Izumi}(2001)}]{VESTA}%
  \BibitemOpen
  \bibfield  {author} {\bibinfo {author} {\bibfnamefont {K.}~\bibnamefont
  {Momma}}\ and\ \bibinfo {author} {\bibfnamefont {F.}~\bibnamefont {Izumi}},\
  }\bibfield  {title} {\bibinfo {title} {$\textsc{vesta}$ 3 for
  three-dimensional visualization of crystal, volumetric and morphology data},\
  }\href {https://doi.org/10.1107/S0021889811038970} {\bibfield  {journal}
  {\bibinfo  {journal} {J. Appl. Crystallogr.}\ }\textbf {\bibinfo {volume}
  {44}},\ \bibinfo {pages} {1272} (\bibinfo {year} {2001})}\BibitemShut
  {NoStop}%
\bibitem [{\citenamefont {Bulaevskii}(1963)}]{Bulaevskii2}%
  \BibitemOpen
  \bibfield  {author} {\bibinfo {author} {\bibfnamefont {L.~N.}\ \bibnamefont
  {Bulaevskii}},\ }\bibfield  {title} {\bibinfo {title} {Theory of non-uniform
  antiferromagnetic spin chains},\ }\href
  {http://www.jetp.ras.ru/cgi-bin/e/index/e/17/3/p684?a=list} {\bibfield
  {journal} {\bibinfo  {journal} {Sov. Phys. JETP}\ }\textbf {\bibinfo {volume}
  {17}},\ \bibinfo {pages} {684} (\bibinfo {year} {1963})}\BibitemShut
  {NoStop}%
\bibitem [{\citenamefont {Duffy}\ and\ \citenamefont {Barr}(1968)}]{DuffyBarr}%
  \BibitemOpen
  \bibfield  {author} {\bibinfo {author} {\bibfnamefont {W.}~\bibnamefont
  {Duffy}}\ and\ \bibinfo {author} {\bibfnamefont {K.~P.}\ \bibnamefont
  {Barr}},\ }\bibfield  {title} {\bibinfo {title} {Theory of alternating
  antiferromagnetic {H}eisenberg linear chains},\ }\href
  {https://doi.org/10.1103/PhysRev.165.647} {\bibfield  {journal} {\bibinfo
  {journal} {Phys. Rev.}\ }\textbf {\bibinfo {volume} {165}},\ \bibinfo {pages}
  {647} (\bibinfo {year} {1968})}\BibitemShut {NoStop}%
\bibitem [{\citenamefont {Bonner}\ \emph {et~al.}(1979)\citenamefont {Bonner},
  \citenamefont {Bl\"{o}te}, \citenamefont {Bray},\ and\ \citenamefont
  {Jacobs}}]{BonnerChi}%
  \BibitemOpen
  \bibfield  {author} {\bibinfo {author} {\bibfnamefont {J.~C.}\ \bibnamefont
  {Bonner}}, \bibinfo {author} {\bibfnamefont {H.~W.~J.}\ \bibnamefont
  {Bl\"{o}te}}, \bibinfo {author} {\bibfnamefont {J.~W.}\ \bibnamefont
  {Bray}},\ and\ \bibinfo {author} {\bibfnamefont {I.~S.}\ \bibnamefont
  {Jacobs}},\ }\bibfield  {title} {\bibinfo {title} {Susceptibility
  calculations for alternating antiferromagnetic chains},\ }\href
  {https://doi.org/10.1063/1.327177} {\bibfield  {journal} {\bibinfo  {journal}
  {J. Appl. Phys.}\ }\textbf {\bibinfo {volume} {50}},\ \bibinfo {pages} {1810}
  (\bibinfo {year} {1979})}\BibitemShut {NoStop}%
\bibitem [{\citenamefont {Diederix}\ \emph {et~al.}(1979)\citenamefont
  {Diederix}, \citenamefont {Bl\"{o}te}, \citenamefont {Groen}, \citenamefont
  {Klaassen},\ and\ \citenamefont {Poulis}}]{Diederix}%
  \BibitemOpen
  \bibfield  {author} {\bibinfo {author} {\bibfnamefont {K.~M.}\ \bibnamefont
  {Diederix}}, \bibinfo {author} {\bibfnamefont {H.~W.~J.}\ \bibnamefont
  {Bl\"{o}te}}, \bibinfo {author} {\bibfnamefont {J.~P.}\ \bibnamefont
  {Groen}}, \bibinfo {author} {\bibfnamefont {T.~O.}\ \bibnamefont
  {Klaassen}},\ and\ \bibinfo {author} {\bibfnamefont {N.~J.}\ \bibnamefont
  {Poulis}},\ }\bibfield  {title} {\bibinfo {title} {Theoretical and
  experimental study of the magnetic properties of the singlet-ground-state
  system \ce{Cu(NO3)2}$\cdot 2.5$\ce{H2O}: An alternating linear {H}eisenberg
  antiferromagnet},\ }\href {https://doi.org/10.1103/PhysRevB.19.420}
  {\bibfield  {journal} {\bibinfo  {journal} {Phys. Rev. B}\ }\textbf {\bibinfo
  {volume} {19}},\ \bibinfo {pages} {420} (\bibinfo {year} {1979})}\BibitemShut
  {NoStop}%
\bibitem [{\citenamefont {Hyman}\ \emph {et~al.}(1996)\citenamefont {Hyman},
  \citenamefont {Yang}, \citenamefont {Bhatt},\ and\ \citenamefont
  {Girvin}}]{Hyman}%
  \BibitemOpen
  \bibfield  {author} {\bibinfo {author} {\bibfnamefont {R.~A.}\ \bibnamefont
  {Hyman}}, \bibinfo {author} {\bibfnamefont {K.}~\bibnamefont {Yang}},
  \bibinfo {author} {\bibfnamefont {R.~N.}\ \bibnamefont {Bhatt}},\ and\
  \bibinfo {author} {\bibfnamefont {S.~M.}\ \bibnamefont {Girvin}},\ }\bibfield
   {title} {\bibinfo {title} {Random bonds and topological stability in gapped
  quantum spin chains},\ }\href {https://doi.org/10.1103/PhysRevLett.76.839}
  {\bibfield  {journal} {\bibinfo  {journal} {Phys. Rev. Lett.}\ }\textbf
  {\bibinfo {volume} {76}},\ \bibinfo {pages} {839} (\bibinfo {year}
  {1996})}\BibitemShut {NoStop}%
\bibitem [{\citenamefont {Singh}()}]{SinghAppendix}%
  \BibitemOpen
  \bibfield  {author} {\bibinfo {author} {\bibfnamefont {R.~R.~P.}\
  \bibnamefont {Singh}},\ }\href {https://doi.org/10.48550/arXiv.1003.0138}
  {\bibinfo {title} {Valence bond glass phase in dilute kagome
  antiferromagnets}},\ \Eprint {https://arxiv.org/abs/arXiv:1003.0138}
  {arXiv:1003.0138} \BibitemShut {NoStop}%
\bibitem [{\citenamefont {Gegenwart}\ \emph {et~al.}(1999)\citenamefont
  {Gegenwart}, \citenamefont {Kromer}, \citenamefont {Lang}, \citenamefont
  {Sparn}, \citenamefont {Geibel},\ and\ \citenamefont {Steglich}}]{Gegenwart}%
  \BibitemOpen
  \bibfield  {author} {\bibinfo {author} {\bibfnamefont {P.}~\bibnamefont
  {Gegenwart}}, \bibinfo {author} {\bibfnamefont {F.}~\bibnamefont {Kromer}},
  \bibinfo {author} {\bibfnamefont {M.}~\bibnamefont {Lang}}, \bibinfo {author}
  {\bibfnamefont {G.}~\bibnamefont {Sparn}}, \bibinfo {author} {\bibfnamefont
  {C.}~\bibnamefont {Geibel}},\ and\ \bibinfo {author} {\bibfnamefont
  {F.}~\bibnamefont {Steglich}},\ }\bibfield  {title} {\bibinfo {title}
  {Non-{F}ermi-liquid effects at ambient pressure in a stoichiometric
  heavy-fermion compound with very low disorder: {C}e{N}i$_{2}${G}e$_{2}$},\
  }\href {https://doi.org/10.1103/PhysRevLett.82.1293} {\bibfield  {journal}
  {\bibinfo  {journal} {Phys. Rev. Lett.}\ }\textbf {\bibinfo {volume} {82}},\
  \bibinfo {pages} {1293} (\bibinfo {year} {1999})}\BibitemShut {NoStop}%
\bibitem [{\citenamefont {von Löhneysen}(1996)}]{Lohneysen}%
  \BibitemOpen
  \bibfield  {author} {\bibinfo {author} {\bibfnamefont {H.}~\bibnamefont {von
  Löhneysen}},\ }\bibfield  {title} {\bibinfo {title} {Non-{F}ermi-liquid
  behaviour in the heavy-fermion system},\ }\href
  {https://doi.org/10.1088/0953-8984/8/48/003} {\bibfield  {journal} {\bibinfo
  {journal} {J. Phys.: Condens. Matter}\ }\textbf {\bibinfo {volume} {8}},\
  \bibinfo {pages} {9689} (\bibinfo {year} {1996})}\BibitemShut {NoStop}%
\bibitem [{\citenamefont {Fritsch}\ \emph {et~al.}(2014)\citenamefont
  {Fritsch}, \citenamefont {Bagrets}, \citenamefont {Goll}, \citenamefont
  {Kittler}, \citenamefont {Wolf}, \citenamefont {Grube}, \citenamefont
  {Huang},\ and\ \citenamefont {L\"ohneysen}}]{Fritsch}%
  \BibitemOpen
  \bibfield  {author} {\bibinfo {author} {\bibfnamefont {V.}~\bibnamefont
  {Fritsch}}, \bibinfo {author} {\bibfnamefont {N.}~\bibnamefont {Bagrets}},
  \bibinfo {author} {\bibfnamefont {G.}~\bibnamefont {Goll}}, \bibinfo {author}
  {\bibfnamefont {W.}~\bibnamefont {Kittler}}, \bibinfo {author} {\bibfnamefont
  {M.~J.}\ \bibnamefont {Wolf}}, \bibinfo {author} {\bibfnamefont
  {K.}~\bibnamefont {Grube}}, \bibinfo {author} {\bibfnamefont {C.-L.}\
  \bibnamefont {Huang}},\ and\ \bibinfo {author} {\bibfnamefont {H.~v.}\
  \bibnamefont {L\"ohneysen}},\ }\bibfield  {title} {\bibinfo {title}
  {Approaching quantum criticality in a partially geometrically frustrated
  heavy-fermion metal},\ }\href {https://doi.org/10.1103/PhysRevB.89.054416}
  {\bibfield  {journal} {\bibinfo  {journal} {Phys. Rev. B}\ }\textbf {\bibinfo
  {volume} {89}},\ \bibinfo {pages} {054416} (\bibinfo {year}
  {2014})}\BibitemShut {NoStop}%
\bibitem [{\citenamefont {Zhao}\ \emph {et~al.}(2019)\citenamefont {Zhao},
  \citenamefont {Zhang}, \citenamefont {Lyu}, \citenamefont {Bachus},
  \citenamefont {Tokiwa}, \citenamefont {Gegenwart}, \citenamefont {Zhang},
  \citenamefont {Cheng}, \citenamefont {Yang}, \citenamefont {Isikawa},
  \citenamefont {Si}, \citenamefont {Steglich},\ and\ \citenamefont
  {Sun}}]{NatPhysZhao}%
  \BibitemOpen
  \bibfield  {author} {\bibinfo {author} {\bibfnamefont {H.}~\bibnamefont
  {Zhao}}, \bibinfo {author} {\bibfnamefont {J.}~\bibnamefont {Zhang}},
  \bibinfo {author} {\bibfnamefont {M.}~\bibnamefont {Lyu}}, \bibinfo {author}
  {\bibfnamefont {S.}~\bibnamefont {Bachus}}, \bibinfo {author} {\bibfnamefont
  {Y.}~\bibnamefont {Tokiwa}}, \bibinfo {author} {\bibfnamefont
  {P.}~\bibnamefont {Gegenwart}}, \bibinfo {author} {\bibfnamefont
  {S.}~\bibnamefont {Zhang}}, \bibinfo {author} {\bibfnamefont
  {J.}~\bibnamefont {Cheng}}, \bibinfo {author} {\bibfnamefont {Y.-f.}\
  \bibnamefont {Yang}}, \bibinfo {author} {\bibfnamefont {Y.}~\bibnamefont
  {Isikawa}}, \bibinfo {author} {\bibfnamefont {Q.}~\bibnamefont {Si}},
  \bibinfo {author} {\bibfnamefont {F.}~\bibnamefont {Steglich}},\ and\
  \bibinfo {author} {\bibfnamefont {P.}~\bibnamefont {Sun}},\ }\bibfield
  {title} {\bibinfo {title} {Quantum-critical phase from frustrated magnetism
  in a strongly correlated metal},\ }\href
  {https://doi.org/10.1038/s41567-019-0666-6} {\bibfield  {journal} {\bibinfo
  {journal} {Nat. Phys.}\ }\textbf {\bibinfo {volume} {15}},\ \bibinfo {pages}
  {1261} (\bibinfo {year} {2019})}\BibitemShut {NoStop}%
\bibitem [{\citenamefont {Yang}\ \emph {et~al.}(2017)\citenamefont {Yang},
  \citenamefont {Tsuda}, \citenamefont {Umeo}, \citenamefont {Yamane},
  \citenamefont {Onimaru}, \citenamefont {Takabatake}, \citenamefont
  {Kikugawa}, \citenamefont {Terashima},\ and\ \citenamefont {Uji}}]{YangCL}%
  \BibitemOpen
  \bibfield  {author} {\bibinfo {author} {\bibfnamefont {C.~L.}\ \bibnamefont
  {Yang}}, \bibinfo {author} {\bibfnamefont {S.}~\bibnamefont {Tsuda}},
  \bibinfo {author} {\bibfnamefont {K.}~\bibnamefont {Umeo}}, \bibinfo {author}
  {\bibfnamefont {Y.}~\bibnamefont {Yamane}}, \bibinfo {author} {\bibfnamefont
  {T.}~\bibnamefont {Onimaru}}, \bibinfo {author} {\bibfnamefont
  {T.}~\bibnamefont {Takabatake}}, \bibinfo {author} {\bibfnamefont
  {N.}~\bibnamefont {Kikugawa}}, \bibinfo {author} {\bibfnamefont
  {T.}~\bibnamefont {Terashima}},\ and\ \bibinfo {author} {\bibfnamefont
  {S.}~\bibnamefont {Uji}},\ }\bibfield  {title} {\bibinfo {title} {Quantum
  criticality and development of antiferromagnetic order in the quasikagome
  {K}ondo lattice
  $\mathrm{CeR}{\mathrm{h}}_{1\ensuremath{-}x}\mathrm{P}{\mathrm{d}}_{x}\mathrm{Sn}$},\
  }\href {https://doi.org/10.1103/PhysRevB.96.045139} {\bibfield  {journal}
  {\bibinfo  {journal} {Phys. Rev. B}\ }\textbf {\bibinfo {volume} {96}},\
  \bibinfo {pages} {045139} (\bibinfo {year} {2017})}\BibitemShut {NoStop}%
\bibitem [{\citenamefont {Tripathi}\ \emph {et~al.}(2022)\citenamefont
  {Tripathi}, \citenamefont {Adroja}, \citenamefont {Ritter}, \citenamefont
  {Sharma}, \citenamefont {Yang}, \citenamefont {Hillier}, \citenamefont
  {Koza}, \citenamefont {Demmel}, \citenamefont {Sundaresan}, \citenamefont
  {Langridge}, \citenamefont {Higemoto}, \citenamefont {Ito}, \citenamefont
  {Strydom}, \citenamefont {Stenning}, \citenamefont {Bhattacharyya},
  \citenamefont {Keen}, \citenamefont {Walker}, \citenamefont {Perry},
  \citenamefont {Pratt}, \citenamefont {Si},\ and\ \citenamefont
  {Takabatake}}]{Tripathi}%
  \BibitemOpen
  \bibfield  {author} {\bibinfo {author} {\bibfnamefont {R.}~\bibnamefont
  {Tripathi}}, \bibinfo {author} {\bibfnamefont {D.~T.}\ \bibnamefont
  {Adroja}}, \bibinfo {author} {\bibfnamefont {C.}~\bibnamefont {Ritter}},
  \bibinfo {author} {\bibfnamefont {S.}~\bibnamefont {Sharma}}, \bibinfo
  {author} {\bibfnamefont {C.}~\bibnamefont {Yang}}, \bibinfo {author}
  {\bibfnamefont {A.~D.}\ \bibnamefont {Hillier}}, \bibinfo {author}
  {\bibfnamefont {M.~M.}\ \bibnamefont {Koza}}, \bibinfo {author}
  {\bibfnamefont {F.}~\bibnamefont {Demmel}}, \bibinfo {author} {\bibfnamefont
  {A.}~\bibnamefont {Sundaresan}}, \bibinfo {author} {\bibfnamefont
  {S.}~\bibnamefont {Langridge}}, \bibinfo {author} {\bibfnamefont
  {W.}~\bibnamefont {Higemoto}}, \bibinfo {author} {\bibfnamefont {T.~U.}\
  \bibnamefont {Ito}}, \bibinfo {author} {\bibfnamefont {A.~M.}\ \bibnamefont
  {Strydom}}, \bibinfo {author} {\bibfnamefont {G.~B.~G.}\ \bibnamefont
  {Stenning}}, \bibinfo {author} {\bibfnamefont {A.}~\bibnamefont
  {Bhattacharyya}}, \bibinfo {author} {\bibfnamefont {D.}~\bibnamefont {Keen}},
  \bibinfo {author} {\bibfnamefont {H.~C.}\ \bibnamefont {Walker}}, \bibinfo
  {author} {\bibfnamefont {R.~S.}\ \bibnamefont {Perry}}, \bibinfo {author}
  {\bibfnamefont {F.}~\bibnamefont {Pratt}}, \bibinfo {author} {\bibfnamefont
  {Q.}~\bibnamefont {Si}},\ and\ \bibinfo {author} {\bibfnamefont
  {T.}~\bibnamefont {Takabatake}},\ }\bibfield  {title} {\bibinfo {title}
  {Quantum critical spin-liquid-like behavior in the $s=\frac{1}{2}$
  quasikagome-lattice compound
  $\mathrm{CeR}{\mathrm{h}}_{1\ensuremath{-}x}\mathrm{P}{\mathrm{d}}_{x}\mathrm{Sn}$
  investigated using muon spin relaxation and neutron scattering},\ }\href
  {https://doi.org/10.1103/PhysRevB.106.064436} {\bibfield  {journal} {\bibinfo
   {journal} {Phys. Rev. B}\ }\textbf {\bibinfo {volume} {106}},\ \bibinfo
  {pages} {064436} (\bibinfo {year} {2022})}\BibitemShut {NoStop}%
\bibitem [{\citenamefont {Shen}\ \emph {et~al.}(2020)\citenamefont {Shen},
  \citenamefont {Zhang}, \citenamefont {Komijani}, \citenamefont {Nicklas},
  \citenamefont {Borth}, \citenamefont {Wang}, \citenamefont {Chen},
  \citenamefont {Nie}, \citenamefont {Li}, \citenamefont {Lu}, \citenamefont
  {Lee}, \citenamefont {Smidman}, \citenamefont {Steglich}, \citenamefont
  {Coleman},\ and\ \citenamefont {Yuan}}]{NatureShen}%
  \BibitemOpen
  \bibfield  {author} {\bibinfo {author} {\bibfnamefont {B.}~\bibnamefont
  {Shen}}, \bibinfo {author} {\bibfnamefont {Y.}~\bibnamefont {Zhang}},
  \bibinfo {author} {\bibfnamefont {Y.}~\bibnamefont {Komijani}}, \bibinfo
  {author} {\bibfnamefont {M.}~\bibnamefont {Nicklas}}, \bibinfo {author}
  {\bibfnamefont {R.}~\bibnamefont {Borth}}, \bibinfo {author} {\bibfnamefont
  {A.}~\bibnamefont {Wang}}, \bibinfo {author} {\bibfnamefont {Y.}~\bibnamefont
  {Chen}}, \bibinfo {author} {\bibfnamefont {Z.}~\bibnamefont {Nie}}, \bibinfo
  {author} {\bibfnamefont {R.}~\bibnamefont {Li}}, \bibinfo {author}
  {\bibfnamefont {X.}~\bibnamefont {Lu}}, \bibinfo {author} {\bibfnamefont
  {H.}~\bibnamefont {Lee}}, \bibinfo {author} {\bibfnamefont {M.}~\bibnamefont
  {Smidman}}, \bibinfo {author} {\bibfnamefont {F.}~\bibnamefont {Steglich}},
  \bibinfo {author} {\bibfnamefont {P.}~\bibnamefont {Coleman}},\ and\ \bibinfo
  {author} {\bibfnamefont {H.}~\bibnamefont {Yuan}},\ }\bibfield  {title}
  {\bibinfo {title} {Strange-metal behaviour in a pure ferromagnetic {K}ondo
  lattice},\ }\href {https://doi.org/10.1038/s41586-020-2052-z} {\bibfield
  {journal} {\bibinfo  {journal} {Nature (London)}\ }\textbf {\bibinfo {volume}
  {579}},\ \bibinfo {pages} {51} (\bibinfo {year} {2020})}\BibitemShut
  {NoStop}%
\bibitem [{\citenamefont {Singh}\ \emph {et~al.}(2014)\citenamefont {Singh},
  \citenamefont {Yadam}, \citenamefont {Venkateshwarlu}, \citenamefont
  {Gangrade}, \citenamefont {Samatham},\ and\ \citenamefont {Ganesan}}]{Singh}%
  \BibitemOpen
  \bibfield  {author} {\bibinfo {author} {\bibfnamefont {D.}~\bibnamefont
  {Singh}}, \bibinfo {author} {\bibfnamefont {S.}~\bibnamefont {Yadam}},
  \bibinfo {author} {\bibfnamefont {D.}~\bibnamefont {Venkateshwarlu}},
  \bibinfo {author} {\bibfnamefont {M.}~\bibnamefont {Gangrade}}, \bibinfo
  {author} {\bibfnamefont {S.~S.}\ \bibnamefont {Samatham}},\ and\ \bibinfo
  {author} {\bibfnamefont {V.}~\bibnamefont {Ganesan}},\ }\bibfield  {title}
  {\bibinfo {title} {Magnetic field driven quantum critical phase transition in
  {C}e$_{3}${A}l},\ }\href {https://doi.org/10.1088/2053-1591/1/4/046114}
  {\bibfield  {journal} {\bibinfo  {journal} {Mater. Res. Express}\ }\textbf
  {\bibinfo {volume} {1}},\ \bibinfo {pages} {046114} (\bibinfo {year}
  {2014})}\BibitemShut {NoStop}%
\bibitem [{\citenamefont {Freeman}\ \emph {et~al.}(1998)\citenamefont
  {Freeman}, \citenamefont {de~Andrade}, \citenamefont {Dickey}, \citenamefont
  {Dilley},\ and\ \citenamefont {Maple}}]{Freeman}%
  \BibitemOpen
  \bibfield  {author} {\bibinfo {author} {\bibfnamefont {E.~J.}\ \bibnamefont
  {Freeman}}, \bibinfo {author} {\bibfnamefont {M.~C.}\ \bibnamefont
  {de~Andrade}}, \bibinfo {author} {\bibfnamefont {R.~P.}\ \bibnamefont
  {Dickey}}, \bibinfo {author} {\bibfnamefont {N.~R.}\ \bibnamefont {Dilley}},\
  and\ \bibinfo {author} {\bibfnamefont {M.~B.}\ \bibnamefont {Maple}},\
  }\bibfield  {title} {\bibinfo {title} {Non-{F}ermi-liquid behavior and
  magnetic order in the {U}$_{1\ensuremath{-}x}${Y}$_{x}${P}d$_{2}${A}l$_{3}$
  system},\ }\href {https://doi.org/10.1103/PhysRevB.58.16027} {\bibfield
  {journal} {\bibinfo  {journal} {Phys. Rev. B}\ }\textbf {\bibinfo {volume}
  {58}},\ \bibinfo {pages} {16027} (\bibinfo {year} {1998})}\BibitemShut
  {NoStop}%
\bibitem [{\citenamefont {Bernal}\ \emph {et~al.}(1995)\citenamefont {Bernal},
  \citenamefont {MacLaughlin}, \citenamefont {Lukefahr},\ and\ \citenamefont
  {Andraka}}]{Bernal}%
  \BibitemOpen
  \bibfield  {author} {\bibinfo {author} {\bibfnamefont {O.~O.}\ \bibnamefont
  {Bernal}}, \bibinfo {author} {\bibfnamefont {D.~E.}\ \bibnamefont
  {MacLaughlin}}, \bibinfo {author} {\bibfnamefont {H.~G.}\ \bibnamefont
  {Lukefahr}},\ and\ \bibinfo {author} {\bibfnamefont {B.}~\bibnamefont
  {Andraka}},\ }\bibfield  {title} {\bibinfo {title} {Copper {NMR} and
  thermodynamics of {UC}u$_{5\ensuremath{-}\mathit{x}}${P}d$_{\mathit{x}}$:
  Evidence for {K}ondo disorder},\ }\href
  {https://doi.org/10.1103/PhysRevLett.75.2023} {\bibfield  {journal} {\bibinfo
   {journal} {Phys. Rev. Lett.}\ }\textbf {\bibinfo {volume} {75}},\ \bibinfo
  {pages} {2023} (\bibinfo {year} {1995})}\BibitemShut {NoStop}%
\bibitem [{\citenamefont {Weber}\ \emph {et~al.}(2001)\citenamefont {Weber},
  \citenamefont {K\"orner}, \citenamefont {Scheidt}, \citenamefont {Kehrein},\
  and\ \citenamefont {Stewart}}]{Weber}%
  \BibitemOpen
  \bibfield  {author} {\bibinfo {author} {\bibfnamefont {A.}~\bibnamefont
  {Weber}}, \bibinfo {author} {\bibfnamefont {S.}~\bibnamefont {K\"orner}},
  \bibinfo {author} {\bibfnamefont {E.-W.}\ \bibnamefont {Scheidt}}, \bibinfo
  {author} {\bibfnamefont {S.}~\bibnamefont {Kehrein}},\ and\ \bibinfo {author}
  {\bibfnamefont {G.~R.}\ \bibnamefont {Stewart}},\ }\bibfield  {title}
  {\bibinfo {title} {Order and non-{F}ermi-liquid behavior in
  {UC}u$_{4}${P}d},\ }\href {https://doi.org/10.1103/PhysRevB.63.205116}
  {\bibfield  {journal} {\bibinfo  {journal} {Phys. Rev. B}\ }\textbf {\bibinfo
  {volume} {63}},\ \bibinfo {pages} {205116} (\bibinfo {year}
  {2001})}\BibitemShut {NoStop}%
\bibitem [{\citenamefont {Bauer}\ \emph {et~al.}(2005)\citenamefont {Bauer},
  \citenamefont {Zapf}, \citenamefont {Ho}, \citenamefont {Butch},
  \citenamefont {Freeman}, \citenamefont {Sirvent},\ and\ \citenamefont
  {Maple}}]{Bauer}%
  \BibitemOpen
  \bibfield  {author} {\bibinfo {author} {\bibfnamefont {E.~D.}\ \bibnamefont
  {Bauer}}, \bibinfo {author} {\bibfnamefont {V.~S.}\ \bibnamefont {Zapf}},
  \bibinfo {author} {\bibfnamefont {P.-C.}\ \bibnamefont {Ho}}, \bibinfo
  {author} {\bibfnamefont {N.~P.}\ \bibnamefont {Butch}}, \bibinfo {author}
  {\bibfnamefont {E.~J.}\ \bibnamefont {Freeman}}, \bibinfo {author}
  {\bibfnamefont {C.}~\bibnamefont {Sirvent}},\ and\ \bibinfo {author}
  {\bibfnamefont {M.~B.}\ \bibnamefont {Maple}},\ }\bibfield  {title} {\bibinfo
  {title} {Non-{F}ermi-liquid behavior within the ferromagnetic phase in
  {UR}u$_{2\ensuremath{-}x}${R}e$_{x}${S}i$_{2}$},\ }\href
  {https://doi.org/10.1103/PhysRevLett.94.046401} {\bibfield  {journal}
  {\bibinfo  {journal} {Phys. Rev. Lett.}\ }\textbf {\bibinfo {volume} {94}},\
  \bibinfo {pages} {046401} (\bibinfo {year} {2005})}\BibitemShut {NoStop}%
\bibitem [{\citenamefont {Huy}\ \emph {et~al.}(2007)\citenamefont {Huy},
  \citenamefont {Gasparini}, \citenamefont {Klaasse}, \citenamefont
  {de~Visser}, \citenamefont {Sakarya},\ and\ \citenamefont {van Dijk}}]{Huy}%
  \BibitemOpen
  \bibfield  {author} {\bibinfo {author} {\bibfnamefont {N.~T.}\ \bibnamefont
  {Huy}}, \bibinfo {author} {\bibfnamefont {A.}~\bibnamefont {Gasparini}},
  \bibinfo {author} {\bibfnamefont {J.~C.~P.}\ \bibnamefont {Klaasse}},
  \bibinfo {author} {\bibfnamefont {A.}~\bibnamefont {de~Visser}}, \bibinfo
  {author} {\bibfnamefont {S.}~\bibnamefont {Sakarya}},\ and\ \bibinfo {author}
  {\bibfnamefont {N.~H.}\ \bibnamefont {van Dijk}},\ }\bibfield  {title}
  {\bibinfo {title} {Ferromagnetic quantum critical point in {UR}h{G}e doped
  with {R}u},\ }\href {https://doi.org/10.1103/PhysRevB.75.212405} {\bibfield
  {journal} {\bibinfo  {journal} {Phys. Rev. B}\ }\textbf {\bibinfo {volume}
  {75}},\ \bibinfo {pages} {212405} (\bibinfo {year} {2007})}\BibitemShut
  {NoStop}%
\bibitem [{\citenamefont {Matsumoto}\ \emph {et~al.}(2011)\citenamefont
  {Matsumoto}, \citenamefont {Nakatsuji}, \citenamefont {Kuga}, \citenamefont
  {Karaki}, \citenamefont {Horie}, \citenamefont {Shimura}, \citenamefont
  {Sakakibara}, \citenamefont {Nevidomskyy},\ and\ \citenamefont
  {Coleman}}]{Matsumoto}%
  \BibitemOpen
  \bibfield  {author} {\bibinfo {author} {\bibfnamefont {Y.}~\bibnamefont
  {Matsumoto}}, \bibinfo {author} {\bibfnamefont {S.}~\bibnamefont
  {Nakatsuji}}, \bibinfo {author} {\bibfnamefont {K.}~\bibnamefont {Kuga}},
  \bibinfo {author} {\bibfnamefont {Y.}~\bibnamefont {Karaki}}, \bibinfo
  {author} {\bibfnamefont {N.}~\bibnamefont {Horie}}, \bibinfo {author}
  {\bibfnamefont {Y.}~\bibnamefont {Shimura}}, \bibinfo {author} {\bibfnamefont
  {T.}~\bibnamefont {Sakakibara}}, \bibinfo {author} {\bibfnamefont {A.~H.}\
  \bibnamefont {Nevidomskyy}},\ and\ \bibinfo {author} {\bibfnamefont
  {P.}~\bibnamefont {Coleman}},\ }\bibfield  {title} {\bibinfo {title} {Quantum
  criticality without tuning in the mixed valence compound
  $\beta$-{Y}b{A}l{B}$_4$},\ }\href {https://doi.org/10.1126/science.1197531}
  {\bibfield  {journal} {\bibinfo  {journal} {Science}\ }\textbf {\bibinfo
  {volume} {331}},\ \bibinfo {pages} {316} (\bibinfo {year}
  {2011})}\BibitemShut {NoStop}%
\bibitem [{\citenamefont {Nicklas}\ \emph {et~al.}(1999)\citenamefont
  {Nicklas}, \citenamefont {Brando}, \citenamefont {Knebel}, \citenamefont
  {Mayr}, \citenamefont {Trinkl},\ and\ \citenamefont {Loidl}}]{Nicklas}%
  \BibitemOpen
  \bibfield  {author} {\bibinfo {author} {\bibfnamefont {M.}~\bibnamefont
  {Nicklas}}, \bibinfo {author} {\bibfnamefont {M.}~\bibnamefont {Brando}},
  \bibinfo {author} {\bibfnamefont {G.}~\bibnamefont {Knebel}}, \bibinfo
  {author} {\bibfnamefont {F.}~\bibnamefont {Mayr}}, \bibinfo {author}
  {\bibfnamefont {W.}~\bibnamefont {Trinkl}},\ and\ \bibinfo {author}
  {\bibfnamefont {A.}~\bibnamefont {Loidl}},\ }\bibfield  {title} {\bibinfo
  {title} {Non-{F}ermi-liquid behavior at a ferromagnetic quantum critical
  point in {N}i$_{x}${P}d$_{1\ensuremath{-}x}$},\ }\href
  {https://doi.org/10.1103/PhysRevLett.82.4268} {\bibfield  {journal} {\bibinfo
   {journal} {Phys. Rev. Lett.}\ }\textbf {\bibinfo {volume} {82}},\ \bibinfo
  {pages} {4268} (\bibinfo {year} {1999})}\BibitemShut {NoStop}%
\bibitem [{\citenamefont {Yang}\ \emph {et~al.}(2011)\citenamefont {Yang},
  \citenamefont {Chen}, \citenamefont {Ohta}, \citenamefont {Michioka},
  \citenamefont {Yoshimura}, \citenamefont {Wang},\ and\ \citenamefont
  {Fang}}]{YangJ}%
  \BibitemOpen
  \bibfield  {author} {\bibinfo {author} {\bibfnamefont {J.}~\bibnamefont
  {Yang}}, \bibinfo {author} {\bibfnamefont {B.}~\bibnamefont {Chen}}, \bibinfo
  {author} {\bibfnamefont {H.}~\bibnamefont {Ohta}}, \bibinfo {author}
  {\bibfnamefont {C.}~\bibnamefont {Michioka}}, \bibinfo {author}
  {\bibfnamefont {K.}~\bibnamefont {Yoshimura}}, \bibinfo {author}
  {\bibfnamefont {H.}~\bibnamefont {Wang}},\ and\ \bibinfo {author}
  {\bibfnamefont {M.}~\bibnamefont {Fang}},\ }\bibfield  {title} {\bibinfo
  {title} {Spin fluctuations on the verge of a ferromagnetic quantum phase
  transition in {N}i$_3${A}l$_{1\ensuremath{-}x}${G}a$_x$},\ }\href
  {https://doi.org/10.1103/PhysRevB.83.134433} {\bibfield  {journal} {\bibinfo
  {journal} {Phys. Rev. B}\ }\textbf {\bibinfo {volume} {83}},\ \bibinfo
  {pages} {134433} (\bibinfo {year} {2011})}\BibitemShut {NoStop}%
\bibitem [{\citenamefont {Huang}\ \emph {et~al.}(2020)\citenamefont {Huang},
  \citenamefont {Hallas}, \citenamefont {Grube}, \citenamefont {Kuntz},
  \citenamefont {Spie\ss{}}, \citenamefont {Bayliff}, \citenamefont {Besara},
  \citenamefont {Siegrist}, \citenamefont {Cai}, \citenamefont {Beare},
  \citenamefont {Luke},\ and\ \citenamefont {Morosan}}]{Huang}%
  \BibitemOpen
  \bibfield  {author} {\bibinfo {author} {\bibfnamefont {C.-L.}\ \bibnamefont
  {Huang}}, \bibinfo {author} {\bibfnamefont {A.~M.}\ \bibnamefont {Hallas}},
  \bibinfo {author} {\bibfnamefont {K.}~\bibnamefont {Grube}}, \bibinfo
  {author} {\bibfnamefont {S.}~\bibnamefont {Kuntz}}, \bibinfo {author}
  {\bibfnamefont {B.}~\bibnamefont {Spie\ss{}}}, \bibinfo {author}
  {\bibfnamefont {K.}~\bibnamefont {Bayliff}}, \bibinfo {author} {\bibfnamefont
  {T.}~\bibnamefont {Besara}}, \bibinfo {author} {\bibfnamefont
  {T.}~\bibnamefont {Siegrist}}, \bibinfo {author} {\bibfnamefont
  {Y.}~\bibnamefont {Cai}}, \bibinfo {author} {\bibfnamefont {J.}~\bibnamefont
  {Beare}}, \bibinfo {author} {\bibfnamefont {G.~M.}\ \bibnamefont {Luke}},\
  and\ \bibinfo {author} {\bibfnamefont {E.}~\bibnamefont {Morosan}},\
  }\bibfield  {title} {\bibinfo {title} {Quantum critical point in the
  itinerant ferromagnet {N}i$_{1\ensuremath{-}x}${R}h$_x$},\ }\href
  {https://doi.org/10.1103/PhysRevLett.124.117203} {\bibfield  {journal}
  {\bibinfo  {journal} {Phys. Rev. Lett.}\ }\textbf {\bibinfo {volume} {124}},\
  \bibinfo {pages} {117203} (\bibinfo {year} {2020})}\BibitemShut {NoStop}%
\bibitem [{\citenamefont {Brando}\ \emph {et~al.}(2008)\citenamefont {Brando},
  \citenamefont {Duncan}, \citenamefont {Moroni-Klementowicz}, \citenamefont
  {Albrecht}, \citenamefont {Gr\"uner}, \citenamefont {Ballou},\ and\
  \citenamefont {Grosche}}]{Brando}%
  \BibitemOpen
  \bibfield  {author} {\bibinfo {author} {\bibfnamefont {M.}~\bibnamefont
  {Brando}}, \bibinfo {author} {\bibfnamefont {W.~J.}\ \bibnamefont {Duncan}},
  \bibinfo {author} {\bibfnamefont {D.}~\bibnamefont {Moroni-Klementowicz}},
  \bibinfo {author} {\bibfnamefont {C.}~\bibnamefont {Albrecht}}, \bibinfo
  {author} {\bibfnamefont {D.}~\bibnamefont {Gr\"uner}}, \bibinfo {author}
  {\bibfnamefont {R.}~\bibnamefont {Ballou}},\ and\ \bibinfo {author}
  {\bibfnamefont {F.~M.}\ \bibnamefont {Grosche}},\ }\bibfield  {title}
  {\bibinfo {title} {Logarithmic {F}ermi-liquid breakdown in {N}b{F}e$_{2}$},\
  }\href {https://doi.org/10.1103/PhysRevLett.101.026401} {\bibfield  {journal}
  {\bibinfo  {journal} {Phys. Rev. Lett.}\ }\textbf {\bibinfo {volume} {101}},\
  \bibinfo {pages} {026401} (\bibinfo {year} {2008})}\BibitemShut {NoStop}%
\bibitem [{\citenamefont {Moroni-Klementowicz}\ \emph
  {et~al.}(2009)\citenamefont {Moroni-Klementowicz}, \citenamefont {Brando},
  \citenamefont {Albrecht}, \citenamefont {Duncan}, \citenamefont {Grosche},
  \citenamefont {Gr\"uner},\ and\ \citenamefont {Kreiner}}]{Moroni}%
  \BibitemOpen
  \bibfield  {author} {\bibinfo {author} {\bibfnamefont {D.}~\bibnamefont
  {Moroni-Klementowicz}}, \bibinfo {author} {\bibfnamefont {M.}~\bibnamefont
  {Brando}}, \bibinfo {author} {\bibfnamefont {C.}~\bibnamefont {Albrecht}},
  \bibinfo {author} {\bibfnamefont {W.~J.}\ \bibnamefont {Duncan}}, \bibinfo
  {author} {\bibfnamefont {F.~M.}\ \bibnamefont {Grosche}}, \bibinfo {author}
  {\bibfnamefont {D.}~\bibnamefont {Gr\"uner}},\ and\ \bibinfo {author}
  {\bibfnamefont {G.}~\bibnamefont {Kreiner}},\ }\bibfield  {title} {\bibinfo
  {title} {Magnetism in {N}b$_{1\ensuremath{-}y}${F}e$_{2+y}$: Composition and
  magnetic field dependence},\ }\href
  {https://doi.org/10.1103/PhysRevB.79.224410} {\bibfield  {journal} {\bibinfo
  {journal} {Phys. Rev. B}\ }\textbf {\bibinfo {volume} {79}},\ \bibinfo
  {pages} {224410} (\bibinfo {year} {2009})}\BibitemShut {NoStop}%
\bibitem [{\citenamefont {Waki}\ \emph {et~al.}(2010)\citenamefont {Waki},
  \citenamefont {Terazawa}, \citenamefont {Tabata}, \citenamefont {Oba},
  \citenamefont {Michioka}, \citenamefont {Yoshimura}, \citenamefont {Ikeda},
  \citenamefont {Kobayashi}, \citenamefont {Ohoyama},\ and\ \citenamefont
  {Nakamura}}]{Waki}%
  \BibitemOpen
  \bibfield  {author} {\bibinfo {author} {\bibfnamefont {T.}~\bibnamefont
  {Waki}}, \bibinfo {author} {\bibfnamefont {S.}~\bibnamefont {Terazawa}},
  \bibinfo {author} {\bibfnamefont {Y.}~\bibnamefont {Tabata}}, \bibinfo
  {author} {\bibfnamefont {F.}~\bibnamefont {Oba}}, \bibinfo {author}
  {\bibfnamefont {C.}~\bibnamefont {Michioka}}, \bibinfo {author}
  {\bibfnamefont {K.}~\bibnamefont {Yoshimura}}, \bibinfo {author}
  {\bibfnamefont {S.}~\bibnamefont {Ikeda}}, \bibinfo {author} {\bibfnamefont
  {H.}~\bibnamefont {Kobayashi}}, \bibinfo {author} {\bibfnamefont
  {K.}~\bibnamefont {Ohoyama}},\ and\ \bibinfo {author} {\bibfnamefont
  {H.}~\bibnamefont {Nakamura}},\ }\bibfield  {title} {\bibinfo {title}
  {Non-{F}ermi-liquid behavior on an iron-based itinerant electron magnet
  {F}e$_{3}${M}o$_{3}${N}},\ }\href {https://doi.org/10.1143/JPSJ.79.043701}
  {\bibfield  {journal} {\bibinfo  {journal} {J. Phys. Soc. Jpn.}\ }\textbf
  {\bibinfo {volume} {79}},\ \bibinfo {pages} {043701} (\bibinfo {year}
  {2010})}\BibitemShut {NoStop}%
\bibitem [{\citenamefont {Brando}\ \emph {et~al.}(2016)\citenamefont {Brando},
  \citenamefont {Kerkau}, \citenamefont {Todorova}, \citenamefont {Yamada},
  \citenamefont {Khuntia}, \citenamefont {F\"{o}rster}, \citenamefont
  {Burkhard}, \citenamefont {Baenitz},\ and\ \citenamefont
  {Kreiner}}]{Brando2}%
  \BibitemOpen
  \bibfield  {author} {\bibinfo {author} {\bibfnamefont {M.}~\bibnamefont
  {Brando}}, \bibinfo {author} {\bibfnamefont {A.}~\bibnamefont {Kerkau}},
  \bibinfo {author} {\bibfnamefont {A.}~\bibnamefont {Todorova}}, \bibinfo
  {author} {\bibfnamefont {Y.}~\bibnamefont {Yamada}}, \bibinfo {author}
  {\bibfnamefont {P.}~\bibnamefont {Khuntia}}, \bibinfo {author} {\bibfnamefont
  {T.}~\bibnamefont {F\"{o}rster}}, \bibinfo {author} {\bibfnamefont
  {U.}~\bibnamefont {Burkhard}}, \bibinfo {author} {\bibfnamefont
  {M.}~\bibnamefont {Baenitz}},\ and\ \bibinfo {author} {\bibfnamefont
  {G.}~\bibnamefont {Kreiner}},\ }\bibfield  {title} {\bibinfo {title} {Quantum
  phase transitions and multicriticality in
  {T}a({F}e$_{1\ensuremath{-}x}${V}$_x$)$_2$},\ }\href
  {https://doi.org/10.7566/JPSJ.85.084707} {\bibfield  {journal} {\bibinfo
  {journal} {J. Phys. Soc. Jpn.}\ }\textbf {\bibinfo {volume} {85}},\ \bibinfo
  {pages} {084707} (\bibinfo {year} {2016})}\BibitemShut {NoStop}%
\bibitem [{\citenamefont {Wu}\ \emph {et~al.}(2014)\citenamefont {Wu},
  \citenamefont {Kim}, \citenamefont {Park}, \citenamefont {Tsvelik},\ and\
  \citenamefont {Aronson}}]{Wu}%
  \BibitemOpen
  \bibfield  {author} {\bibinfo {author} {\bibfnamefont {L.~S.}\ \bibnamefont
  {Wu}}, \bibinfo {author} {\bibfnamefont {M.~S.}\ \bibnamefont {Kim}},
  \bibinfo {author} {\bibfnamefont {K.}~\bibnamefont {Park}}, \bibinfo {author}
  {\bibfnamefont {A.~M.}\ \bibnamefont {Tsvelik}},\ and\ \bibinfo {author}
  {\bibfnamefont {M.~C.}\ \bibnamefont {Aronson}},\ }\bibfield  {title}
  {\bibinfo {title} {Quantum critical fluctuations in layered
  {YF}e$_2${A}l$_{10}$},\ }\href {https://doi.org/10.1073/pnas.1413112111}
  {\bibfield  {journal} {\bibinfo  {journal} {Proc. Natl. Acad. Sci. USA}\
  }\textbf {\bibinfo {volume} {111}},\ \bibinfo {pages} {14088} (\bibinfo
  {year} {2014})}\BibitemShut {NoStop}%
\bibitem [{\citenamefont {Schoop}\ \emph {et~al.}(2014)\citenamefont {Schoop},
  \citenamefont {Hirschberger}, \citenamefont {Tao}, \citenamefont {Felser},
  \citenamefont {Ong},\ and\ \citenamefont {Cava}}]{Schoop}%
  \BibitemOpen
  \bibfield  {author} {\bibinfo {author} {\bibfnamefont {L.}~\bibnamefont
  {Schoop}}, \bibinfo {author} {\bibfnamefont {M.}~\bibnamefont
  {Hirschberger}}, \bibinfo {author} {\bibfnamefont {J.}~\bibnamefont {Tao}},
  \bibinfo {author} {\bibfnamefont {C.}~\bibnamefont {Felser}}, \bibinfo
  {author} {\bibfnamefont {N.~P.}\ \bibnamefont {Ong}},\ and\ \bibinfo {author}
  {\bibfnamefont {R.~J.}\ \bibnamefont {Cava}},\ }\bibfield  {title} {\bibinfo
  {title} {Paramagnetic to ferromagnetic phase transition in lightly {F}e-doped
  {C}r$_2${B}},\ }\href {https://doi.org/10.1103/PhysRevB.89.224417} {\bibfield
   {journal} {\bibinfo  {journal} {Phys. Rev. B}\ }\textbf {\bibinfo {volume}
  {89}},\ \bibinfo {pages} {224417} (\bibinfo {year} {2014})}\BibitemShut
  {NoStop}%
\bibitem [{\citenamefont {Jia}\ \emph {et~al.}(2011)\citenamefont {Jia},
  \citenamefont {Jiramongkolchai}, \citenamefont {Suchomel}, \citenamefont
  {Toby}, \citenamefont {Checkelsky}, \citenamefont {Ong},\ and\ \citenamefont
  {Cava}}]{Jia}%
  \BibitemOpen
  \bibfield  {author} {\bibinfo {author} {\bibfnamefont {S.}~\bibnamefont
  {Jia}}, \bibinfo {author} {\bibfnamefont {P.}~\bibnamefont
  {Jiramongkolchai}}, \bibinfo {author} {\bibfnamefont {M.~R.}\ \bibnamefont
  {Suchomel}}, \bibinfo {author} {\bibfnamefont {B.~H.}\ \bibnamefont {Toby}},
  \bibinfo {author} {\bibfnamefont {J.~G.}\ \bibnamefont {Checkelsky}},
  \bibinfo {author} {\bibfnamefont {N.~P.}\ \bibnamefont {Ong}},\ and\ \bibinfo
  {author} {\bibfnamefont {R.~J.}\ \bibnamefont {Cava}},\ }\bibfield  {title}
  {\bibinfo {title} {Ferromagnetic quantum critical point induced by
  dimer-breaking in {S}r{C}o$_2$({G}e$_{1\ensuremath{-}x}${P}$_x$)$_2$},\
  }\href {https://doi.org/10.1038/nphys1868} {\bibfield  {journal} {\bibinfo
  {journal} {Nat. Phys.}\ }\textbf {\bibinfo {volume} {7}},\ \bibinfo {pages}
  {207} (\bibinfo {year} {2011})}\BibitemShut {NoStop}%
\bibitem [{\citenamefont {Michon}\ \emph {et~al.}(2019)\citenamefont {Michon},
  \citenamefont {Girod}, \citenamefont {Badoux}, \citenamefont {Kačmarčík},
  \citenamefont {Ma}, \citenamefont {Dragomir}, \citenamefont {Dabkowska},
  \citenamefont {Gaulin}, \citenamefont {Zhou}, \citenamefont {Pyon},
  \citenamefont {Takayama}, \citenamefont {Takagi}, \citenamefont {Verret},
  \citenamefont {Doiron-Leyraud}, \citenamefont {Marcenat}, \citenamefont
  {Taillefer},\ and\ \citenamefont {Klein}}]{Michon}%
  \BibitemOpen
  \bibfield  {author} {\bibinfo {author} {\bibfnamefont {B.}~\bibnamefont
  {Michon}}, \bibinfo {author} {\bibfnamefont {C.}~\bibnamefont {Girod}},
  \bibinfo {author} {\bibfnamefont {S.}~\bibnamefont {Badoux}}, \bibinfo
  {author} {\bibfnamefont {J.}~\bibnamefont {Kačmarčík}}, \bibinfo {author}
  {\bibfnamefont {Q.}~\bibnamefont {Ma}}, \bibinfo {author} {\bibfnamefont
  {M.}~\bibnamefont {Dragomir}}, \bibinfo {author} {\bibfnamefont {H.~A.}\
  \bibnamefont {Dabkowska}}, \bibinfo {author} {\bibfnamefont {B.~D.}\
  \bibnamefont {Gaulin}}, \bibinfo {author} {\bibfnamefont {J.-S.}\
  \bibnamefont {Zhou}}, \bibinfo {author} {\bibfnamefont {S.}~\bibnamefont
  {Pyon}}, \bibinfo {author} {\bibfnamefont {T.}~\bibnamefont {Takayama}},
  \bibinfo {author} {\bibfnamefont {H.}~\bibnamefont {Takagi}}, \bibinfo
  {author} {\bibfnamefont {S.}~\bibnamefont {Verret}}, \bibinfo {author}
  {\bibfnamefont {N.}~\bibnamefont {Doiron-Leyraud}}, \bibinfo {author}
  {\bibfnamefont {C.}~\bibnamefont {Marcenat}}, \bibinfo {author}
  {\bibfnamefont {L.}~\bibnamefont {Taillefer}},\ and\ \bibinfo {author}
  {\bibfnamefont {T.}~\bibnamefont {Klein}},\ }\bibfield  {title} {\bibinfo
  {title} {Thermodynamic signatures of quantum criticality in cuprate
  superconductors},\ }\href {https://doi.org/10.1038/s41586-019-0932-x}
  {\bibfield  {journal} {\bibinfo  {journal} {Nature (London)}\ }\textbf
  {\bibinfo {volume} {567}},\ \bibinfo {pages} {218} (\bibinfo {year}
  {2019})}\BibitemShut {NoStop}%
\bibitem [{\citenamefont {Girod}\ \emph {et~al.}(2021)\citenamefont {Girod},
  \citenamefont {LeBoeuf}, \citenamefont {Demuer}, \citenamefont {Seyfarth},
  \citenamefont {Imajo}, \citenamefont {Kindo}, \citenamefont {Kohama},
  \citenamefont {Lizaire}, \citenamefont {Legros}, \citenamefont {Gourgout},
  \citenamefont {Takagi}, \citenamefont {Kurosawa}, \citenamefont {Oda},
  \citenamefont {Momono}, \citenamefont {Chang}, \citenamefont {Ono},
  \citenamefont {Zheng}, \citenamefont {Marcenat}, \citenamefont {Taillefer},\
  and\ \citenamefont {Klein}}]{Girod}%
  \BibitemOpen
  \bibfield  {author} {\bibinfo {author} {\bibfnamefont {C.}~\bibnamefont
  {Girod}}, \bibinfo {author} {\bibfnamefont {D.}~\bibnamefont {LeBoeuf}},
  \bibinfo {author} {\bibfnamefont {A.}~\bibnamefont {Demuer}}, \bibinfo
  {author} {\bibfnamefont {G.}~\bibnamefont {Seyfarth}}, \bibinfo {author}
  {\bibfnamefont {S.}~\bibnamefont {Imajo}}, \bibinfo {author} {\bibfnamefont
  {K.}~\bibnamefont {Kindo}}, \bibinfo {author} {\bibfnamefont
  {Y.}~\bibnamefont {Kohama}}, \bibinfo {author} {\bibfnamefont
  {M.}~\bibnamefont {Lizaire}}, \bibinfo {author} {\bibfnamefont
  {A.}~\bibnamefont {Legros}}, \bibinfo {author} {\bibfnamefont
  {A.}~\bibnamefont {Gourgout}}, \bibinfo {author} {\bibfnamefont
  {H.}~\bibnamefont {Takagi}}, \bibinfo {author} {\bibfnamefont
  {T.}~\bibnamefont {Kurosawa}}, \bibinfo {author} {\bibfnamefont
  {M.}~\bibnamefont {Oda}}, \bibinfo {author} {\bibfnamefont {N.}~\bibnamefont
  {Momono}}, \bibinfo {author} {\bibfnamefont {J.}~\bibnamefont {Chang}},
  \bibinfo {author} {\bibfnamefont {S.}~\bibnamefont {Ono}}, \bibinfo {author}
  {\bibfnamefont {G.-q.}\ \bibnamefont {Zheng}}, \bibinfo {author}
  {\bibfnamefont {C.}~\bibnamefont {Marcenat}}, \bibinfo {author}
  {\bibfnamefont {L.}~\bibnamefont {Taillefer}},\ and\ \bibinfo {author}
  {\bibfnamefont {T.}~\bibnamefont {Klein}},\ }\bibfield  {title} {\bibinfo
  {title} {Normal state specific heat in the cuprate superconductors
  {L}$_{2\ensuremath{-}x}${S}r$_x${C}u{O}$_4$ and
  {B}$_{2+y}${S}r$_{2\ensuremath{-}x\ensuremath{-}y}${L}a$_x${C}u{O}$_{6+\ensuremath{\delta}}$
  near the critical point of the pseudogap phase},\ }\href
  {https://doi.org/10.1103/PhysRevB.103.214506} {\bibfield  {journal} {\bibinfo
   {journal} {Phys. Rev. B}\ }\textbf {\bibinfo {volume} {103}},\ \bibinfo
  {pages} {214506} (\bibinfo {year} {2021})}\BibitemShut {NoStop}%
\bibitem [{\citenamefont {Brühwiler}\ \emph {et~al.}(2006)\citenamefont
  {Brühwiler}, \citenamefont {Batlogg}, \citenamefont {Kazakov}, \citenamefont
  {Niedermayer},\ and\ \citenamefont {Karpinski}}]{Bruhwiler}%
  \BibitemOpen
  \bibfield  {author} {\bibinfo {author} {\bibfnamefont {M.}~\bibnamefont
  {Brühwiler}}, \bibinfo {author} {\bibfnamefont {B.}~\bibnamefont {Batlogg}},
  \bibinfo {author} {\bibfnamefont {S.}~\bibnamefont {Kazakov}}, \bibinfo
  {author} {\bibfnamefont {C.}~\bibnamefont {Niedermayer}},\ and\ \bibinfo
  {author} {\bibfnamefont {J.}~\bibnamefont {Karpinski}},\ }\bibfield  {title}
  {\bibinfo {title} {{N}a$_x${C}o{O}$_2$: {E}nhanced low-energy excitations of
  electrons on a 2d triangular lattice},\ }\href
  {https://doi.org/10.1016/j.physb.2006.01.422} {\bibfield  {journal} {\bibinfo
   {journal} {Physica B Condens. Matter}\ }\textbf {\bibinfo {volume}
  {378-380}},\ \bibinfo {pages} {630} (\bibinfo {year} {2006})}\BibitemShut
  {NoStop}%
\bibitem [{\citenamefont {Balicas}\ \emph {et~al.}(2008)\citenamefont
  {Balicas}, \citenamefont {Jo}, \citenamefont {Shu}, \citenamefont {Chou},\
  and\ \citenamefont {Lee}}]{Balicas}%
  \BibitemOpen
  \bibfield  {author} {\bibinfo {author} {\bibfnamefont {L.}~\bibnamefont
  {Balicas}}, \bibinfo {author} {\bibfnamefont {Y.~J.}\ \bibnamefont {Jo}},
  \bibinfo {author} {\bibfnamefont {G.~J.}\ \bibnamefont {Shu}}, \bibinfo
  {author} {\bibfnamefont {F.~C.}\ \bibnamefont {Chou}},\ and\ \bibinfo
  {author} {\bibfnamefont {P.~A.}\ \bibnamefont {Lee}},\ }\bibfield  {title}
  {\bibinfo {title} {Local moment, itinerancy, and deviation from
  {F}ermi-liquid behavior in {N}a$_x${C}o{O}$_2$ for
  $0.71\ensuremath{\le}x\ensuremath{\le}0.84$},\ }\href
  {https://doi.org/10.1103/PhysRevLett.100.126405} {\bibfield  {journal}
  {\bibinfo  {journal} {Phys. Rev. Lett.}\ }\textbf {\bibinfo {volume} {100}},\
  \bibinfo {pages} {126405} (\bibinfo {year} {2008})}\BibitemShut {NoStop}%
\bibitem [{\citenamefont {Okamoto}\ \emph {et~al.}(2010)\citenamefont
  {Okamoto}, \citenamefont {Nishio},\ and\ \citenamefont {Hiroi}}]{Okamoto}%
  \BibitemOpen
  \bibfield  {author} {\bibinfo {author} {\bibfnamefont {Y.}~\bibnamefont
  {Okamoto}}, \bibinfo {author} {\bibfnamefont {A.}~\bibnamefont {Nishio}},\
  and\ \bibinfo {author} {\bibfnamefont {Z.}~\bibnamefont {Hiroi}},\ }\bibfield
   {title} {\bibinfo {title} {Discontinuous {L}ifshitz transition achieved by
  band-filling control in {N}a$_x${C}o{O}$_2$},\ }\href
  {https://doi.org/10.1103/PhysRevB.81.121102} {\bibfield  {journal} {\bibinfo
  {journal} {Phys. Rev. B}\ }\textbf {\bibinfo {volume} {81}},\ \bibinfo
  {pages} {121102} (\bibinfo {year} {2010})}\BibitemShut {NoStop}%
\bibitem [{\citenamefont {Rost}\ \emph {et~al.}(2011)\citenamefont {Rost},
  \citenamefont {Grigera}, \citenamefont {Bruin}, \citenamefont {Perry},
  \citenamefont {Tian}, \citenamefont {Raghu}, \citenamefont {Kivelson},\ and\
  \citenamefont {Mackenzie}}]{PNASQCrit11}%
  \BibitemOpen
  \bibfield  {author} {\bibinfo {author} {\bibfnamefont {A.~W.}\ \bibnamefont
  {Rost}}, \bibinfo {author} {\bibfnamefont {S.~A.}\ \bibnamefont {Grigera}},
  \bibinfo {author} {\bibfnamefont {J.~A.~N.}\ \bibnamefont {Bruin}}, \bibinfo
  {author} {\bibfnamefont {R.~S.}\ \bibnamefont {Perry}}, \bibinfo {author}
  {\bibfnamefont {D.}~\bibnamefont {Tian}}, \bibinfo {author} {\bibfnamefont
  {S.}~\bibnamefont {Raghu}}, \bibinfo {author} {\bibfnamefont {S.~A.}\
  \bibnamefont {Kivelson}},\ and\ \bibinfo {author} {\bibfnamefont {A.~P.}\
  \bibnamefont {Mackenzie}},\ }\bibfield  {title} {\bibinfo {title}
  {Thermodynamics of phase formation in the quantum critical metal
  \ce{Sr3Ru2O7}},\ }\href {https://doi.org/10.1073/pnas.1112775108} {\bibfield
  {journal} {\bibinfo  {journal} {Proc. Natl. Acad. Sci. USA}\ }\textbf
  {\bibinfo {volume} {108}},\ \bibinfo {pages} {16549} (\bibinfo {year}
  {2011})}\BibitemShut {NoStop}%
\bibitem [{\citenamefont {Sun}\ \emph {et~al.}(2018)\citenamefont {Sun},
  \citenamefont {Rost}, \citenamefont {Perry}, \citenamefont {Mackenzie},\ and\
  \citenamefont {Brando}}]{PRBQCrit18}%
  \BibitemOpen
  \bibfield  {author} {\bibinfo {author} {\bibfnamefont {D.}~\bibnamefont
  {Sun}}, \bibinfo {author} {\bibfnamefont {A.~W.}\ \bibnamefont {Rost}},
  \bibinfo {author} {\bibfnamefont {R.~S.}\ \bibnamefont {Perry}}, \bibinfo
  {author} {\bibfnamefont {A.~P.}\ \bibnamefont {Mackenzie}},\ and\ \bibinfo
  {author} {\bibfnamefont {M.}~\bibnamefont {Brando}},\ }\bibfield  {title}
  {\bibinfo {title} {Low temperature thermodynamic investigation of the phase
  diagram of \ce{Sr3Ru2O7}},\ }\href
  {https://doi.org/10.1103/PhysRevB.97.115101} {\bibfield  {journal} {\bibinfo
  {journal} {Phys. Rev. B}\ }\textbf {\bibinfo {volume} {97}},\ \bibinfo
  {pages} {115101} (\bibinfo {year} {2018})}\BibitemShut {NoStop}%
\end{thebibliography}%


\begin{thebibliography}{11}%
\makeatletter
\providecommand \@ifxundefined [1]{%
 \@ifx{#1\undefined}
}%
\providecommand \@ifnum [1]{%
 \ifnum #1\expandafter \@firstoftwo
 \else \expandafter \@secondoftwo
 \fi
}%
\providecommand \@ifx [1]{%
 \ifx #1\expandafter \@firstoftwo
 \else \expandafter \@secondoftwo
 \fi
}%
\providecommand \natexlab [1]{#1}%
\providecommand \enquote  [1]{``#1''}%
\providecommand \bibnamefont  [1]{#1}%
\providecommand \bibfnamefont [1]{#1}%
\providecommand \citenamefont [1]{#1}%
\providecommand \href@noop [0]{\@secondoftwo}%
\providecommand \href [0]{\begingroup \@sanitize@url \@href}%
\providecommand \@href[1]{\@@startlink{#1}\@@href}%
\providecommand \@@href[1]{\endgroup#1\@@endlink}%
\providecommand \@sanitize@url [0]{\catcode `\\12\catcode `\$12\catcode
  `\&12\catcode `\#12\catcode `\^12\catcode `\_12\catcode `\%12\relax}%
\providecommand \@@startlink[1]{}%
\providecommand \@@endlink[0]{}%
\providecommand \url  [0]{\begingroup\@sanitize@url \@url }%
\providecommand \@url [1]{\endgroup\@href {#1}{\urlprefix }}%
\providecommand \urlprefix  [0]{URL }%
\providecommand \Eprint [0]{\href }%
\providecommand \doibase [0]{https://doi.org/}%
\providecommand \selectlanguage [0]{\@gobble}%
\providecommand \bibinfo  [0]{\@secondoftwo}%
\providecommand \bibfield  [0]{\@secondoftwo}%
\providecommand \translation [1]{[#1]}%
\providecommand \BibitemOpen [0]{}%
\providecommand \bibitemStop [0]{}%
\providecommand \bibitemNoStop [0]{.\EOS\space}%
\providecommand \EOS [0]{\spacefactor3000\relax}%
\providecommand \BibitemShut  [1]{\csname bibitem#1\endcsname}%
\let\auto@bib@innerbib\@empty
\bibitem [{\citenamefont {Malo}\ \emph {et~al.}(2014)\citenamefont {Malo},
  \citenamefont {Abakumov}, \citenamefont {Daturi}, \citenamefont {Pelloquin},
  \citenamefont {Van~Tendeloo}, \citenamefont {Guesdon},\ and\ \citenamefont
  {Hervieu}}]{Malo}%
  \BibitemOpen
  \bibfield  {author} {\bibinfo {author} {\bibfnamefont {S.}~\bibnamefont
  {Malo}}, \bibinfo {author} {\bibfnamefont {A.~M.}\ \bibnamefont {Abakumov}},
  \bibinfo {author} {\bibfnamefont {M.}~\bibnamefont {Daturi}}, \bibinfo
  {author} {\bibfnamefont {D.}~\bibnamefont {Pelloquin}}, \bibinfo {author}
  {\bibfnamefont {G.}~\bibnamefont {Van~Tendeloo}}, \bibinfo {author}
  {\bibfnamefont {A.}~\bibnamefont {Guesdon}},\ and\ \bibinfo {author}
  {\bibfnamefont {M.}~\bibnamefont {Hervieu}},\ }\bibfield  {title} {\bibinfo
  {title} {\ce{Sr21Bi8Cu2(CO3)2O41}, a \(\mathrm{Bi}^{5+}\) oxycarbonate with
  an original 10{L} structure},\ }\href {https://doi.org/10.1021/ic501322w}
  {\bibfield  {journal} {\bibinfo  {journal} {Inorg. Chem.}\ }\textbf {\bibinfo
  {volume} {53}},\ \bibinfo {pages} {10266} (\bibinfo {year}
  {2014})}\BibitemShut {NoStop}%
\bibitem [{\citenamefont {Scherrer}(1918)}]{scherrer}%
  \BibitemOpen
  \bibfield  {author} {\bibinfo {author} {\bibfnamefont {P.}~\bibnamefont
  {Scherrer}},\ }\bibfield  {title} {\bibinfo {title} {Bestimmung der {G}röße
  und der inneren {S}truktur von {K}olloidteilchen mittels
  {R}öntgenstrahlen},\ }\href {http://eudml.org/doc/59018} {\bibfield
  {journal} {\bibinfo  {journal} {Nachr. Ges. Wiss. Göttingen}\ }\textbf
  {\bibinfo {volume} {1918}},\ \bibinfo {pages} {98} (\bibinfo {year}
  {1918})}\BibitemShut {NoStop}%
\bibitem [{\citenamefont {Langford}\ and\ \citenamefont
  {Wilson}(1978)}]{langford}%
  \BibitemOpen
  \bibfield  {author} {\bibinfo {author} {\bibfnamefont {J.~I.}\ \bibnamefont
  {Langford}}\ and\ \bibinfo {author} {\bibfnamefont {A.~J.~C.}\ \bibnamefont
  {Wilson}},\ }\bibfield  {title} {\bibinfo {title} {{Scherrer after sixty
  years: A survey and some new results in the determination of crystallite
  size}},\ }\href {https://doi.org/10.1107/S0021889878012844} {\bibfield
  {journal} {\bibinfo  {journal} {J. Appl. Cryst.}\ }\textbf {\bibinfo {volume}
  {11}},\ \bibinfo {pages} {102} (\bibinfo {year} {1978})}\BibitemShut
  {NoStop}%
\bibitem [{\citenamefont {Bain}\ and\ \citenamefont {Berry}(2008)}]{Pascal}%
  \BibitemOpen
  \bibfield  {author} {\bibinfo {author} {\bibfnamefont {G.~A.}\ \bibnamefont
  {Bain}}\ and\ \bibinfo {author} {\bibfnamefont {J.~F.}\ \bibnamefont
  {Berry}},\ }\bibfield  {title} {\bibinfo {title} {Diamagnetic corrections and
  {P}ascal's constants},\ }\href {https://doi.org/10.1021/ed085p532} {\bibfield
   {journal} {\bibinfo  {journal} {J. Chem. Educ.}\ }\textbf {\bibinfo {volume}
  {85}},\ \bibinfo {pages} {532} (\bibinfo {year} {2008})}\BibitemShut
  {NoStop}%
\bibitem [{\citenamefont {Abragam}(1961)}]{Abragam}%
  \BibitemOpen
  \bibfield  {author} {\bibinfo {author} {\bibfnamefont {A.}~\bibnamefont
  {Abragam}},\ }\href@noop {} {\emph {\bibinfo {title} {Principles of Nuclear
  Magnetism}}}\ (\bibinfo  {publisher} {Oxford University Press},\ \bibinfo
  {address} {Oxford},\ \bibinfo {year} {1961})\BibitemShut {NoStop}%
\bibitem [{\citenamefont {Pyykkö}(2008)}]{Pyykko}%
  \BibitemOpen
  \bibfield  {author} {\bibinfo {author} {\bibfnamefont {P.}~\bibnamefont
  {Pyykkö}},\ }\bibfield  {title} {\bibinfo {title} {Year-2008 nuclear
  quadrupole moments},\ }\href {https://doi.org/10.1080/00268970802018367}
  {\bibfield  {journal} {\bibinfo  {journal} {Mol. Phys.}\ }\textbf {\bibinfo
  {volume} {106}},\ \bibinfo {pages} {1965} (\bibinfo {year}
  {2008})}\BibitemShut {NoStop}%
\bibitem [{\citenamefont {Andres}\ \emph {et~al.}(1981)\citenamefont {Andres},
  \citenamefont {Bhatt}, \citenamefont {Goalwin}, \citenamefont {Rice},\ and\
  \citenamefont {Walstedt}}]{Andres}%
  \BibitemOpen
  \bibfield  {author} {\bibinfo {author} {\bibfnamefont {K.}~\bibnamefont
  {Andres}}, \bibinfo {author} {\bibfnamefont {R.~N.}\ \bibnamefont {Bhatt}},
  \bibinfo {author} {\bibfnamefont {P.}~\bibnamefont {Goalwin}}, \bibinfo
  {author} {\bibfnamefont {T.~M.}\ \bibnamefont {Rice}},\ and\ \bibinfo
  {author} {\bibfnamefont {R.~E.}\ \bibnamefont {Walstedt}},\ }\bibfield
  {title} {\bibinfo {title} {Low-temperature magnetic susceptibility of {S}i:
  {P} in the nonmetallic region},\ }\href
  {https://doi.org/10.1103/PhysRevB.24.244} {\bibfield  {journal} {\bibinfo
  {journal} {Phys. Rev. B}\ }\textbf {\bibinfo {volume} {24}},\ \bibinfo
  {pages} {244} (\bibinfo {year} {1981})}\BibitemShut {NoStop}%
\bibitem [{\citenamefont {Singh}()}]{SinghAppendix}%
  \BibitemOpen
  \bibfield  {author} {\bibinfo {author} {\bibfnamefont {R.~R.~P.}\
  \bibnamefont {Singh}},\ }\href {https://doi.org/10.48550/arXiv.1003.0138}
  {\bibinfo {title} {Valence bond glass phase in dilute kagome
  antiferromagnets}},\ \Eprint {https://arxiv.org/abs/arXiv:1003.0138}
  {arXiv:1003.0138} \BibitemShut {NoStop}%
\bibitem [{\citenamefont {Kimchi}\ \emph {et~al.}(2018)\citenamefont {Kimchi},
  \citenamefont {Nahum},\ and\ \citenamefont {Senthil}}]{KimchiPRX}%
  \BibitemOpen
  \bibfield  {author} {\bibinfo {author} {\bibfnamefont {I.}~\bibnamefont
  {Kimchi}}, \bibinfo {author} {\bibfnamefont {A.}~\bibnamefont {Nahum}},\ and\
  \bibinfo {author} {\bibfnamefont {T.}~\bibnamefont {Senthil}},\ }\bibfield
  {title} {\bibinfo {title} {Valence bonds in random quantum magnets: Theory
  and application to \ce{YbMgGaO4}},\ }\href
  {https://doi.org/10.1103/PhysRevX.8.031028} {\bibfield  {journal} {\bibinfo
  {journal} {Phys. Rev. X}\ }\textbf {\bibinfo {volume} {8}},\ \bibinfo {pages}
  {031028} (\bibinfo {year} {2018})}\BibitemShut {NoStop}%
\bibitem [{\citenamefont {Scheven}\ \emph {et~al.}(1997)\citenamefont
  {Scheven}, \citenamefont {Hannahs}, \citenamefont {Immer},\ and\
  \citenamefont {Chaikin}}]{SchevenMagcal}%
  \BibitemOpen
  \bibfield  {author} {\bibinfo {author} {\bibfnamefont {U.~M.}\ \bibnamefont
  {Scheven}}, \bibinfo {author} {\bibfnamefont {S.~T.}\ \bibnamefont
  {Hannahs}}, \bibinfo {author} {\bibfnamefont {C.}~\bibnamefont {Immer}},\
  and\ \bibinfo {author} {\bibfnamefont {P.~M.}\ \bibnamefont {Chaikin}},\
  }\bibfield  {title} {\bibinfo {title} {Thermodynamics in the high-field
  phases of ${(\mathrm{TMTSF})}_{2}$\ce{ClO4}},\ }\href
  {https://doi.org/10.1103/PhysRevB.56.7804} {\bibfield  {journal} {\bibinfo
  {journal} {Phys. Rev. B}\ }\textbf {\bibinfo {volume} {56}},\ \bibinfo
  {pages} {7804} (\bibinfo {year} {1997})}\BibitemShut {NoStop}%
\bibitem [{\citenamefont {Fortune}\ \emph {et~al.}(2009)\citenamefont
  {Fortune}, \citenamefont {Hannahs}, \citenamefont {Yoshida}, \citenamefont
  {Sherline}, \citenamefont {Ono}, \citenamefont {Tanaka},\ and\ \citenamefont
  {Takano}}]{FortuneMagcal}%
  \BibitemOpen
  \bibfield  {author} {\bibinfo {author} {\bibfnamefont {N.~A.}\ \bibnamefont
  {Fortune}}, \bibinfo {author} {\bibfnamefont {S.~T.}\ \bibnamefont
  {Hannahs}}, \bibinfo {author} {\bibfnamefont {Y.}~\bibnamefont {Yoshida}},
  \bibinfo {author} {\bibfnamefont {T.~E.}\ \bibnamefont {Sherline}}, \bibinfo
  {author} {\bibfnamefont {T.}~\bibnamefont {Ono}}, \bibinfo {author}
  {\bibfnamefont {H.}~\bibnamefont {Tanaka}},\ and\ \bibinfo {author}
  {\bibfnamefont {Y.}~\bibnamefont {Takano}},\ }\bibfield  {title} {\bibinfo
  {title} {Cascade of magnetic-field-induced quantum phase transitions in a
  spin-$\frac{1}{2}$ triangular-lattice antiferromagnet},\ }\href
  {https://doi.org/10.1103/PhysRevLett.102.257201} {\bibfield  {journal}
  {\bibinfo  {journal} {Phys. Rev. Lett.}\ }\textbf {\bibinfo {volume} {102}},\
  \bibinfo {pages} {257201} (\bibinfo {year} {2009})}\BibitemShut {NoStop}%
\end{thebibliography}%
\end{document}


\preprint{APS/123-QED}

\title{Formation of random singlets in the nanocrystalline quasi-one-dimensional spin-1/2 antiferromagnet \ce{Sr21Bi8Cu2(CO3)2O41}: Supplemental Material}

\author{Yanbo~Guo}
\author{Xinzhe~Hu}
\altaffiliation[Present address: ]{Institute of Physics, Chinese Academy of Sciences, Beijing 100190, China.}
\affiliation{Department of Physics, University of Florida, Gainesville, Florida 32611-8440, USA}
\author{Hasan~Siddiquee}
\altaffiliation[Present address: ]{Nokia Bell Labs, New Providence, New Jersey 07974, USA.}
\author{Kapila Kumarasinghe}
\affiliation{Department of Physics, University of Central Florida, Orlando, Florida 32816, USA}
\author{Swapnil~M.~Yadav}
\altaffiliation[Present address: ]{Walmart Global Tech, Sunnyvale, California 94086, USA.}
\affiliation{Department of Physics, University of Florida, Gainesville, Florida 32611-8440, USA}
\author{Eun Sang~Choi}
\affiliation{National High Magnetic Field Laboratory, Tallahassee, Florida 32310, USA}
\author{Yasuyuki~Nakajima}
\affiliation{Department of Physics, University of Central Florida, Orlando, Florida 32816, USA}
\author{Yasumasa~Takano}
\affiliation{Department of Physics, University of Florida, Gainesville, Florida 32611-8440, USA}
\date{\today}

\maketitle

\begin{center}
\bf{1. X-ray diffraction}
\end{center}

\begin{figure}[h!]
	\centering
		\includegraphics[width=0.8\linewidth]{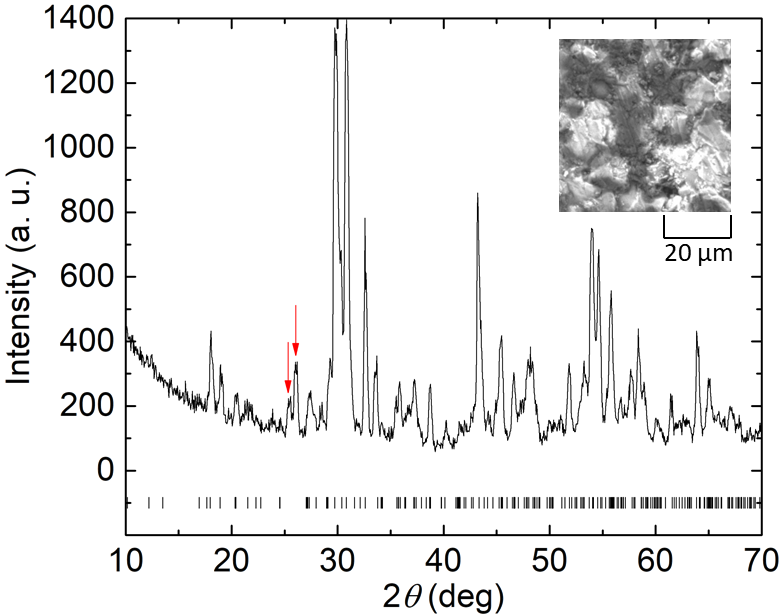}
	\caption{\label{fig:sg1}
		Powder X-ray diffraction pattern of the sample. Tick marks represent peak positions according to the crystal structure of \ce{Sr21Bi8Cu2(CO3)2O41}, reported in Ref.~\cite{Malo}. Red arrows indicate 
		peaks 
		associated with an unidentified impurity phase, also reported in Ref.~\cite{Malo}. Inset: scanning electron micrograph of the sample.}
\end{figure}

Figure \ref{fig:sg1} shows powder X-ray diffraction data collected at room temperature by using monochromatic Cu\,K\(\alpha\) (\(\lambda\) = \SI{1.541874}{\AA}) X-rays. We estimate the average crystallite size from 36 peaks in our X-ray diffraction data using Scherrer's formula \cite{scherrer,langford}, obtaining the value given in the main text. The contribution of the instrument to the peak broadening was determined by a \ce{LaB6} standard.

\begin{center}
	\bf{2. Curie-Weiss fit of the magnetic susceptibility}
\end{center}

\begin{figure}[h!]
	\centering
	\includegraphics[width=0.9\linewidth]{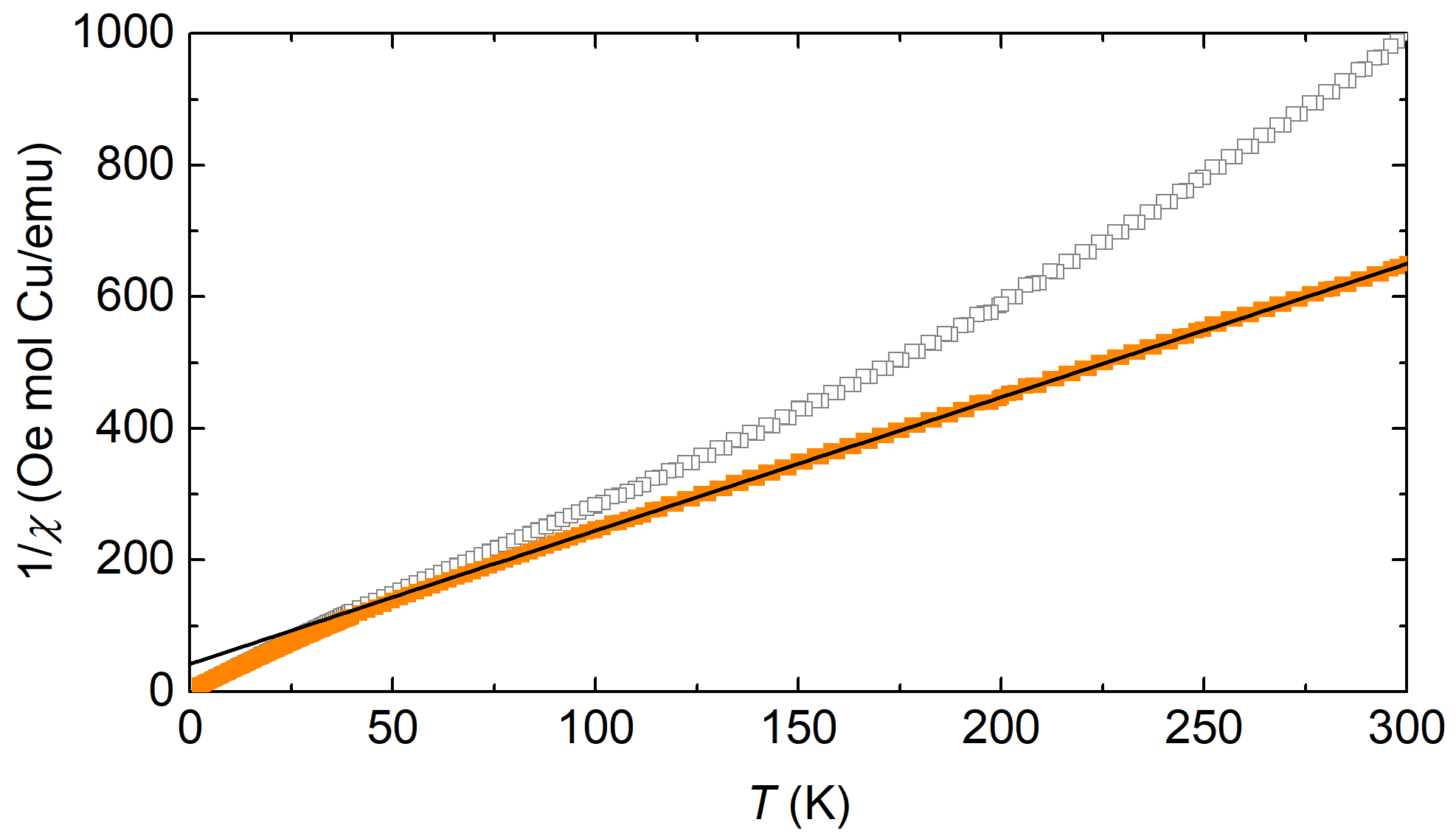}
	\caption{\label{fig:sg2}
		Inverse magnetic susceptibility as a function of temperature. In the data shown as open squares, contribution of ferromagnetic impurities has been subtracted, as described in the supplemental text. Data shown as solid squares are obtained by further adding \(5.39\times 10^{-4}\)\,emu/mol\,Cu/Oe, which is consistent with an estimated correction for diamagnetism. Solid line is a Curie-Weiss fit with parameters reported 
		in the main text.}
\end{figure}

Magnetization, \(M\), was measured as a function of temperature between 1.8 and \SI{300}{K} in a commercial SQUID magnetometer at \(H=0.1\), 3.5, 5, and \SI{7}{T}. Comparison of \(M/H\) at these fields indicate the presence of a temperature-independent ferromagnetic-impurity contribution, 0.83(7)\,emu/mol\,Cu, which can be accounted for by about 35\,ppm of Fe\(^{2+}\) impurities. This contribution has been subtracted from all magnetization data presented in the main text and in Figs.~\ref{fig:sg2} and \ref{fig:sg4}. 

Figure \ref{fig:sg2} shows the inverse magnetic susceptibility, \((M/H)^{-1}\), as a function of temperature, after the ferromagnetic-impurity contribution is subtracted. Here, the \SI{3.5}{T} data are used at temperatures above \SI{100}{K} and, to avoid nonlinearity, the \SI{0.1}{T} data below that temperature. In order to make the data obey the Curie-Weiss law at temperatures above \SI{100}{K}, we have found that it is necessary to add \(5.39(18)\times 10^{-4}\)\,emu/mol\,Cu/Oe to \(M/H\), 
consistent with a diamagnetic contribution expected from Pascal's constants~\cite{Pascal}, \(-5.77\times 10^{-4}\)\,emu/mol\,Cu/Oe. The difference between the two values, about 7\%, may be due to the Van Vleck paramagnetism of Cu\(^{2+}\) ions. Subsequently, \(5.39\times 10^{-4}\)\,emu/mol\,Cu/Oe has been added 
to all magnetization data presented in the main text and in Fig.~\ref{fig:sg4}. 
The Curie-Weiss fit yields the Curie constant and Curie-Weiss temperature reported in the main text. 

\begin{figure}[h]
	\centering
	\includegraphics[width=0.9\linewidth]{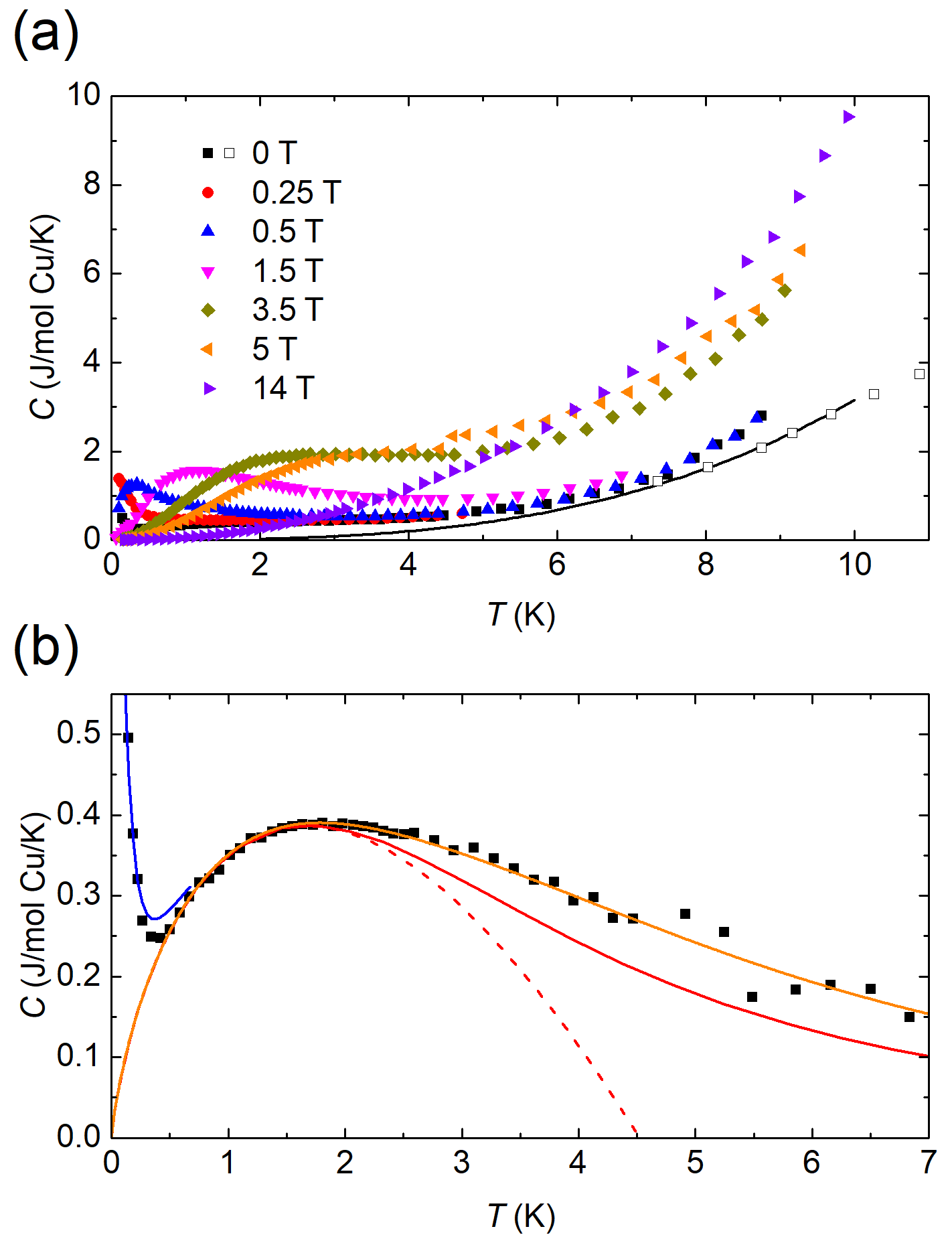} 
	\caption{\label{fig:sg3}(a) Total specific heat 
		at zero field and various magnetic fields. Solid symbols are data taken in a custom-built relaxation calorimeter for temperatures below \SI{10}{K}. 
		Open squares are zero-field data taken in another custom-built relaxation calorimeter that has been optimized for high-temperature measurements. (b) Zero-field specific-heat data after subtraction of the phonon contribution. Dashed line is an extrapolation of the fit \(C=T\ln(T_0/T)\), shown in Fig.\,2(b) of the main text. Solid lines are calculations using a spin-pair approximation described in the supplemental text. 
	}
\end{figure}

\begin{center}
	\bf{3. Total specific heat and subtraction of the phonon contribution}
\end{center}

Figure~\ref{fig:sg3}(a) shows the total specific heat of our sample 
as a function of temperature. 
As described in the main text, the specific heat at zero field was measured by using two calorimeters. 
Of the two sets of zero-field data, the one represented by open squares were taken with the calorimeter optimized 
to determine the phonon contribution with improved accuracy. The solid line is the best fit of 
those data to a $T^3$ dependence. 
This fit has been subtracted 
in the magnetic speific-heat data presented in the main text and in Fig.~\ref{fig:sg3}(b). 

\begin{center}
	\bf{4. Estimate of the nuclear-quadrupole specific heat}
\end{center}

As described in the main text, excess specific heat proportional to \(1/T^2\) 
is found at temperatures below about \SI{0.5}{K} in zero magnetic field. Here we show that this anomaly is too large to be the tail of a Schottky contribution of nuclear quadrupoles.  

The Hamiltonian for a nucleus with an electric quadrupole moment, \(Q\), is \cite{Abragam}
\begin{equation}
	\mathcal{H}=\frac{eQV_{zz}}{4I(2I-1)} [3I_z^2-I(I+1)+\frac{1}{2} \eta(I_+^2+I_-^2)],	
\end{equation}
where \(e\) is the elementary charge, \(V_{zz}\) a principal value of the electric field gradient (EFG) tensor at the nuclear site, \(I\) the nuclear spin quantum number, \(I_z\) the \(z\) component of the nuclear spin, \(\eta\) the anisotropy factor \(\eta=\frac{V_{xx}-V_{yy}}{V_{zz}}\), and \(I_+=I_x+iI_y\) and \(I_-=I_x-iI_y\). \(V_{xx}\) and \(V_{yy}\) are the two other principal values of the EFG tensor, and \(I_x\) and \(I_y\) are the \(x\) and \(y\) components of the nuclear spins. By convention, the three principal axes are chosen such that \(|V_{zz}|\geq|V_{yy}|\geq|V_{xx}|\). 

The nuclear-quadrupole specific heat of \ce{Sr21Bi8Cu2(CO3)2O41} is expected to be dominated by \(^{209}\)Bi because of its large \(Q\) of \SI{-516}{mb}, large \(I\) of \(9/2\), and high natural abundance of 100\% \cite{Pyykko}. In this compound, there are two nonequivalent Bi sites, Bi1 and Bi2, each surrounded by six O\(^{2-}\) to form a \ce{BiO6} octahedron \cite{Malo}. The \ce{BiO6} octahedron enclosing Bi1 is highly distorted, whereas that enclosing Bi2 is undistorted except for a \(21^{\circ}\) twist about a threefold axis. Out of the eight Bi in the chemical formula of the compound, six are located at Bi1 sites, and two at Bi2 sites. We have calculated the EFG at each Bi site due to the six O\(^{2-}\) by using a point-charge approximation. We find \(V_{zz}=4.99\times10^{20}\)\,V/m\(^2\) 
and \(\eta=0.80\) for Bi1 sites, and
\(V_{zz}=-1.34\times10^{20}\)\,V/m\(^2\) and 
\(\eta=0\) for Bi2 sites.  

These EFG parameters yield nuclear-quadrupole specific heats \(C_n=2.07 \times 10^{-5} /T^2\)\,mJ/K per mole of site-1 Bi and \(C_n=0.12 \times 10^{-5} /T^2\)\,mJ/K per mole of site-2 Bi, together resulting in \(C_n=6.33 \times 10^{-5} /T^2\)\,mJ/K per mole of Cu. Here, temperature \(T\) is in K. This \(C_n\) is five orders of magnitude smaller than the excess specific heat observed 
at temperatures below about \SI{0.5}{K} [see Fig.~\ref{fig:sg3}(b)], ruling out the possibility that the excess specific heat is due to a nuclear-quadrupole contribution.  

\begin{center}
	\bf{4. Calculations of magnetic specific heat, magnetic susceptibility, and magnetization}
\end{center}

Here we present calculations of magnetic specific heat, magnetic susceptibility, and magnetization using a spin-pair approximation \cite{Andres,SinghAppendix,KimchiPRX}. Treating a random-singlet state as a product state comprising independent spin pairs that experience only intra-pair exchanges, \(J\), with probability distribution \(P(J)\), this approximation allows calculations of these quantities by using canonical-ensemble formulas. For instance, the magnetic specific heat at zero field is given by 
\begin{equation}
	C=nR\int_{0}^{J_0}P(J)(\frac{J}{k_B T})^2\frac{3e^{J/k_B T}}{2(3+e^{J/k_B T})}dJ,	
\end{equation}
where \(n\) is the mole number of random-singlet spin pairs per mole of Cu, \(R\) the gas constant, and \(k_B\) the Boltzmann constant. \(J_0\) is a cutoff exchange energy. 

Shown in Fig.~\ref{fig:sg3}(b) are the specific-heat data at zero field, after subtracting the phonon contribution. The solid red line is the best fit calculated for \(P(J)=\ln (J_0/J)\), with \(J_0\)\,=\,14.6(3)\,K and \(n\)\,=\,0.210(3), that reproduces the \(T\ln (T_0/T)\) specific heat observed at temperatures between 0.5 and \SI{1.9}{K} [see Fig.~2(b) in the main text]. The calculation underestimates the specific heat at temperatures above \SI{1.9}{K}. To account for the excess specific heat in this temperature region, we find that a Schottky specific heat with a single \(J\) value of 14.5(5)\,K, with \(n\)\,=\,0.015(2), is required. The orange line in the figure indicates this contribution added to the 
\(\ln (J_0/J)\) contribution.

Excess specific heat is present also below \SI{0.5}{K}, as described in the main text. This excess fits the best to another Schottky contribution with a single \(J\) value of 0.150(2)\,K, which is two orders of magnitude smaller than the other single \(J\) value. The blue line indicates this third contribution added to the 
\(\ln (J_0/J)\) contribution. For this group of random singlets, \(n\) is 0.362(9), which is not a fitting parameter. Instead, it is automatically determined from the other two \(n\)'s and the total molar fraction 
of random singlets found from the vibrating-sample-magnetometer (VSM) data, as described in the main text. 

\begin{figure}[h]
	\centering
	\includegraphics[width=0.9\linewidth]{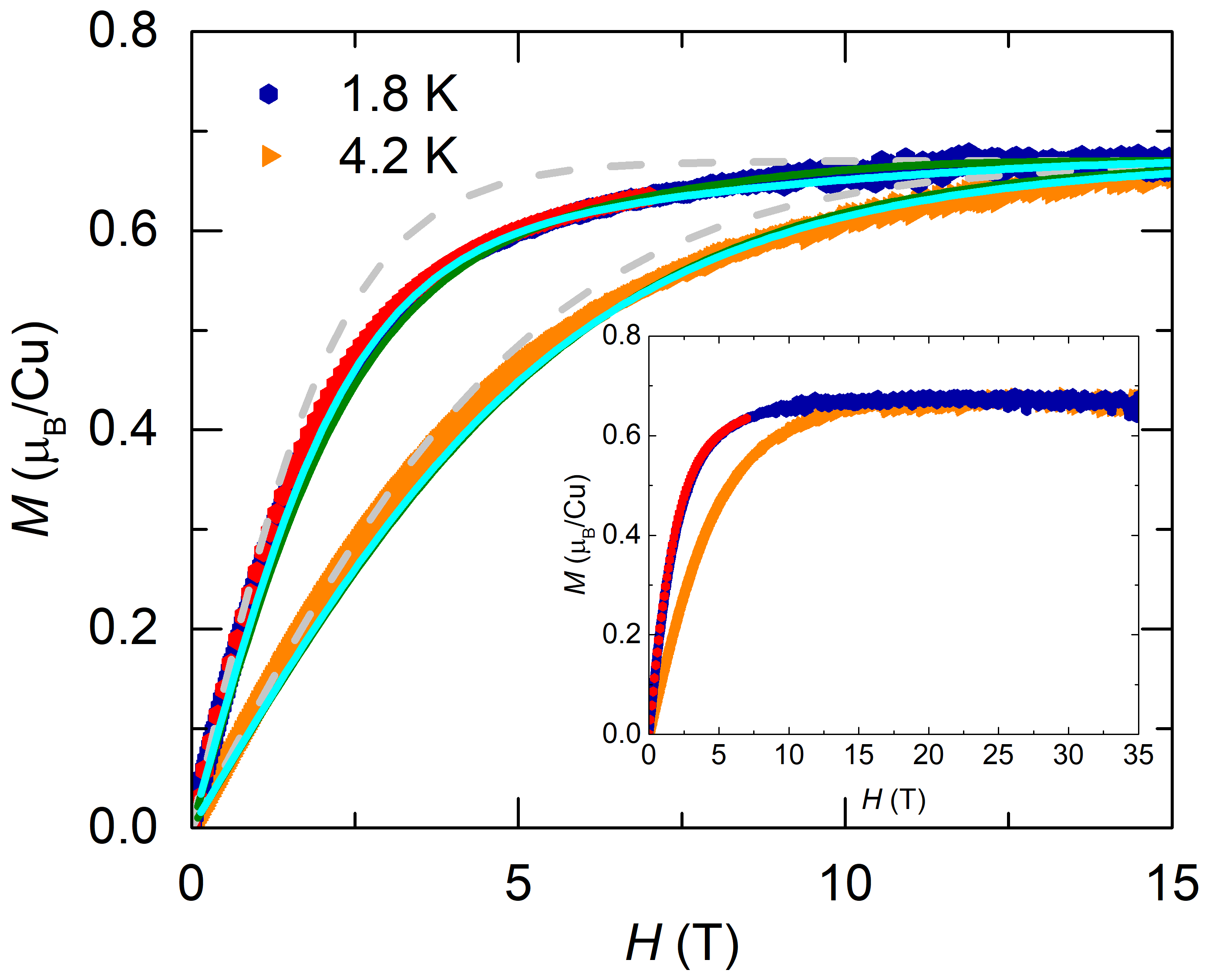} 
	\caption{\label{fig:sg4} Comparison of the VSM 
		data with calculations using the spin-pair approximation. Solid cyan lines are for \(P(J)\propto 1/J\), as described in the main text. Solid dark green lines are for \(P(J)\) extracted from zero-field specific heat data, with no adjustable parameters. Dashed lines are curves for free spins for comparison. Red circles are SQUID-magnetometer data at \SI{1.8}{K}, to which the VSM data have been scaled to improve accuracy. 
	}
\end{figure}

Calculated magnetization using the spin-pair approximation is compared with the VSM data in Fig.~\ref{fig:sg4}. The data and solid cyan lines, the best fit calculated for \(P(J)\)\,\(\propto\)\,\(1/J\), are from Fig.~1(b) in the main text.
Solid dark green lines are alternative calculations using the spin-pair approximation for \(P(J)\) extracted from the zero-field specific-heat data as described above---21.0(3)\% of spins with \(P(J)\)\,=\,\(\ln(J_0/J)\), 1.5(2)\% of spins with \(J\)\,=\,14.5(5)\,K, and 36.2(9)\% of spins with \(J\)\,=\,0.150(2)\,K---with no adjustable parameters. The calculations agree well with the data, albeit slightly less so 
than those for \(P(J)\propto 1/J\). 
Although, at high magnetic fields, the spin-pair approximation may not be valid for either \(P(J)\), the agreement between the data and calculations suggests consistency between the specific-heat data and the VSM magnetization data. 

\begin{figure}[h]
	\centering
	\includegraphics[width=0.9\linewidth]{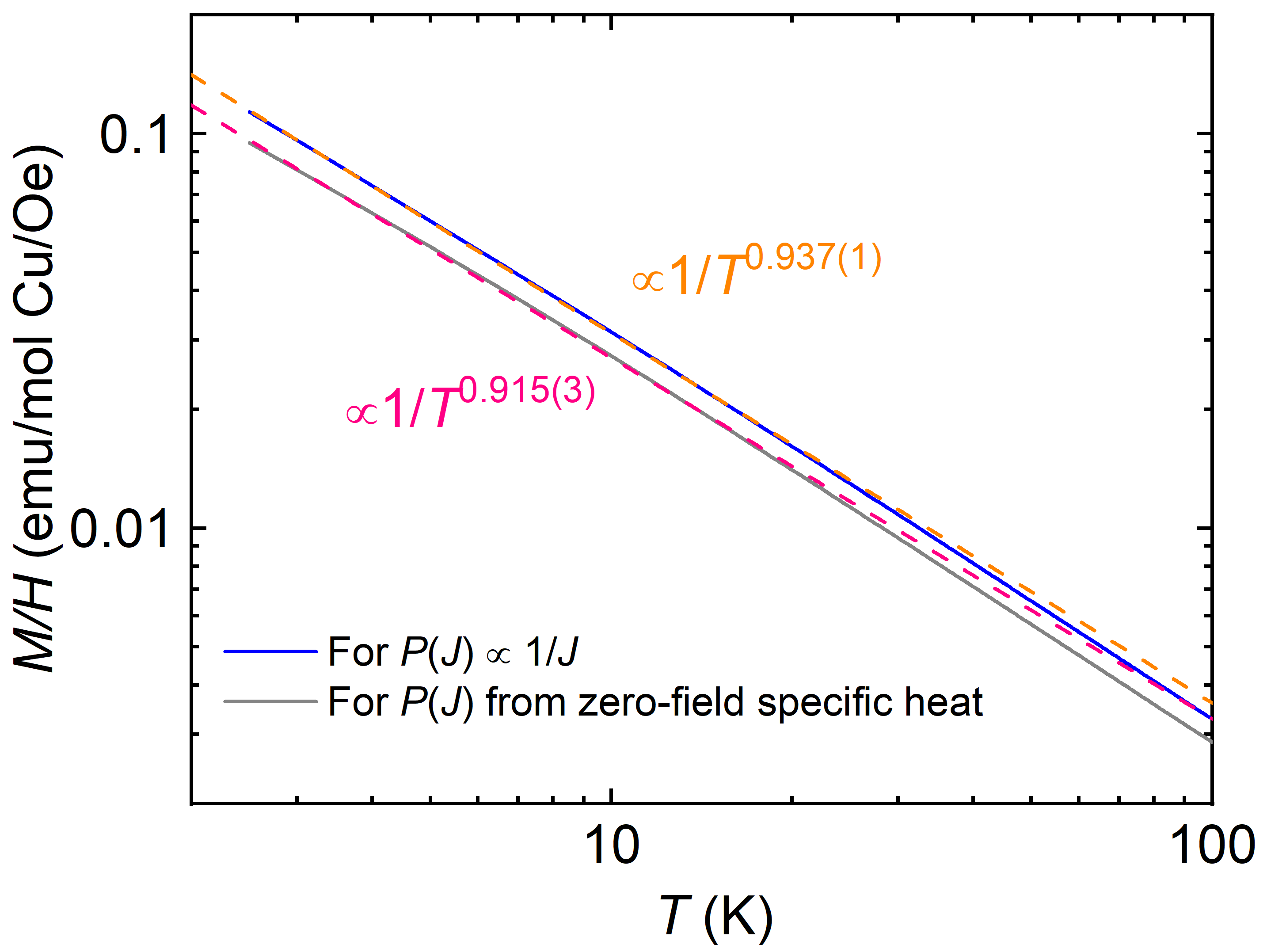}
	\caption{\label{fig:sg5}Calculated magnetic susceptibility, \(M/H\), as a function of temperature using the spin-pair approximation with experimentally determined \(P(J)\). The calculation is for \SI{0.1}{T}, the field at which the susceptibility was measured. Solid blue line: for \(P(J)\propto 1/J\) inferred from the scaling of high-field specific-heat data, with cutoffs determined by fitting calculated \(M\) to the VSM 
		data as a function of \(H\), as described in the main text. 
		Solid gray line: for \(P(J)\) extracted from zero-field specific-heat data. Dashed lines are power-law fits to 
		compare with a corresponding 
		fit to the experimental data, shown in Fig.~1(c) in the main text.}
\end{figure}

The spin-pair approximation is also used to calculate the magnetic susceptibility, \(M/H\), at \SI{0.1}{T} for comparison with the data shown in Fig.~1(c) in the main text. As shown in Fig.~\ref{fig:sg5}, for both \(P(J)\propto 1/J\) and \(P(J)\) extracted from zero-field specific-heat data, consisting of the logarithmic part and two \(\delta\)-function parts, the calculated susceptibility can be fit to a power law over one decade in temperature, with an exponents close to the experimental value, \(\gamma\)\,=\,0.955(1), 
supporting the notion that the magnetic-susceptibility data do not at all imply that \(P(J)\propto 1/J^{\gamma}\).

\begin{figure}[h]
	\centering
	\includegraphics[width=0.9\linewidth]{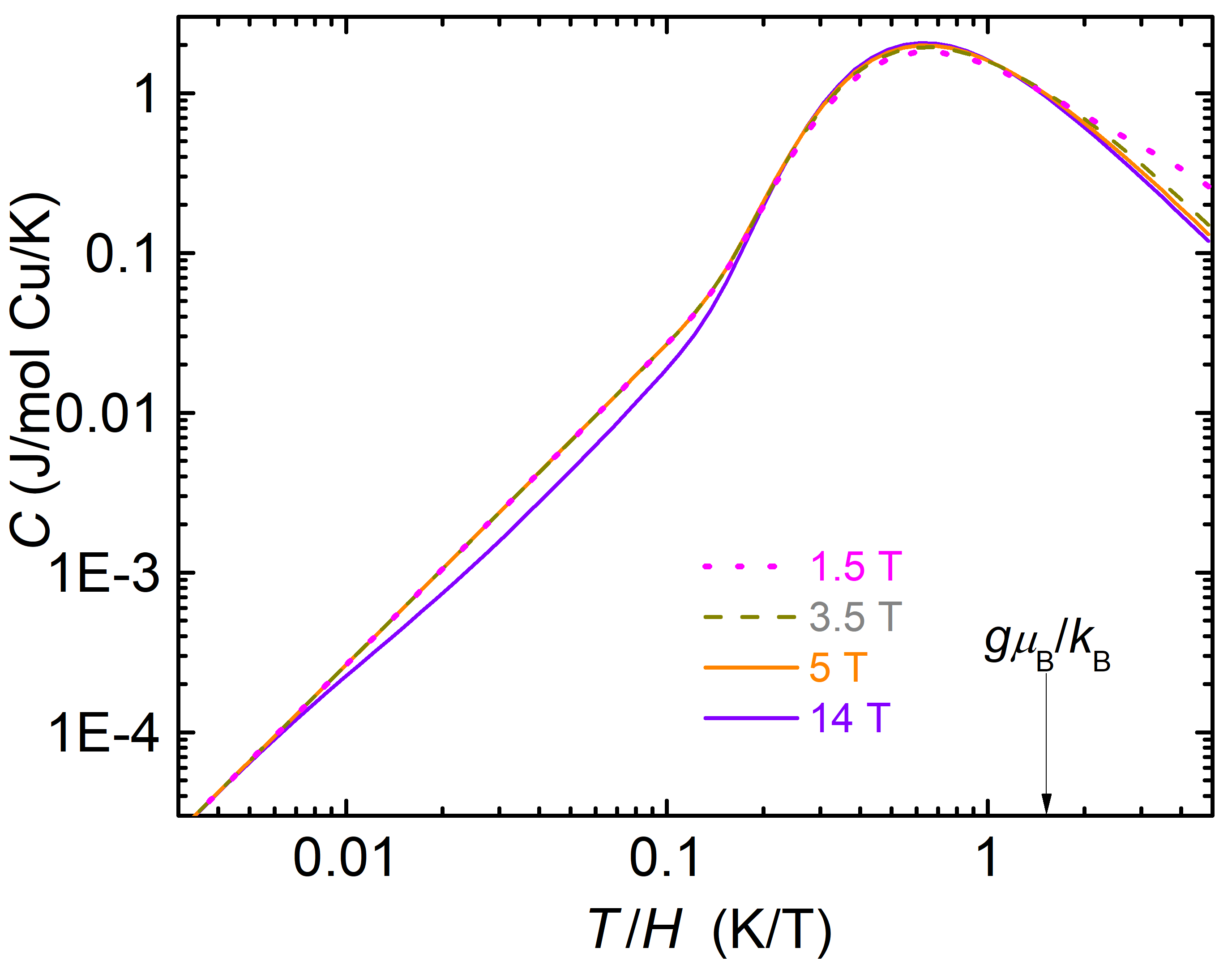} 
	\caption{\label{fig:sg6} Calculated magnetic specific heat, in magnetic fields, using the spin-pair approximation. The vertical arrow, labeled \(g\mu_B/k_B\), indicates the location of \(T/H\)\,=\,1 in the dimensionless unit. In this unit, the specific heat peaks at about \(T/H\)\,=\,0.41, similar to the specific heat of free spins.
	}
\end{figure}

Calculated specific heat in magnetic fields is shown in Fig.~\ref{fig:sg6}, using the spin-pair approximation for \(P(J)\propto 1/J\), with cutoffs \(J_H\)\,=\,22.0(5)\,K and \(\lambda\)\,=\,10\(^{-6}J_H\) determined from the best fit to the VSM data. To include the Dzyaloshinskii-Moriya (DM) interaction, we have assumed that the sample is a single crystal 
with the DM vector parallel to the \(x\) axis and the magnetic field parallel to the \(z\) axis, so that we obtain \(q\)\,=\,1 as was found in the specific-heat data, without performing an powder average. We have also assumed that the DM interaction has a Gaussian distribution centered around zero, with a standard deviation \(\sigma\)\,=\,\(0.5J\) and cutoffs at \(\pm J\).

The calculated specific heat collapses to a single scaling curve for \(T/H\)\,\(\leq\)\,\SI{0.2}{K/T} and becomes proportional to \((T/H)^2\) for \(T/H\)\,\(\leq\)\,\SI{0.1}{K/T}, except at \SI{14}{T}. The breakdown of scaling at this field arises from its proximity to the critical field \(J_H/g \mu_B\)\,=\,14.3(5)\,T, associated with the cutoff \(J_H\)\,=\,22.0(5)\,K. The calculated specific heat becomes proportional to \((T/H)^2\) only at about three times smaller  \(T/H\) than the experimental data do, with a value as small as about 23\% of the experimental value. These discrepancies we take to be a limitation of the spin-pair approximation. 

\begin{center}
	\bf{5. Magnetocaloric effect}
\end{center}

\begin{figure}[h]
	\centering
	\includegraphics[width=0.9\linewidth]{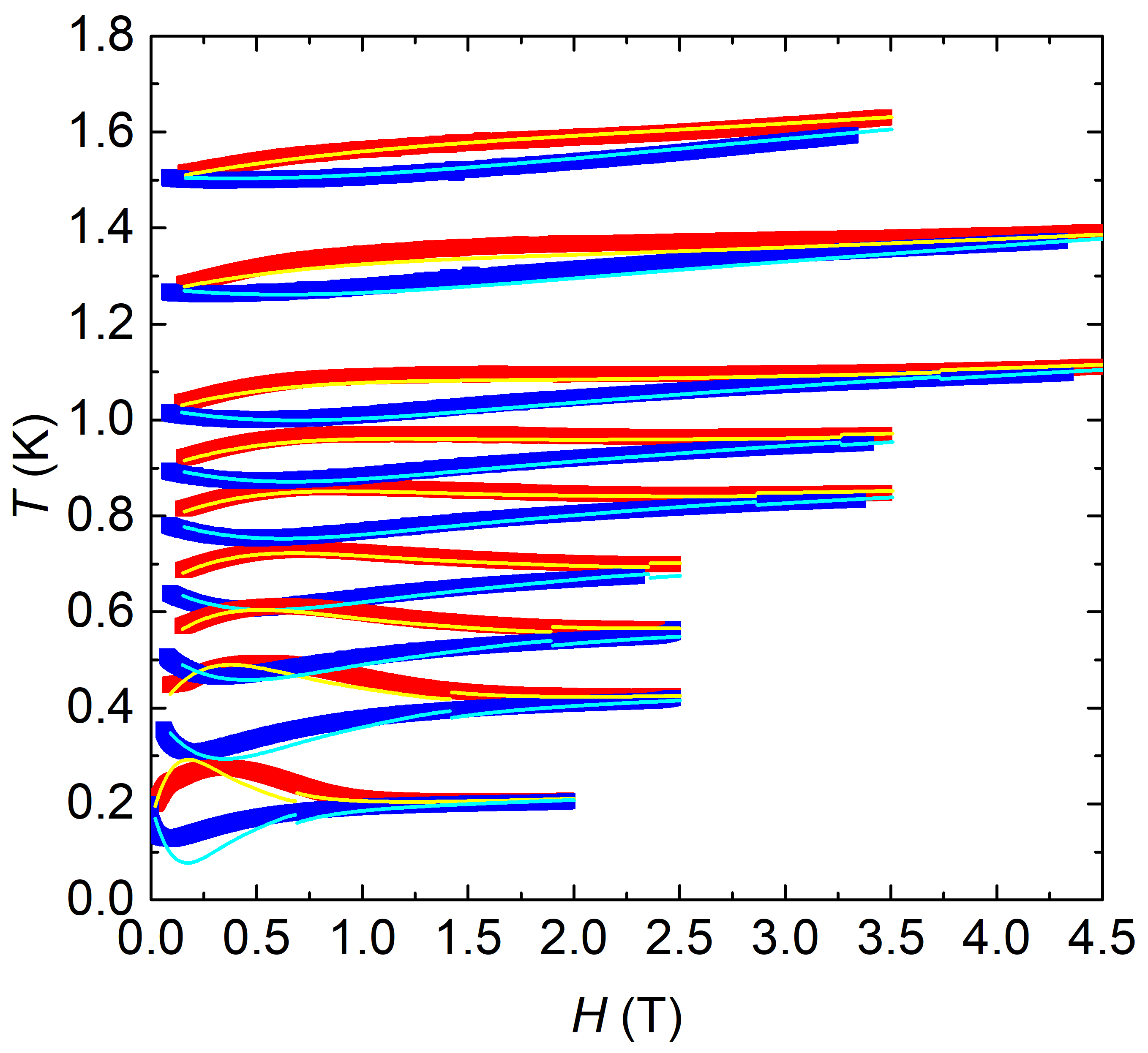}
	\caption{\label{fig:sg8}Magnetocaloric-effect data. The temperature of the sample during field sweeps 
		is plotted against the magnetic field. Solid red lines are for upward field sweeps, and solid blue lines for 
		downward field sweeps. Thin lines are calculations described in the supplemental text.}
\end{figure}

Magnetocaloric-effect data are shown in Fig.~\ref{fig:sg8}, where the sample temperature is plotted as a function of the magnetic field during field sweeps while the temperature of the thermal reservoir was roughly held constant by applying a constant electric current to a heater attached to the reservoir. The magnetic field was swept at the rate
of \SI{0.1}{T/min} except at the lowest temperature of about \SI{0.2}{K}, where the rate was reduced 
to \SI{0.05}{T/min} in order to minimize eddy-current heating of the calorimeter. 

In magnetocaloric-effect measurements, the temperature difference \(\Delta T\) between the sample and the thermal reservoir to which it is weakly coupled is given by \cite{SchevenMagcal,FortuneMagcal}
\begin{equation}\label{eq2}
	\Delta T = - \left[\frac{T}{\kappa}\left(\frac{\partial M}{\partial T}\right)_H +\frac{C}{\kappa} \frac{d(\Delta
		T)}{dH} \right]\dot H.
\end{equation}
where \(\kappa\) is the thermal conductance of the weak link between the sample and the thermal reservoir, \(C\) the total heat capacity of the sample and addenda, and \(\dot H\) the field-sweep rate. This relation allows us to calculate \(\Delta T\) and compare it with our magnetocaloric-effect data, with no adjustable parameters, since \((\partial M/\partial T)_H\) can be evaluated from our magnetization data or, equivalently, \((\partial S/\partial H)_T\) from our specific-heat data and \(\kappa\) has been separately measured. In the magnetocaloric-effect data, the second term, which produces a time-constant effect, is negligible except at the two lowest reservoir temperatures, 0.2 and \SI{0.4}{K}, where it manifests itself as asymmetry between an upward-sweep curve and a corresponding downward-sweep curve (see Fig.~\ref{fig:sg8}). 
We therefore ignore the second term.

For \(H/T\) less than \SI{3.5}{T/K}, where we have a power-polynomial fit to the scaled magnetization data, as described in the main text, we evaluate \((\partial M/\partial T)_H\) from the fit. For larger \(H/T\), we instead evaluate \((\partial S/\partial H)_T\) from the fit to \(C\) proportional to \((T/H)^2\) [see Fig.~2(c) in the main text]. The magnetization data from which \((\partial M/\partial T)_H\) is evaluated through the scaling fit are limited to temperatures down to \SI{2}{K}. On the other hand, the specific-heat data from which \((\partial S/\partial H)_T\) is evaluated through scaling deviate from the \((T/H)^2\) fit at and below \SI{1.5}{T}. Given that we have evaluated \((\partial M/\partial T)_H\) entirely below \SI{2}{K}, and \((\partial S/\partial H)_T\) down to \SI{1.5}{T} at about \SI{0.4}{K} and to \SI{0.7}{T} at about \SI{0.2}{K}, 
the calculation is in good agreement with the magnetocaloric-effect data.

\bibliography{Ref}